\documentclass[JHEP,12pt]{article}
\pdfoutput=1
\usepackage{tikzFigures}
\usepackage{floatrow}   
\usepackage{subcaption}
\usepackage{caption}
\usepackage[multiple]{footmisc}

\usepackage{graphics,color,soul}
\usepackage{graphicx}
\usepackage{amssymb}
\usepackage{jheppub}
\usepackage{lmodern,mathtools}
\usepackage{caption}
\usepackage {subcaption}
\usepackage{booktabs}
\usepackage[english]{babel}
\usepackage{amsmath,amssymb,amsbsy,amstext, amsthm, simplewick}
\usepackage{hyperref}
\usepackage{tikz}
\usetikzlibrary{decorations.markings}
\usetikzlibrary{decorations.pathmorphing}
\tikzset{snake it/.style={decorate, decoration=snake}}
\usepackage{xcolor}

\usepackage{caption}

\usepackage{amsfonts}
\usepackage{amssymb}
\usepackage{upgreek}
\usepackage{simplewick}
 \usepackage{exscale,relsize}
\usepackage{mathtools}
\usepackage{comment}

\usepackage[margin=1cm,labelfont={sf,bf,scriptsize},textfont={sf,scriptsize}]{caption}

\def \la {\langle}
\def \ra {\rangle}

\def\nn{{\nonumber}}

\DeclareRobustCommand{\Fig}[1]{Fig.~\ref{#1}}

\def\be#1\ee{\begin{align}#1\end{align}}

\begin{document}

\title{
Energy correlations and Planckian collisions
}

\author{Hao Chen$^{a}$,}
\author{Robin Karlsson$^{b}$,}
\author{Alexander Zhiboedov$^{b}$}

\affiliation{$^a$Zhejiang Institute of Modern Physics, School of Physics, Zhejiang University, Hangzhou, 310027, China}
\affiliation{$^b$CERN, Theoretical Physics Department, CH-1211 Geneva 23, Switzerland}

\abstract{Energy correlations characterize the energy flux through detectors at infinity produced in a collision event. Remarkably, in holographic conformal field theories, they probe high-energy gravitational scattering in the dual anti-de Sitter geometry. We use known properties of high-energy gravitational scattering and its unitarization to explore the leading quantum-gravity correction to the energy-energy correlator at strong coupling. 
We find that it includes a part originating from large impact parameter scattering that is non-analytic in the angle between detectors and is $\log N_c$ enhanced compared to the standard $1/N_c$ perturbative expansion. It is sensitive to the full bulk geometry, including the internal manifold, providing a refined probe of the emergent holographic spacetime. Similarly, scattering at small impact parameters leads to contributions that are further enhanced by extra powers of the 't Hooft coupling assuming it is corrected by stringy effects. We conclude that energy correlations are sensitive to the UV properties of the dual gravitational theory and thus provide a promising target for the conformal bootstrap.
}

\begin{flushleft}
\hfill \parbox[c]{40mm}{CERN-TH-2024-050}
\end{flushleft}
\maketitle



\section{Introduction}
In a collider experiment, energy correlations are familiar observables that characterize energy flux at infinity \cite{Basham:1978zq,Basham:1978bw}. Originally computed using scattering amplitudes, they were later expressed in terms of integrated correlation functions \cite{Sveshnikov:1995vi,Korchemsky:1997sy,Korchemsky:1999kt,Belitsky:2001ij}. This latter definition in terms of correlation functions readily generalizes to theories where the amplitude approach is not available, such as in strongly coupled holographic CFTs \cite{Hofman:2008ar}, which are also the subject of the present paper.

From a more formal point of view, energy correlations present an interesting example of the matrix elements of light-ray operators \cite{Hofman:2008ar,Kravchuk:2018htv}, allowing exploration of their properties, such as positivity \cite{Hartman:2016lgu,Faulkner:2016mzt,Hofman:2016awc,Komargodski:2016gci,Hartman:2023qdn}, the operator product expansion \cite{Kologlu:2019mfz,Chang:2020qpj}, renormalization and mixing \cite{Caron-Huot:2022eqs}, connection to dispersion relations \cite{Kologlu:2019bco,Caron-Huot:2021enk,Chang:2023szz}. They also received much attention recently in QCD as they occupy a middle ground between scattering amplitudes and correlation functions exhibiting properties of both \cite{Dixon:2018qgp,Luo:2019nig,Chicherin:2020azt,Dixon:2019uzg,Moult:2018jzp}. One particularly interesting property of the energy correlators is their universal scaling behavior in the collinear limit, which is the consequence of QCD factorization \cite{Dixon:2019uzg,Chen:2023zzh} and the light-ray OPE \cite{Hofman:2008ar,Kologlu:2019mfz,Chang:2020qpj}. This theoretical understanding and further phenomenological studies \cite{Komiske:2022enw,Chen:2020vvp,Chen:2023zlx} have recently led to an experimental measurement of energy correlations inside a jet at the LHC \cite{CMS:2023wcp,CMS:2024mlf,Tamis:2023guc}.

\begin{figure}
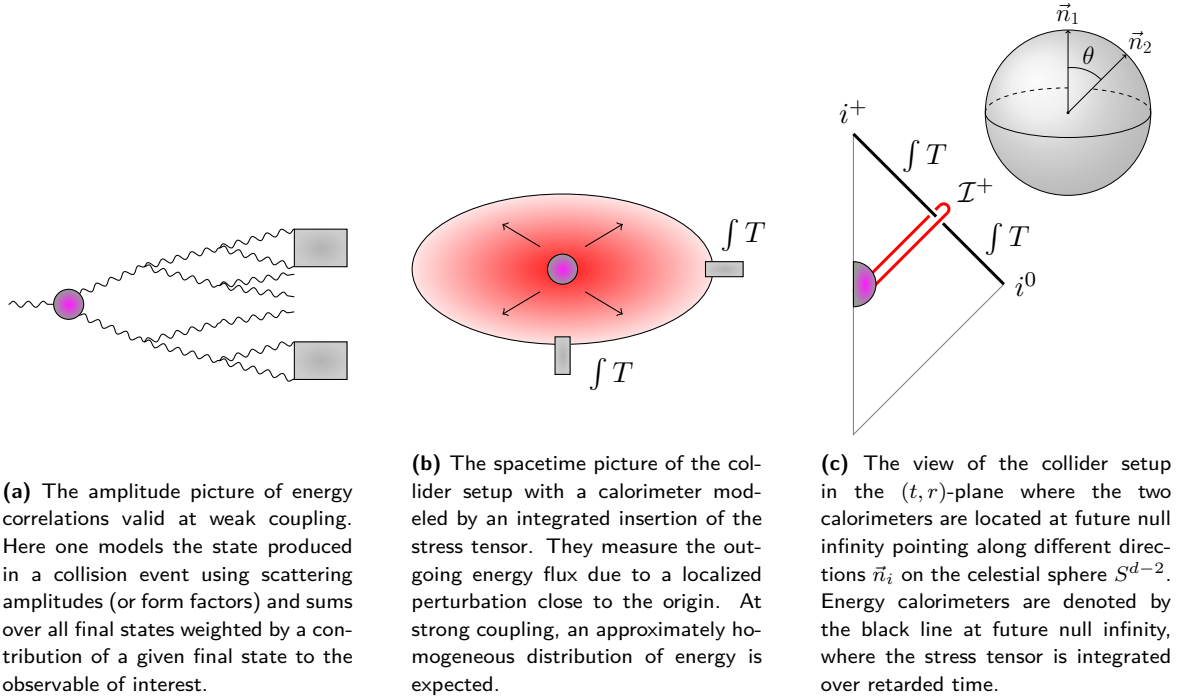
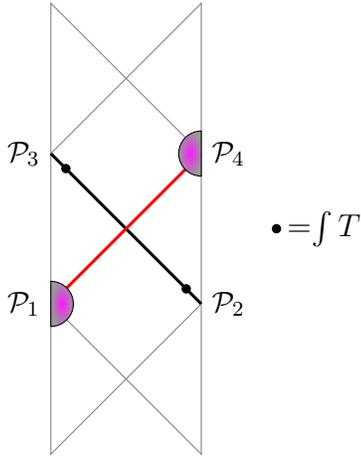
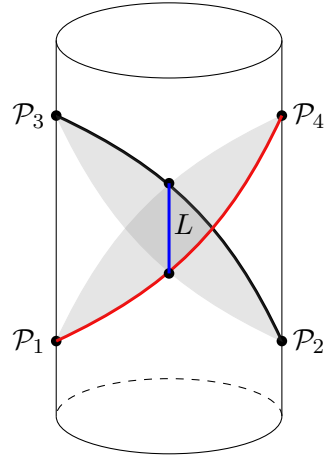

\centering
\begin{subfigure}{.3\linewidth}
    \centering
    \amplitudes
    \caption{The amplitude picture of energy correlations valid at weak coupling. Here one models the state produced in a collision event using scattering amplitudes (or form factors) and sums over all final states weighted by a contribution of a given final state to the observable of interest.}\label{fig:image1}
\end{subfigure}
    \hfill
\begin{subfigure}{.3\linewidth}
    \centering
    \collider
    \caption{The spacetime picture of the collider setup with a calorimeter modeled by an integrated insertion of the stress tensor. They measure the outgoing energy flux due to a localized perturbation close to the origin. At strong coupling, an approximately homogeneous distribution of energy is expected.}\label{fig:image12}
\end{subfigure}
   \hfill
\begin{subfigure}{.3\linewidth}
    \centering
    \fold
    \caption{ 
    The view of the collider setup in the $(t,r)$-plane where the two calorimeters are located at future null infinity pointing along different directions $\vec{n}_i$ on the celestial sphere $S^{d-2}$. Energy calorimeters are denoted by the black line at future null infinity, where the stress tensor is integrated over retarded time. }\label{fig:image13}
\end{subfigure}

\bigskip
\begin{subfigure}{0.45\linewidth}
  \centering
  \regge
  \caption{In CFTs, where the source and the sink are modeled by a local operator acting on the vacuum, we can `unfold' the collider setup in (c) to relate it to the Regge limit of a four-point function where the source and sink are inserted close to ${\cal P}_{1,4}$, whereas detectors approach ${\cal P}_{2,3}$. The stress tensors are denoted by black dots and are integrated over the black line. Here we suppress $(d-2)$ transverse dimensions.}\label{fig:image3}
\end{subfigure} 
\hspace{10mm}
\begin{subfigure}{0.38\linewidth}
  \centering
  \Cyltwo
  \caption{In holographic CFTs the Regge limit is further captured by high-energy ($S$) and fixed impact parameter ($L$) scattering in the AdS bulk. In particular, at large impact parameters, it is universally determined by the graviton exchange with a characteristic linear growth of the bulk phase shift $\delta(S,L)$ in energy $S$.}\label{fig:image33}
\end{subfigure} 
\RawCaption{\caption{From a collider experiment to high-energy gravitational scattering in AdS. Here we present different ways to view the energy-energy correlator in  CFTs and their relationship to each other.}
\label{fig:EEC}}
\end{figure}

In this paper, we explore a fascinating connection between energy correlations and high-energy gravitational scattering (as well as  its unitarization) \cite{Amati:1987uf,Cornalba:2006xk,Cornalba:2006xm,Cornalba:2007zb,Giddings:2007bw,Giddings:2009gj,DiVecchia:2023frv} in the context of holography \cite{Maldacena:1997re,Gubser:1998bc,Witten:1998qj}. Throughout the paper, we focus on the simplest example of this observable, namely the two-point energy correlator, which involves two detectors at infinity, see Figure \ref{fig:EEC}. Weakly coupled dynamics in this context typically produces jets \cite{Sterman:1977wj}, or energy fluxes collimated in a small angular cone, while a hallmark of strong coupling dynamics is a homogeneous energy distribution  \cite{Hofman:2008ar}.\footnote{One can get approximately homogeneous energy distribution at weak coupling as well by considering states with many particles \cite{Chicherin:2023gxt,Firat:2023lbp}. It was also observed in QCD data at extremely small angles \cite{Komiske:2022enw,CMS:2023wcp,CMS:2024mlf}.} There is another possibly less familiar feature of energy correlators at strong coupling, something we can call \emph{the enhancement property}. It states that perturbative corrections to the leading homogeneous distribution of energy are \emph{larger} than naively expected from the perturbative expansion of the correlator of local operators from which the energy correlators are computed.  Let us illustrate this with a simple example of the four-point function used to compute the two-point energy correlator (EEC) in ${\cal N}=4$ SYM. At strong coupling, we schematically have for the connected part
\be
\label{eq:correlator}
\langle {\cal O}^\dagger T_{\mu \nu} T_{\rho \sigma} {\cal O} \rangle_c = \text{Tree-level Sugra} +  \textcolor{blue}{{1 \over \lambda^{3/2}}} \text{Stringy} + \textcolor{red}{{1 \over c_T}} \text{One-loop Sugra} + ... \ ,
\ee
where $\lambda \gg 1$ is the 't Hooft coupling, and $c_T\gg1$ is the central charge.\footnote{We define $c_T\sim 1/G_N$, where $G_N$ is Newton's constant of the dual gravitational theory in AdS, in Appendix \ref{App:Conventions}. For $SU(N_c)$ ${\cal N}=4$ SYM it is $c_T=40(N_c^2-1)$.}$^{,}$\footnote{The $1/c_T$ correction that we wrote is universal, but not entirely precise. In the explicit examples dual to string/M-theory UV completions of gravity, the leading $1/c_T$ correction comes from an AdS contact term, see e.g.\ \cite{Binder:2019mpb,Chester:2019pvm,Chester:2020dja}. We will focus in this paper on the universal corrections that capture scattering at non-zero impact parameters in AdS for which \eqref{eq:correlator}  applies.} 

Let us next introduce the detector operator which is given by the light-transform of the stress tensor at future null infinity
\be\label{eq:introEDef}
{\cal E}(n) \equiv {1 \over 4}\int_{-\infty}^{\infty} d u  \lim_{r \to \infty} r^{d-2} T_{\mu \nu}(u,r \vec n) \bar n^\mu \bar n^{\nu},
\ee
where $u = t-r$ is the retarded time at future null infinity, and we defined the null vectors $n^\mu = (1, \vec n)$, $\bar n^\mu = (1, - \vec n)$, and the unit vector $\vec n^2 = 1$ to specify the position of the detector on the celestial sphere $S^{d-2}$, see Figure \ref{fig:EEC}.

The main object studied in this paper is the energy-energy correlator in a unit-normalized momentum eigenstate (source) $|p\ra = \frac{1}{\sqrt{{\cal N}_p}}\int d^dx e^{ip \cdot x}{\cal O}(x)| \Omega \rangle$:
\be
\label{eq:defEEC}
\la {\cal E}(n_1) {\cal E}(n_2)  \ra \equiv \la p|{\cal E}(n_1){\cal E}(n_2)|p\ra.
\ee 
Compared to four-point function of local operators \eqref{eq:correlator}, the leading correction to the energy-energy correlator, computed from \eqref{eq:correlator}, takes the form \cite{Hofman:2008ar,Goncalves:2014ffa}
\be
\label{eq:grstringy}
\la {\cal E}(n_1) {\cal E}(n_2)  \ra = \left({p^0 \over 4 \pi} \right)^2 \Big[ 1 + \textcolor{blue}{{1 \over \lambda}} 4 \pi^2 (1-6 z + 6 z^2) + ... \Big],
\ee
where $z = {1 - \cos \theta \over 2}$ encodes the angle between the energy calorimeters $\cos \theta \equiv \vec n_1 \cdot \vec n_2$ and we assumed that the source carries momentum $p^\mu = (p^0, \vec 0)$ for simplicity. We see that the leading stringy correction to the energy-energy correlator is enhanced by an extra factor of $\sqrt{\lambda}$.\footnote{There are further nonperturbative stringy corrections to the energy-energy correlator of the type $\sim z^{\# \lambda^{1/4}}$ which are exponentially suppressed at strong coupling $\lambda \gg 1$.}

The enhancement of corrections to the energy correlations originates from the fact that they probe \emph{high-energy scattering} in the dual AdS geometry. Sensitivity to the high-energy behavior of the scattering amplitude in the bulk makes the computation of the correction to the energy correlations at strong coupling an interesting and subtle problem. By naively using the first stringy correction to the correlator \eqref{eq:correlator} to compute the energy-energy correlator, one  gets infinity! To get a finite result, one must effectively perform the computation at finite $\lambda$ first --- it makes the high-energy behavior of the scattering amplitude in AdS softer --- and only then perform the large $\lambda$ expansion \cite{Hofman:2008ar,Goncalves:2014ffa}. We can call it \emph{stringy unitarization}, and it is related to scattering at small impact parameters. In contrast, this paper explores the enhancement phenomenon in the context of \emph{gravitational unitarization}, i.e.\ unitarization of scattering at large impact parameters by resummation of gravitational loops in AdS.

At the level of the correlation function of local operators, gravitational loops are suppressed by the inverse number of colors in the gauge theory or, equivalently, the central charge $1/c_T$ \cite{tHooft:1973alw,Aharony:2016dwx}. Again, trying to use the leading $1/c_T$ correction to the correlator \eqref{eq:correlator} recently found in \cite{Aprile:2017bgs,Alday:2017xua} to compute the leading $1/c_T$ correction to the energy-energy correlator leads to infinity due to the high-energy behavior of gravitational scattering in AdS. One can trace the divergence to the integral over energies of the squared tree-level phase shift $\delta^2_{\text{tree}}(S,L)$, i.e.\ the integral $\int^\infty {dS \over S^3} \delta_{\text{tree}}^2(S,L)$, where $S$ is the scattering energy and $L$ is the impact parameter. Indeed, it is a characteristic feature of gravitational theories that $\delta_{\text{tree}}(S,L) \sim G_N S$ which produces the logarithmic divergence in the integral over energies.\footnote{For general detectors of spin $J_1$ and $J_2$ the measure becomes $\int {d S \over S^{J_1 + J_2 -1}}$.} It is, however, a well-known fact that such a divergence is an artifact of perturbation theory, and in the simplest case of elastic scattering regime the amplitude `unitarizes' such that we instead get $\int^\infty {dS \over S^3} \sin^2 \Big( {\delta_{\text{tree}}(S,L) \over 2} \Big)$. Performing the integral first and expanding the result as $G_N \to 0$, we get a finite contribution $G_N^2 \log G_N$. 

The main subtlety of the argument above comes from the fact that it is a \emph{small}, but \emph{finite} $G_N$ computation. We present arguments based on unitarity and expected properties of a gravitational EFT that the computation is still under control. Another important point is that the computation of the energy-energy correlator involves integration over \emph{all} impact parameters (including zero impact parameter which is sensitive to the details of the UV completion of a gravitational theory) for which the simple model of elastic scattering does not apply. We will see that in general the small impact parameter contribution leads to an additional enhancement of the correction to the energy-energy correlator, which we model using stringy effects. Scattering at zero impact parameter instead contributes a polynomial in $z$ correction to the energy-energy correlator (similar to the stringy corrections in \eqref{eq:grstringy}). In contrast,  large impact parameter scattering leads to non-analytic at small $z$ contributions.

Performing the computation, we get for the universal quantum-gravity correction to the energy-energy correlator at small angles $z\approx \theta^2/4$: 
\be
\label{eq:universalAsy}
\la {\cal E}(\theta) {\cal E}(0)  \ra_{\text{QG}} =\gamma_d \left(\frac{p^0}{\Omega_{d-1}}\right)^2
\textcolor{red}{{\log c_T \over c_T}} \log {1 \over \theta^2} ,  ~~~ \theta \to 0, 
\ee
where $\Omega_d={2\pi^{d/2}\over \Gamma(d/2)}$ is the volume of a $(d-1)$-dimensional unit sphere and $\gamma_d$ is given in \eqref{eq:finalGR}.  
Compared to the stringy correction \eqref{eq:grstringy}, it is notably non-analytic around $\theta=0$. As we explain below, this is related to the fact that it comes from \emph{large impact parameter} scattering in the bulk and the universality of the eikonal phase shift in gravity \cite{Amati:1987uf}. Notice also that the strong coupling expression above does not depend on the source, expressing the universal nature of gravity. 

Making further assumptions about the dual geometry, we can compute the leading non-planar correction to the energy-energy correlator exactly. For example, for pure Einstein gravity in 
$AdS_{4}$ dual to a putative $CFT_3$ we get
\be\label{eq:GRAds4}
\langle {\cal E}(n_1){\cal E}(n_2) \rangle_{AdS_4}=& - \frac{96}{\pi^2} \left({p^0 \over 2 \pi} \right)^2{\log c_T \over c_T}\Big[(5-8 z) \log (4 z) + 8 (1-z)\Big] .
\ee 
In the same way, we get for pure Einstein gravity in $AdS_5$ dual to a a putative $CFT_4$
\be
\label{eq:GRAds5}
&\langle {\cal E}(n_1){\cal E}(n_2)\rangle_{AdS_5}=\frac{45}{64} \left( \frac{p^0}{4 \pi} \right)^2 {\log c_T \over c_T} \\
&\times {1 \over z^2}\left(6 \left(-3+z-65 z^2+67 z^3\right)+\left(9-z-\frac{1475 z^2}{3}+1225
   z^3\right) \log z \right. \cr
   &\left. +\frac{\left(9-4z+54 z^2-900 z^3+1225 z^4\right)
   \left(\text{Li}_2\left(\sqrt{z}\right) -\text{Li}_2\left(-\sqrt{z}\right)+\frac{1}{2} \log
   \left(\frac{1-\sqrt{z}}{1+\sqrt{z}}\right) \log (z)\right)}{\sqrt{z}} \right) \ , \nn
\ee
where the $\sqrt{z}$ dependence is spurious and the small $z$ dependence is again of the type $\sum_n z^n(a_n+b_n \log z)$. Let us note that the underlying four-point correlation functions of local operators which could be used to compute \eqref{eq:GRAds4} and \eqref{eq:GRAds5} are not known explicitly, rather we obtained the results above by arguing that they come solely from the regime dual to high-energy scattering at fixed impact parameter in the bulk.

The result in general $d$ for $AdS_{d+1}/CFT_d$ can be found in Section \ref{Sec:Einstein}, see \eqref{eq:GRres} and the discussion around it. An important novelty in $d>4$ is that in addition to the high-energy enhancement, there is an extra enhancement that comes from physics at small impact parameters. In this paper, we model it using stringy effects which, similar to \eqref{eq:correlator} and \eqref{eq:grstringy}, lead to contributions that are enhanced by the string scale $\Delta_{\text{gap}} \sim {R_{AdS} \over \ell_s}$. For $D$-dimensional bulk geometries of the type $AdS_{d+1}\times X$, the same statement about the additional enhancement coming from unitarization of small impact parameter scattering holds as long as $D>5$. In particular, it holds for string/M-theory UV completions of gravity for which we have $D=10$ and $D=11$ correspondingly.  

For the case of ${\cal N}=4$ SYM, for which the dual geometry is $AdS_5 \times S^5$ we instead get
\be
\label{eq:N4res}
\la {\cal E}(n_1) {\cal E}(n_2)  \ra_{AdS_5 \times S^5} &= -384 \left({p^0 \over 4 \pi} \right)^2 {\log c_T \over c_T} \log z \Big( 1-36 z+ 216 z^2 - 400 z^3 + 225 z^4 \Big) + \cdots \ ,  
\ee
which up to the precise coefficient was anticipated in \cite{Korchemsky:2021okt}. In addition to the `non-local' $\sim \log z$ contribution \eqref{eq:N4res}, there are also polynomial corrections denoted by $\cdots$, enhanced by powers of $\sqrt{\lambda}$, similar to the stringy corrections \eqref{eq:grstringy}, which we omit here and discuss in the main text. The leading $\log z$ term of \eqref{eq:N4res} as $z\to0$ agrees with the one of \eqref{eq:GRAds5} and \eqref{eq:universalAsy} due to the universality of the graviton exchange in the bulk at large impact parameters. The terms $z^{n} \log z$ are, on the other hand, different due to the fact that in the case of ${\cal N}=4$ there is an additional contribution from the $S^5$ KK modes. Therefore, by measuring the coefficient of ${z^n \log z \log c_T \over c_T}$ term in the energy correlator we can probe the size of the internal manifold.

To make the connection to the problem of high-energy scattering in the bulk more explicit, we notice that there is an exact symmetry of the correlator that relates the energy correlation computation to the bulk scattering computation
\be
\label{eq:sourceunfold}
\la  {\cal E}(n_1) {\cal E}(n_2) \ra \equiv {\la {\cal O}^\dagger {\cal E}(n_1) {\cal E}(n_2) {\cal O} \ra_{\text{fold}} \over \la {\cal O}^\dagger {\cal O} \ra_{\text{fold}} } = {\la {\cal O}^\dagger {\cal E}(n_1) {\cal E}(n_2) {\cal O} \ra_{\text{unfold}} \over \la {\cal O}^\dagger {\cal O} \ra_{\text{unfold}} } ,
\ee
as depicted in \Fig{fig:EEC}. Here by `fold' we mean that the correlator is evaluated on the Schwinger-Keldysh contour, or a time-fold, see \Fig{fig:image13}. The `unfold' stands for the normal time-ordered correlator, see \Fig{fig:image3}. Unfolding, where an operator acting on the vacuum is moved a Poincaré patch forward leaves the conformal cross-ratios invariant. The conformal tensor structures that multiply the non-trivial functions of conformal cross-ratios in CFT correlation functions \cite{Costa:2011mg}, on the other hand, acquire a nontrivial overall phase which however cancels between the numerator and the denominator in \eqref{eq:sourceunfold} leaving the ratio invariant, see \cite{Kravchuk:2018htv,Caron-Huot:2022lff} and Appendix \ref{app:fold}.\footnote{More precisely, \eqref{eq:sourceunfold} holds in even dimensions and for conformal parity-even structures in odd dimensions. For parity-odd conformal structures in odd dimensions, there could be an extra minus sign under the unfold transformation, see Appendix \ref{app:fold}. In this paper, we will only consider parity-even structures.} We then notice that there is a region of the retarded time integration in \eqref{eq:introEDef} that is sensitive to high-energy bulk scattering. It corresponds to the limit when the detector integration times $u_i \to \pm \infty$ with opposite signs, see \Fig{fig:image3} and Figure 6 in \cite{Kologlu:2019bco}. For weakly coupled theories, the contribution from the high-energy scattering region cannot be easily isolated from the rest of the energy correlator. However, as we show in this paper, in holographic theories, high-energy bulk scattering produces a distinctive contribution to the energy-energy correlator which is enhanced compared to the naive perturbative expansion of the correlator.

Conceptually, the difference between our work and the computation of the leading stringy correction is that  \eqref{eq:grstringy} requires considering the high-energy behavior of string tree-level amplitudes, whereas the leading ${1 \over c_T}$ correction requires considering high-energy gravitational scattering at finite $G_N$. A convenient way to characterize this kinematical regime is via the bulk phase shift $\delta(S,L)$ \cite{Cornalba:2006xm,Cornalba:2008sp,Cornalba:2009ax,Kulaxizi:2017ixa}, where $S$ is the scattering energy and $L$ is the impact parameter. In terms of the phase shift, the high-energy scattering contribution to the energy-energy correlator takes the following form\footnote{We use mostly-plus signature.}
\be
\label{eq:correctionQG}
\la {\cal E}(n_1){\cal E}(n_2)\ra = \frac{8\rho_T c_T\Gamma(d+1)^2 (-p^2)^d}{(-p\cdot n_1)^{d-1}(-p\cdot n_2)^{d-1}} \int_0^1 d u {\cal K}_d (u,z) {\rm Re} \langle 1 - e^{i \delta}\rangle,
\ee
where $u=\tanh L$ and we introduced the notation 
\be
\label{eq:phaseshiftaverageEEC}
\langle X \rangle  = \int_{S_0}^\infty {d S \over S^3}X(S,L) |_{\tanh L = u} \ .
\ee
Here ${\cal K}_d (u,z)$ is a kinematical kernel that is given in Appendix \ref{App:Kernel} and $\rho_T$ is given in \eqref{eq:GVacuum}.\footnote{Curiously, ${\mathcal K}_d (u,z)$ turns out to be closely related to the Mandelstam kernel that appears in the study of double discontinuity in gapped theories \cite{Correia:2020xtr}.} Our results are independent of the arbitrary fixed lower integration limit $S_0$, which leads to the sub-leading $O(1/c_T)$ contribution to the energy-energy correlator. The correction \eqref{eq:correctionQG} automatically satisfies the  momentum conservation Ward identities \cite{Dixon:2019uzg,Korchemsky:2019nzm,Kologlu:2019mfz}, see Appendix \ref{app:WardId}, thanks to the following properties of the kernel ${\cal K}_d (u,z)$
\be\label{eq:WardIntro}
\int_0^1 d z (z (1-z))^{{d-4 \over 2}}\ {\cal K}_d (u,z) = \int_0^1 d z (z (1-z))^{{d-4 \over 2}} \ z \ {\cal K}_d (u,z) = 0 \ . 
\ee
The structure of the kernel ${\cal K}_d (u,z)$ is such that the expansion of the phase shift at large impact parameter $L \to \infty$ is mapped to terms which are \emph{non-analytic} when $z \to 0$. This is in contrast to the terms that are encoding point-like scattering at zero impact parameter, such as the stringy correction in \eqref{eq:grstringy}, which produce terms that are \emph{polynomials} in $z$. 

At high energies and large impact parameters, the phase shift takes the universal and simple form
\be
\label{eq:eikonal}
\delta (S,L) = 4 \pi G_N S \ \Pi(L) ,
\ee
where $\Pi(L)$ depends on the dual geometry and details of the scattered states. 
Inserting this universal behavior into the formula \eqref{eq:phaseshiftaverageEEC} leads to the $\log c_T/c_T$ correction to the energy-energy correlator reported above. 

Let us also briefly discuss the regime of validity of the results above upon inclusion of the stringy corrections. We can characterize them by the parameter $\Delta_{\text{gap}}$, which is defined as the scaling dimension of the lightest spin-four stringy state \cite{Heemskerk:2009pn}. In this case, the leading high-energy behavior of the phase shift is given by the contribution of the leading Regge pole, see e.g.\ \cite{Cornalba:2008sp,Kulaxizi:2017ixa,Costa:2017twz}. We then find that the relevant parameter for our discussion is a characteristic impact parameter $L_*^2 \equiv {\log c_T \over \Delta_{\text{gap}}^2}$. Our results then apply in the regime
\be
L_* \ll 1 ~~~~ \text{(gravitational unitarization)} .
\ee
In contrast, when $L_* \gg 1$ the leading correction to the energy-energy correlator is mostly unitarized by stringy effects
\be
L_* \gg 1 ~~~~ \text{(stringy unitarization)} .
\ee
In this case, the bulk scattering is effectively unitarized at the string scale, and the leading enhanced quantum-gravity correction to the energy-energy correlator is $\sim {\Delta_{\text{gap}}^2 \over c_T}$ instead of ${\log c_T \over c_T}$. In this paper, we focus on the more universal case of gravitational unitarization. Throughout, we also assumed that the source is a light field, namely that its dimension does not scale with $c_T \gg 1$.

The plan of the paper is the following. In Section \ref{sec:phaseshift}, we derive an expression for the contribution of the high-energy bulk scattering to the energy-energy correlators in holographic CFTs in terms of the bulk phase shift. In Section \ref{Sec:Einstein}, we use this representation to obtain the leading quantum-gravity correction to the EEC in pure Einstein gravity on $AdS_{d+1}$. In Section \ref{Sec:N4SYM}, we consider ${\cal N}=4$ SYM dual to $AdS_5\times S^5$ and derive the relation between the phase shift and the EEC which is in perfect agreement with the general representation in Section \ref{sec:phaseshift}. In Section \ref{Sec:LightRayOPE}, we present the relation between the representation in terms of the phase shift and the light-ray OPE. In Section \ref{sec:corrections}, we discuss in more detail various corrections to our results and their regime of validity.  We conclude in Section \ref{sec:discussion} with a discussion. Appendix \ref{App:Conventions} contains our conventions for the stress tensor two-point function, Appendix \ref{app:fold} reviews properties of correlators on the Lorentzian cylinder, Appendix \ref{App:Kernel} explores the kernel of the representation of the EEC in terms of the phase shift, Appendix \ref{app:ReggeLightRayOPE} relates properties of our representation of the EEC to the light-ray OPE, in Appendix \ref{app:TensorStructures} we comment on the tensor structures that are relevant in the case of pure Einstein gravity and in Appendix \ref{App:StringyPhaseShift} we discuss the stringy phase shift in $d=4$ in detail.

\section{Energy correlators and high-energy scattering in AdS}
\label{sec:phaseshift}

In this section, we derive a representation of the energy-energy correlator in terms of the bulk phase shift $\delta(S,L)$ that characterizes high-energy scattering in AdS, see \Fig{fig:image33}. Here $(S,L)$ is the scattering energy and impact parameter in the bulk respectively, see \Fig{fig:image33}. In holographic CFTs with classical gravity duals, this representation allows us to compute the leading quantum-gravity correction to the energy-energy correlator using known properties of high-energy scattering in gravity. In the following sections, Section \ref{Sec:Einstein} and \ref{Sec:N4SYM}, we will use this representation to derive the EEC in pure Einstein gravity in $AdS_{d+1}$ and ${\cal N}=4$ SYM dual to $AdS_5\times S^5$. We discuss further corrections and the regime of validity of our formulas in Section \ref{sec:corrections}.

\subsection{Two-point energy correlator and bulk scattering}

We want to study the conformal collider setup of \cite{Hofman:2008ar} where we probe a localized perturbation using energy detectors $\mathcal{E}(\vec{n})$ sitting far away and measuring the energy flux in the direction $\vec{n}\in S^{d-2}$ on the celestial sphere, see \Fig{fig:EEC}. The energy detectors can be defined in terms of the stress tensor operator integrated over retarded time, see 
\eqref{eq:introEDef}. Equivalently, it can be defined as the light transform of the stress tensor ${\cal E}(n)=2{\mathbf L}[T](x_{\infty},n)$ at spatial infinity $x_\infty$ defined in \cite{Kravchuk:2018htv}. It is clear from \Fig{fig:image3} that spatial infinity $x_\infty$ of the Poincaré patch ${\cal P}_1$ coincides with the origin of the Poincaré patch ${\cal P}_2$. We can then write the following equivalent expression for the energy detector
\be\label{eq:LTrans}
{\cal E}(n) = 2\int_{-\infty}^{\infty}\frac{d\alpha}{(-\alpha)^{d+2}}T(-n/\alpha,n) ,
\ee
where we have introduced index-free notation $T(-n/\alpha,n)= T_{\mu\nu}(-n/\alpha)n^\mu n^\nu$, and $T_{\mu\nu}(0)$ on the right-hand side corresponds to the origin of the Poincaré patch ${\cal P}_2$.

We now consider the energy-energy correlator in a momentum eigenstate $|p\ra$ of unit norm produced by a primary operator acting on the vacuum
\be 
|p\ra &={1 \over \sqrt{{\cal N}_p}} \int d^dx e^{ip \cdot x}{\cal O}(x)| \Omega \rangle,\quad{\cal N}_p = \frac{2\pi^{1+\frac{d}{2}}}{\Gamma(\Delta_{\cal O})\Gamma(\Delta_{\cal O}-\frac{d-2}{2})}\left(\frac{-p^2}{4}\right)^{\Delta_{\cal O}-\frac{d}{2}},
\ee 
with $p$ in the future light cone. In the formula above, we used a scalar primary operator with scaling dimension $\Delta_{\cal O}$ to create the state and we comment on the generalization to the spinning case at the end of the section.

The EEC, defined in \eqref{eq:defEEC}, can be equivalently written as
\be\label{eq:DefEEC} 
\la {\cal E}(n_1){\cal E}(n_2)\ra 
={\int d^dx e^{ip \cdot x} \langle \Omega | {\cal O}^\dagger (0) {\cal E}(n_1) {\cal E}(n_2) {\cal O}(x)| \Omega \rangle \over \int d^dx e^{ip \cdot x} \langle \Omega | {\cal O}^\dagger (0) {\cal O}(x)| \Omega \rangle} ,
\ee 
where operators are ordered as written.\footnote{Note that in this section, $\mathcal{E}$ is represented in the Poincaré patch $\mathcal{P}_2$ (see \eqref{eq:LTrans}), while the coordinates of source operators $\mathcal{O}^\dagger(0), \mathcal{O}(x)$ is in the first Poincaré patch $\mathcal{P}_1$.} Given that the detector operators are inserted at future null infinity, this correlator is effectively evaluated on a time-fold, see Figure \ref{fig:image13}.

By Lorentz invariance and homogeneity with respect to rescaling $n_i^\mu \to \lambda_i n_i^\mu$, which can be easily derived for example from \eqref{eq:LTrans}, we have 
\be 
\la {\cal E}(n_1){\cal E}(n_2)\ra = \frac{(-p^2)^d}{(-p\cdot n_1)^{d-1}(-p\cdot n_2)^{d-1}}{\cal F}(z),
\ee 
with the cross-ratio $0 \leq z \leq 1$ given by
\be\label{eq:zDef}
 z = \frac{p^2 (n_1\cdot n_2)}{2(p\cdot n_1)(p\cdot n_2)}.
\ee 
In the rest frame of the source $z$ is given by $z=\frac{1}{2}(1-\vec{n}_1\cdot \vec{n}_2)=\sin^2(\theta/2)$ where $\theta$ is the angle between the detectors. 

Given the light transform \eqref{eq:LTrans} and \eqref{eq:DefEEC}, the EEC can be written as follows 
\be 
    \langle {\cal E}(n_1){\cal E}(n_2)\rangle = {1 \over {\cal N}_p} \int_{-\infty}^\infty \frac{4d\alpha_1d\alpha_2}{(-\alpha_1)^{d+2}(-\alpha_2)^{d+2}}\langle {\cal O}^\dagger (0) T(-n_1/\alpha_1,n_1)T(-n_2/\alpha_2,n_2) {\cal O} (p)\rangle .
\ee
To proceed it is convenient to introduce the following notation
\be
G(\alpha_1,\alpha_2) \equiv {1 \over {\cal N}_p} \langle \Omega | {\cal O}^\dagger (0) T(n_1/\alpha_1,n_1)T(n_2/\alpha_2,n_2) {\cal O} (p) | \Omega \rangle,
\ee
and we split the integrals into four regions depending on the signs of $\alpha_i$
\be 
\label{eq:splitrep}
    &\langle {\cal E}(n_1){\cal E}(n_2)\rangle = \int_{0}^\infty \frac{4d\alpha_1d\alpha_2}{(\alpha_1)^{d+2}(\alpha_2)^{d+2}} \cr
    &\Big( G_{33}(-\alpha_1,-\alpha_2)+ G_{22}(\alpha_1,\alpha_2) +  G_{23}(\alpha_1,-\alpha_2)+ G_{32}(-\alpha_1,\alpha_2) \Big) ,
\ee
where the index represents the Poincaré patch where the stress tensor is inserted, see Figure \ref{fig:image3}.  To obtain this, we have rewritten the light-transform as $\int_{-\infty}^\infty \frac{d\alpha_1}{(-\alpha_1)^{d+2}} T(-n_1/\alpha_1,n_1) =\int_0^\infty \frac{d\alpha_1}{\alpha_1^{d+2}} \Big( T(0_2+n_1/\alpha_1,n_1) + T(0_3-n_1/\alpha_1,n_1)\Big)$ with $0_i$ denoting the origin of the Poincaré patch ${\cal P}_i$.

We next introduce the following approximation
\be
G_{jk}(\alpha_1,\alpha_2) \simeq \int {d^d q \over (2 
\pi)^d} e^{i q \cdot ({n_1 \over 
\alpha_1} - {n_2 \over 
\alpha_2})} \tilde G_{jk}(p,q,n_1,n_2) + \cdots ,
\ee
which should correctly capture the large $q$-limit of the correlator. 

Let us, for example, consider $G_{22}(\alpha_1,\alpha_2)$. At large $q$ the leading contribution to it is given by the unit operator so that
\be
 \tilde G_{22}(p,q,n_1,n_2) \simeq \rho_T c_T \theta(q^0) \theta(-q^2) G_0(q,n_1,n_2)
\ee 
with the vacuum Wightman two-point function of the stress tensors given by  
\be\label{eq:GVacuum}
    G_0(q,n_1,n_2) &= (q\cdot n_1)^2(q\cdot n_2)^2(-q^2)^{\frac{d-4}{2}}(4\chi^2-4\chi+\frac{d-2}{d-1})\nn\\
    \chi &= \frac{q^2 (n_1\cdot n_2)}{2(q\cdot n_1)(q\cdot n_2)}\\
    \rho_T &= \frac{\pi^{\frac{d}{2}+1}}{2^{d-1}(d+1)\Gamma(d-1)\Gamma(\frac{d+2}{2})\Omega_d^2}.\nn
\ee 
The overall normalization of $\rho_T c_T$ follows from the normalization of the stress tensor two-point function in the vacuum $\la TT\ra\sim c_T$ in momentum space, see Appendix \ref{App:Conventions}. 

Consider next $G_{33}(- \alpha_1, - \alpha_2 )$. This correlator is closely related to $G_{22}(\alpha_1, \alpha_2)$. To see it, we can translate the in- and the out- source one Poincaré patch forward. Following \cite{Kravchuk:2018htv}, we can call this operation ${\cal T}$ and we notice the following transformation property of the correlator
\be
\langle \Omega | {\cal O}^\dagger ({\cal T}^a x {\cal T}^{-a}) \dots {\cal O} ({\cal T}^b y {\cal T}^{-b}) | \Omega \rangle = e^{i \pi \Delta(a-b)} \langle \Omega | {\cal O}^\dagger (x) \dots {\cal O} (y) | \Omega \rangle ,
\ee
that we review in Appendix \ref{app:fold}.

Using this property, we see that in our approximation
\be
\label{eq:firstsym}
G_{33}(- \alpha_1, - \alpha_2 ) = G_{22}^*(\alpha_1, \alpha_2) ,
\ee
where the complex conjugation on the right comes from flipping the phase sign in the Fourier transform and the fact that $G_0(q,n_1,n_2)$ is real.

Next we consider the $G_{23}$ and $G_{32}$ terms. By moving the out-source one Poincaré patch forward we see that these terms are related to bulk scattering \cite{Cornalba:2006xk,Cornalba:2006xm,Kulaxizi:2017ixa}. We therefore modify the formula above by introducing the bulk phase shifts
\be
 \tilde G_{23}(p,q,n_1,n_2) &\simeq  (-1)^d \rho_T c_T \theta(q^0) \theta(-q^2) G_0(q,n_1,n_2) e^{i \delta_{23}(S,L)} , \\
  \tilde G_{32}(p,q,n_1,n_2) &\simeq  (-1)^d \rho_T c_T \theta(q^0) \theta(-q^2) G_0(q,n_1,n_2) e^{i \delta_{32}(S,L)} ,
\ee 
where the factor $(-1)^d = e^{\pm i \pi d}$ comes from the trivial evolution of $G_0$ (recall that for the stress tensor $\Delta = d$), see Appendix \ref{app:fold}, and we introduced
\be\label{eq:SLDef}
S &=\sqrt{-q^2}\sqrt{-p^2}, \nn \\
\cosh L &= -{p \cdot q \over \sqrt{-q^2}\sqrt{-p^2}},
\ee 
which are identified with the scattering energy and the impact parameter in the bulk. Physically, we expect the phase shift formulas above to apply at high energies $S L^2 \gg 1$, such that the Compton wavelength of the scattered particles is much less than the impact parameter.

By exchanging the ingoing and outgoing stress-energy tensors, we get the following crossing relation\footnote{To see it, notice that $ 2 \leftrightarrow 3$ exchange here corresponds to two ways of approaching the Regge limit $z, \bar z \to 1$ of the four-point function. In the space of cross-ratios it corresponds to continuation of $z$ around the origin: clockwise $z \to z e^{2 \pi i}$ and counterclockwise $z \to z e^{-2 \pi i}$. This operation introduces opposite phases in the $\textbf{source} \times \textbf{detector}$ OPE discussed earlier, which then becomes \eqref{eq:crossingPS}. }
\be
\label{eq:crossingPS}
e^{i \delta_{32}(S,L) } = \left( e^{i \delta_{23} (S,L)} \right)^*. 
\ee
Using this relationship, we have 
\be
G_{23}(\alpha_1, - \alpha_2 ) = G_{32}^*(-\alpha_1, \alpha_2) ,
\ee
and we can thus rewrite the energy-energy correlator above as follows
\be 
\label{eq:eecderivA}
    &\langle {\cal E}(n_1){\cal E}(n_2)\rangle = \int_{0}^\infty \frac{4d\alpha_1d\alpha_2}{(\alpha_1)^{d+2}(\alpha_2)^{d+2}} 2 {\rm Re}\Big( G_{22}(\alpha_1,\alpha_2)+ G_{23}(\alpha_1,-\alpha_2)\Big) .
\ee

We next plug the Fourier-transformed expressions above into this formula and perform the $\alpha$-integrals to finally get the following representation for the contribution of the high-energy scattering to the energy-energy correlator
\be 
\label{eq:eecRep1}
{\cal F}(z) = 
8\rho_T c_T\Gamma(d+1)^2\int_0^1 d u {\cal K}_d (u,z) {\rm Re} \langle 1 - e^{i \delta}\rangle,
\ee
where we dropped the index on the phase shift $\delta_{23}(S,L)$ and introduced the notation
\be
\label{eq:psaver}
\langle e^{i \delta} \rangle  = \int_{S_0}^\infty {d S \over S^3} e^{i\delta (S,L)} |_{\tanh L = u} \ ,
\ee
where $S_0$ is an arbitrary fixed energy scale.
The kinematical kernel ${\mathcal K}_d (u,z)$ is defined as follows
\be
{(-p^2)^d \mathcal{K}_d(u,z) \over (p\cdot n_1)^{d-1} (p\cdot n_2)^{d-1}}  \equiv \int_{M^+} {d^dq \over (2\pi)^d} {G_0(q,n_1,n_2) \delta(1-\sqrt{-q^2}\sqrt{-p^2}) \over (q\cdot n_1)^{d+1}(q\cdot n_2)^{d+1}}  \delta\left(u - \sqrt{1-{p^2 q^2 \over (p\cdot q)^2}}\right)
\ee
where $M^+$ is the future Milne wedge defined by $q^2<0$ and $q^0>0$. The explicit expression for the kernel ${\mathcal K}_d (u,z)$ is given in \eqref{eq:kerneld}.

Let us notice that in agreement with the expected momentum conservation Ward identities satisfied by the energy-energy correlator, the kernel has the following properties
\be
\label{eq:CWI}
\int_0^1 dz (z(1-z))^{{d-4 \over 2}} {\cal K}_d (u,z) = \int_0^1 d z (z(1-z))^{{d-4 \over 2}} \ z {\cal K}_d (u,z) = 0 . 
\ee

In this paper we will use \eqref{eq:eecRep1} to explore the quantum gravity or $1/c_T$ corrections to the energy-energy correlator. However, before proceeding with this, we need to discuss the regime of validity of this formula. Physically, we only expect the formula \eqref{eq:eecRep1} to correctly capture the high-energy large enough impact parameter contribution to the energy-energy correlator. As we explain below, this translates to a potential polynomial in $z$ ambiguity in some of our computations related to the small impact parameter physics. More details can be found in Section \ref{sec:corrections}.

A complete treatment of the bulk scattering for the four-point function $\langle {\cal O}^\dagger T_{\mu \nu} T_{\rho \sigma} {\cal O}\rangle$ can be found in \cite{Costa:2017twz}. It involves two additional tensor structures compared to the ones considered above. These additional structures however are suppressed in theories where the stringy modes are heavy, or, equivalently, which have a large gap in the spectrum of single trace higher spin operators $\Delta_{\text{gap}} \gg 1$, see    \cite{Camanho:2014apa,Afkhami-Jeddi:2016ntf,Kulaxizi:2017ixa,Costa:2017twz,Caron-Huot:2021enk,Chang:2023szz}. In particular, in this case the three-point function of gravitons is the one of general relativity. For the same reason, we expect that our formula applies to scattering of spinning states as well because in general relativity the phase shift is diagonal in polarization space. It would be interesting to generalize our formulas to include the `non-minimal' tensor structures.

\subsection{High-energy limit and the enhancement property}

We would like next to understand the properties of the integral over $S$ above which leads to the enhancement property discussed in the introduction. A characteristic feature of the gravitational scattering is that the phase shift exhibits a linear growth with energy
\be
\label{eq:phaseshift1loop}
\delta(S,L) = {1 \over c_T} S \Pi(L) ,
\ee
where we have substituted $G_N$ in terms of the central charge of the CFT dual. 

We therefore need to understand the following integral
\be
\label{eq:finiteintegral}
\int_{S_0}^\infty {d S \over S^3} e^{i\delta (S,L)} \ \ .
\ee 
Now comes the key point of our analysis. Looking at the finite integral \eqref{eq:finiteintegral} we see that the integral over energies and the expansion at large $c_T$ do not commute. Indeed, expanding under the integral we get the UV divergence at order ${1 \over c_T^2}$ (which corresponds to the one-loop supergravity computation)
\be
\int_{S_0}^\infty {d S \over S^3} \Big(1 + i {S \Pi(L) \over c_T} - {1 \over 2} \Big({S \Pi(L) \over c_T}\Big)^2 + ... \Big) ,
\ee
whereas if we do the integral over $S$ first we get
\be\label{eq:Sintegral}
\int_{S_0}^\infty {d S \over S^3} e^{i {1 \over c_T} S \Pi(L)} = c_0 + {c_1 \over c_T} -{1 \over 2} \Pi^2(L) {\log {c_T \over S_0 \Pi(L)} \over c_T^2} + \dots \ , 
\ee
where ${\log c_T \over c_T^2}$ is enhanced compared to the naive expansion of the correlator. This is true as long as $S_0$ does not scale with $c_T$. Physically, this condition translates to the statement that the source operator is light, namely its scaling dimension does not scale with $c_T$. Moreover, we will focus only on the term enhanced in $c_T$ which is correctly captured by the high-energy regime \eqref{eq:phaseshift1loop}. This is substantiated by bounding the contribution of other regimes in Section \ref{sec:corrections}. 

Focusing on the term enhanced in $c_T$ we can perform the integral over $S$ in \eqref{eq:eecRep1} with the result
\be 
\label{eq:finalformula}
{\cal F}(z) = 
 {4 \rho_T \log c_T\over c_T}\Gamma(d+1)^2\int_0^1 d u {\cal K}_d (u,z) \la \Pi^2(u)\ra,
\ee
where we denoted by $\Pi(u) \equiv \Pi(L(u))$ and we hope that this will not cause confusion. Let us also comment that in the presence of extra dimensions the graviton mixes with KK modes at high energies. In this case the eikonal phase operator should be diagonalized before exponentiation and instead of $e^{i \delta}$ we will get $\sum_k |\langle \text{in}| k \rangle|^2 e^{i \delta_k}$ and $\sum_k |\langle \text{in}| k \rangle|^2 \Pi_k^2$ correspondingly. We will see how it works very explicitly in the example of ${\cal N}=4$ SYM.
    
We have concluded that the enhanced term of the EEC is determined by the bulk phase shift according to \eqref{eq:finalformula} and all that is left to be done is the integral over impact parameters. 

\subsection{Integral over impact parameters}

To better understand the consequences of the representation \eqref{eq:finalformula}, it is useful to consider the following integral
\be
\label{eq:Reggeblock2}
R_\Delta(z) \equiv \int_0^1 du {\cal K}_d (u,z) {\cal H}_{\Delta - 1}(L(u)),
\ee
where ${\cal H}_{\Delta-1}(L)$ is the hyperbolic space propagator in $(d-1)$-dimensions
\be\label{eq:hyperbolicProp}
    {\cal H}_{\Delta-1}(L)=\frac{\pi^{1-\frac{d}{2}}\Gamma(\Delta-1)}{2\Gamma(\Delta-\frac{d-2}{2})}e^{-(\Delta-1)L}{}_2F_1\Big(\frac{d}{2}-1,\Delta-1,\Delta-\frac{d}{2}+1;e^{-2L}\Big).
\ee
We show in Appendix \ref{app:ReggeLightRayOPE} that $R_{\Delta}(z)$ is given by
\be\label{eq:identity}
R_{\Delta}(z)&=\frac{\left((d-\Delta )^2-1\right) \Gamma \left(\frac{\Delta -1}{2}\right) \Gamma \left(\frac{\Delta +3}{2}\right)
   \Gamma \left(d-\frac{\Delta }{2}-\frac{1}{2}\right) \Gamma \left(\frac{1}{2} (d+\Delta -1)\right)}{2^{d+2} \pi^d \Gamma (d+1)^2 \Gamma
   \left(-\frac{d}{2}+\Delta +1\right)} f_{\Delta}(z) \cr
   &\quad +\frac{(d-2) (2 \pi )^{-d}}{(d-1) (-2 d+\Delta +1) (d+\Delta -1)}g_{\Delta}(z),
\ee 
where $f_\Delta(z)$ is a celestial conformal block that appears in the light-ray OPE \cite{Kologlu:2019mfz}
\be 
\label{eq:celblock}
f_{\Delta}(z) = z^{{\Delta-(2d-1) \over 2}} \ _2 F_1({\Delta -1 \over 2}, {\Delta -1 \over 2}, \Delta - {d \over 2}+1, z) 
\ee 
and $g_{\Delta}(z)$ is analytic at $z=0$ and is given explicitly in \eqref{eq:defg}.\footnote{The connection between the phase shift formulas and the light-ray OPE will be discussed further in Section \ref{Sec:LightRayOPE}.} 

In our computations we will encounter $\Delta=(2d-1)+2n$ with $n\in\mathbb{Z}$ coming from the graviton exchange in the bulk. In this case the coefficient of $f_\Delta(z)$ in \eqref{eq:identity} develops a pole which cancels with a similar pole coming from $g_\Delta(z)$ producing a $\log z$:
\be\label{eq:LogApp} 
&R_{(2d-1)+2n}(z)=r_{d,n}f_{(2d-1)+2n,d}(z)\log z+\ldots,\cr 
&r_{d,n}=\frac{(-1)^{n+1} \left((d+2 n-1)^2-1\right) \Gamma (d+n-1) \Gamma (d+n+1) \Gamma \left(\frac{3 d}{2}+n-1\right)}{2^{d+2} \pi ^{d} n! \Gamma (d+1)^2 \Gamma \left(\frac{3
   d}{2}+2 n\right)},
\ee 
and the ellipses denote analytic terms. The universal ($n=0$) leading small-angle contribution to the energy-energy correlator takes the form
\be
    \la {\cal E}(n_1) {\cal E}(n_2) \ra_{\text{graviton}} = - \frac{\log c_T}{c_T}\frac{2^{3 d-5} (d-2) d  \Gamma \left(\frac{d-1}{2}\right)^4 \Gamma \left(\frac{d+3}{2}\right) \Gamma \left(\frac{3 d}{2}-1\right)}{\pi ^{d+\frac{1}{2}}\Gamma
   \left(\frac{d}{2}\right)^2 \Gamma (2 d-2)}\log z,
\ee
where the non-analyticity in $z$ arises from the large impact parameter regime.

We thus see that the large impact parameter physics produces small angle non-analyticities at $z=0$. This is in contrast to the contribution coming from finite impact parameters: if we cut off the impact parameter integral in \eqref{eq:finalformula} so that we have $\int_0^{u_0} du$, the result is analytic at small $z$. The conclusion of this discussion is that large impact parameter expansion of the phase shift controls non-analytic in $z$ contributions to the energy-energy correlator. 

\subsection{The OPE perspective}

We can also understand the discussion above starting from the $\textbf{source} \times \textbf{detector}$ OPE that was worked out in \cite{Kologlu:2019bco}. By inserting a complete set of states spanned by primary operators we get the following representation of the energy-energy correlator in the scalar state
\be
\label{eq:sourcedetectorOPE}
{\cal F}(z)&=\sum_{\Delta,J}|\lambda_{\Delta,J}|^2 2 \sin^2(\pi {\Delta - J - d - \Delta_{\cal O} \over 2}) {\cal P}_{\Delta,J}(z) , 
\ee
where ${\cal P}_{\Delta,J}(z)$ is a polynomial of degree $J$ in $z$. In writing the formula above we used the fact that $\langle {\cal O} T_{\mu \nu} {\cal O}_{\Delta, J} \rangle$ admits a single conformal invariant structure. We can rewrite the formula above as follows
\be\label{eq:OPERep}
{\cal F}(z) &= {\rm Re} \sum_{\Delta, J} {|\lambda_{\Delta,J}|^2 \over |\lambda_{\Delta,J}^{GFF}|^2} \left(1 - e^{- i \pi \gamma_{\Delta,J}} \right) \left( |\lambda_{\Delta,J}^{GFF}|^2 {\cal P}_{\Delta,J}(z) \right),
\ee
where we have defined $\gamma_{\Delta,J} \equiv \Delta - J - d - \Delta_{\cal O}$. 

We can now try to connect this expression to the phase shift discussion. The finite $J$ contributions correspond to zero impact parameter scattering in the language of the previous section and they produce polynomial contributions in $z$. To get the non-analytic contributions, we take the limit $\Delta, J \to \infty$ with ${\Delta \over J}$ kept fixed, see e.g. \cite{Cornalba:2006xk,Cornalba:2006xm,Cornalba:2007zb}. 

Exchanging the sum for the integral, and identifying $\delta(S,L) = -\pi \gamma_{\Delta,J}$, see e.g. \cite{Kulaxizi:2017ixa}, we should recover the previous formula, where the kinematical kernel arises from the large $\Delta,J$ limit of $|\lambda_{\Delta,J}^{GFF}|^2 {\cal P}_{\Delta,J}(z)$. It would be interesting to check this statement explicitly. In the case of ${\mathcal N=4}$ SYM we do it in Section \ref{Sec:N4SYM}, where we see in detail how the phase shift formula emerges from the $\textbf{source} \times \textbf{detector}$ OPE. Notice also that the leading contribution to the phase shift \eqref{eq:phaseshift1loop} arises from the contribution of the double trace operators $[{\cal O}T_{\mu \nu}]_{n,J}$ \cite{Cornalba:2006xm,Cornalba:2006xk,Cornalba:2007zb} with $n\sim J \gg1$. 

Finally, let us comment on inelasticity. In the standard eikonal approach to the gravitational scattering inelasticity is encoded in the imaginary part of the phase shift ${\rm Im} \delta(S,L)$. From the OPE point of view this arises from averaging over effectively degenerate operators $e^{i \delta(S,L)} ={1 \over k} \sum_{j=1}^k e^{- i \pi \gamma_j}$.

\subsection{Relation to the quadruple discontinuity}

In the discussion above we saw a finite number of operators in the  $\textbf{source} \times \textbf{detector}$ OPE produces a polynomial contribution to the energy-energy correlator. Equivalently, any non-analyticity in $z$ can only emerge from the sum over infinitely many operators.

The non-analytic terms in $z$ can be conveniently captured by taking the discontinuity around $0$
\be\label{eq:DiscDef}
{\rm Disc}_z {\cal F}(z) \equiv {{\cal F}(z + i 0) - {\cal F}(z - i 0) \over 2 i }.
\ee
The fact that the $s$-channel blocks in \eqref{eq:OPERep} are polynomials in $z$ can be restated by saying that they are annihilated by taking the discontinuity around $z=0$. 

A related and more familiar fact is that taking the $t$-channel \emph{double discontinuity} annihilates the $s$-channel blocks \cite{Caron-Huot:2017vep}. In the formula above, the $s$-channel double discontinuity has already been taken, which can be seen due to the presence of the $\sin^2$ factor in the OPE. It originates from the fact that the energy calorimeters annihilate the vacuum and thus we can equally compute the energy-energy correlator starting from $\langle [{\cal O}, T_{\mu \nu}] [T_{\rho \sigma}, {\cal O}] \rangle$.

\begin{figure}
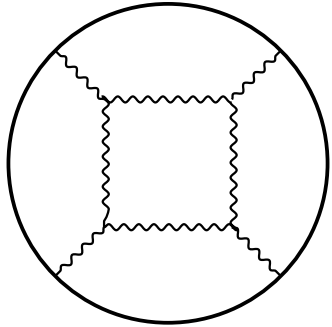

    \centering
    \boxDiagram
    \caption{An AdS box diagram with a non-vanishing quadruple discontinuity. In the phase shift language it arises from an interaction that is non-local in $L$. Such interactions produce a non-analytic in $z$ contribution to the energy-energy correlator.}
    \label{fig:BoxDiagram}
\end{figure}

We conclude that ${\rm Disc}_z {\cal F}(z)$ (the same statement applies to the discontinuity around $z=1$) is captured by \emph{the quadruple discontinuity of the correlator}. This statement is in agreement with the earlier discussion in this section, which related non-analyticity in $z$ to the large impact parameter physics in the bulk. The same bulk diagrams that produce non-trivial large impact parameter behavior of the phase shift, see e.g.\ Figure \ref{fig:BoxDiagram}, are those that have nontrivial double discontinuity in AdS that corresponds to the quadruple discontinuity of the correlator in CFT. 

In the explicit perturbative computations of the energy-energy correlators, see e.g. \cite{Basham:1978bw,Basham:1978zq,Belitsky:2013bja,Belitsky:2013ofa,Dixon:2018qgp,Dixon:2019uzg,Belitsky:2013xxa,Henn:2019gkr}, one can, in principle, write a dispersive representation of the energy-energy correlator by considering complex angles $z$. We see that such a representation is closely related to the Mandelstam representation of the AdS correlators. It would be interesting to explore this connection further.

\section{Einstein gravity}\label{Sec:Einstein}
In the previous section we have developed the general expression for the enhanced 
$\frac{\log c_T}{c_T}$ correction  to the EEC in holographic CFTs. In particular, it comes from high-energy scattering in the bulk, dual to the Regge limit from the CFT point of view. This region is captured by the bulk phase shift \cite{Cornalba:2006xm,Cornalba:2006xk,Cornalba:2007zb} whose characteristic linear growth in energy, intrinsic to gravitational scattering, leads to the logarithmic enhancement in the central charge $c_T\gg1$ that we explained. In this section, we derive the precise prediction for the non-analytic in $c_T$ term in theories dual to pure $AdS_{d+1}$ Einstein gravity.  
\subsection{Bulk phase shift in pure Einstein gravity}
The holographic theories that we consider are dual to $d+1$ pure Einstein gravity
\be\label{eq:GravityAction}
S_{\text{grav}} &=\frac{1}{16\pi G_N}\int d^{d+1}x\sqrt{-g}(R-\Lambda),
\ee 
where $\Lambda=-\frac{d(d-1)}{R_{AdS}^2}$ is the cosmological constant and $R_{AdS}$ the AdS scale which we set to $1$. Here $G_N$ is Newton's constant in $D=d+1$ dimensions and it is related to the central charge of the CFT by (see e.g.\ \cite{Kovtun:2008kw}\footnote{Their definition of $c_T$ differs by a factor of two with ours.})
\be 
c_T =\frac{d+1}{d-1}\frac{\pi^{\frac{d}{2}}\Gamma (d+1)}{\Gamma\left(\frac{d}{2}\right)^3}\frac{R_{AdS}^{d-1}}{2\pi G_N},
\ee 
and see Appendix \ref{App:Conventions} for further conventions. We then want to measure the EEC in a momentum eigenstate $|p\rangle$ created by a local operator ${\cal O}$ such as ${\cal O}\sim \epsilon^{\mu}\epsilon^\nu T_{\mu\nu}$ or another external probe field added to the gravity action \eqref{eq:GravityAction}.

The phase shift in theories dual to pure AdS gravity was first studied in \cite{Cornalba:2006xm,Cornalba:2006xk,Cornalba:2007zb}, see also \cite{Camanho:2014apa,Kulaxizi:2017ixa,Li:2017lmh,Costa:2017twz} for related work on the Regge limit in holographic CFTs. It is determined by the graviton exchange in terms of the scattering energy $S$ and impact parameter $L$ defined by \eqref{eq:SLDef} and it is given by
\be\label{eq:PH}
    \delta^{(GR)}(S,L)=4\pi G_NS\,{\cal H}_{d-1}(L),
\ee 
where ${\cal H}_{d-1}$ is defined in \eqref{eq:hyperbolicProp}. In particular, the large impact parameter limit in any holographic CFT is determined by that of the graviton exchange in \eqref{eq:PH} and will lead to a universal prediction for the small-angle behaviour of the energy-energy correlator. From the CFT point of view, this is due to the universality of the stress tensor exchange and using the relation between the phase shift and the anomalous dimensions of double-trace operators \cite{Cornalba:2006xm,Cornalba:2006xk,Cornalba:2007zb} this can also be found using bootstrap methods, see e.g.\ \cite{Kaviraj:2015xsa,Alday:2017gde,Li:2017lmh,Kulaxizi:2017ixa}. 

To align with our previous representation it is further convenient to define $\Pi^{(GR)}(L)$ by 
\be
\delta^{(GR)}(S,L) = {1 \over c_T} S \Pi^{(GR)}(L)
\ee
which explicitly is given by 
\be\label{eq:PHCT}
    \Pi^{(GR)}(L) = \frac{2\pi^{\frac{d}{2}}\Gamma(d+2)}{(d-1)\Gamma(d/2)^3}{\cal H}_{d-1}(L).
\ee 
The enhanced correction to the energy-energy correlator is given by \eqref{eq:finalformula} and for pure Einstein gravity in $AdS_{d+1}$ we get 
\be\label{eq:finalGR2}
\langle {\cal E}(n_1){\cal E}(n_2) \rangle_{AdS_{d+1}}&= 
\frac{\log c_T}{c_T}{4\rho_T\Gamma(d+1)^2 (-p^2)^d \over (-p \cdot n_1)^{d-1} (-p \cdot n_2)^{d-1}}  \int_0^1 d u {\cal K}_d (u,z)\Pi^{(GR)}(u)^2 .
\ee 
Let us note that in the small impact parameter limit $u\to 0$, the kernel is given by 
\be 
{\cal K}_d(u,z)\propto u^{d-2}\Big(\frac{d-2}{d-1}-4z+4z^2\Big).
\ee 
The pure Einstein gravity phase shift on the other hand behaves in $AdS_{d+1}$ as $(\Pi^{(GR)}(u))^2\sim u^{2(3-d)}$ ($\log^2 u$ in $d=3$). In $d\geq 5$ we therefore see explicitly a small impact parameter divergence which as we discuss in Section \ref{sec:corrections} is regulated by stringy corrections to the phase shift. 

\subsection{$AdS_{4}/CFT_3$}
Consider the phase shift in $d=3$ where from \eqref{eq:PHCT} $\Pi^{(GR)}(L)$ is given by 
\be\label{eq:ph3d} 
    \Pi^{(GR)}(L)= \frac{48 \left( e^{-L} \left(e^{2 L}+1\right) \log \left(\frac{e^L+1}{e^L-1}\right)-2\right)}{\pi }.
\ee 
Besides the phase shift, we need the kernel which can be obtained from Appendix \ref{App:Kernel}. We present it explicitly here for $d=3$:
\be\label{eq:ker3d}
{\cal K}_3(u,z)=&\frac{u \left(1-u^2\right)}{16 \pi ^2 \left(u^2 (z-1)+1\right)^7}\Big[u^8 (z-1)^3 \left(8 z^2+24 z+3\right)\cr
&-u^6 (z-1)^2 \left(16 z^3+136 z^2+30 z-7\right)+3 u^4 \left(48 z^4-24 z^3-62 z^2+39 z-1\right)\cr
&-u^2 \left(144 z^3-184 z^2+42
   z+3\right)+2 \left(8 z^2-8 z+1\right)\Big].
\ee 

We digress slightly to point out explicitly some salient features of the kernel \eqref{eq:ker3d}, already alluded to and which generalizes straightforwardly to general $d$. First, expanding the kernel at small angles $z\to 0$ the behaviour of the kernel at $u\to 1$ gets worse order-by-order in $z$
\be 
{\cal K}_3(u,z)=\frac{u \left(2+3 u^2\right)}{16 \pi ^2 \left(1-u^2\right)^3}-\frac{u \left(4+22
   u^2+9 u^4\right) z}{4 \pi ^2 \left(1-u^2\right)^4}+{\cal O}(z^2).
\ee 
On the other hand the behaviour of the $ (\Pi^{(GR)}(L))^2\sim (1-u)^2$ as $u\to1$. This leads to the non-analytic $\log z$ term in the EEC as was also seen in \eqref{eq:LogApp}. A more detailed asymptotic expansion when $z\to0$ and $u\to 1$ keeping $\frac{1-u}{z}$--fixed is given in Appendix \ref{app:AsymptoticExpansion} in general dimension. 
 
After having showcased some general properties of the kernel, it is clear that the integral in $d=3$ is well-behaved. Given the kernel \eqref{eq:ker3d} and the phase shift \eqref{eq:ph3d} we obtain the enhanced term in $c_T$ in theories dual to pure Einstein gravity in $AdS_4$ by performing the integral in \eqref{eq:finalGR2}:
\be\label{eq:3dRes}
\langle {\cal E}(n_1){\cal E}(n_2) \rangle_{AdS_4}=& - \frac{480}{\pi^2} \left({p^0 \over 2 \pi} \right)^2{\log c_T \over c_T}\Big[(1-\frac{8}{5} z) \log (4 z) + \frac{8}{5} (1-z)\Big].
\ee 
As expected from the general properties of the kernel the result \eqref{eq:3dRes} satisfy the momentum conservation Ward identities 
\be 
&\int_0^1 \frac{dz}{\sqrt{z(1-z)}}\langle {\cal E}(n_1){\cal E}(n_2) \rangle_{AdS_4}=0,\cr
&\int_0^1 \frac{zdz}{\sqrt{z(1-z)}}\langle {\cal E}(n_1){\cal E}(n_2) \rangle_{AdS_4}=0.
\ee 

\subsection{$AdS_{5}/CFT_4$}
Let us now consider instead $d=4$, still in pure Einstein gravity. As for the three-dimensional case, much of the work is laid out for us. To be explicit, the phase shift is determined by $\Pi^{(GR)}(L)$ 
\be 
    \Pi_{d=4}^{(GR)}(L)&=\frac{40 \pi  e^{-3 L}}{1-e^{-2 L}}= \frac{20 \pi  (1-u)^{3/2}}{u \sqrt{1+u}},
\ee 
and we have written it both in terms of $L$ and $u=\tanh(L)$. The kernel on the other hand takes a form similar to the $d=3$ case \eqref{eq:ker3d} but we refrain from writing it down in detail. It can easily be constructed as explained in Appendix \ref{App:Kernel}. 

\begin{figure}
    \centering \includegraphics[width=0.6\textwidth]{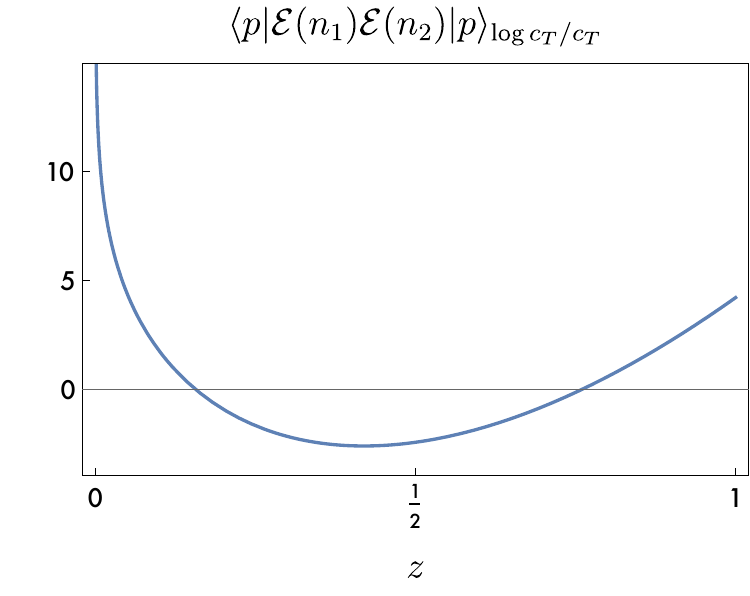}
    \caption{The enhanced $\frac{\log c_T}{c_T}$ part of the energy-energy correlator in $d=4$ computed in pure gravity in $AdS_5$ and given by \eqref{eq:GR4d}.}
    \label{fig:EEC4d}
\end{figure}

Explicitly performing the integral in \eqref{eq:finalGR2}, we find the following result for the term non-analytic in the central charge $c_T$ 
\be
\label{eq:GR4d}
&\langle {\cal E}(n_1){\cal E}(n_2)\rangle_{AdS_5}=\frac{45}{64} \left( \frac{p^0}{4 \pi} \right)^2 {\log c_T \over c_T} \\
&\times {1 \over z^2}\left(6 \left(-3+z-65 z^2+67 z^3\right)+\left(9-z-\frac{1475 z^2}{3}+1225
   z^3\right) \log z \right. \cr
   &\left. +\frac{\left(9-4z+54 z^2-900 z^3+1225 z^4\right)
   \left(\text{Li}_2\left(\sqrt{z}\right) -\text{Li}_2\left(-\sqrt{z}\right)+\frac{1}{2} \log
   \left(\frac{1-\sqrt{z}}{1+\sqrt{z}}\right) \log (z)\right)}{\sqrt{z}} \right) \ , \nn
\ee
where the $\sqrt{z}$ cancels in the small $z$ expansion of the energy-energy correlator and for simplicity we have set $p^\mu=(p^0,\vec{0})$.
The result \eqref{eq:GR4d} is displayed in \Fig{fig:EEC4d} and can be verified explicitly to satisfy the momentum conservation Ward identities \eqref{eq:WI1} and \eqref{eq:WI2}, again as expected by the properties of the kernel \eqref{eq:WardIntro}. 

For later convenience let us isolate the part that is non-analytic in $z$ in \eqref{eq:GR4d}. Its small angle expansion is
\be\label{eq:4dExp}
\langle {\cal E}(n_1){\cal E}(n_2) \rangle_{AdS_5} =- 384 \left(\frac{p^0}{4 \pi} \right)^2 \frac{\log c_T}{c_T}\left(1-\frac{27}{7}z+\frac{12}{7}z^2+\ldots\right)\log z+\ldots.
\ee 
In particular, the leading term in \eqref{eq:4dExp} as $z\to 0$ is universal for theories dual to $AdS_5\times X$ for any internal manifold $X$. This will be verified explicitly in the case of $AdS_5\times S^5$ in Section \ref{Sec:N4SYM}. In the latter case, we will see how subleading terms are altered compared to $AdS_5$ in \eqref{eq:4dExp} due to the presence of KK-modes.

\subsection{$AdS_{d+1}/CFT_d$}
As we have argued, non-analyticities in $z$ arise from the large impact parameter regime and a convenient method to extract these is to study the discontinuity of the kernel ${\cal K}_{d}(u,z)$. A useful way to do this is by utilizing the connection to the Mandelstam kernel \cite{Correia:2020xtr} and we refer to Appendix \ref{App:Kernel} for further details. In particular, the term non-analytic in the central charge is given by 
\eqref{eq:finalGR2}
\be\label{eq:finalformulaDisc}
{\rm Disc}_z \langle {\cal E}(n_1){\cal E}(n_2) \rangle|_{\log c_T} &= \theta(-z)
4\rho_T\frac{\log c_T}{c_T}\Gamma(d+1)^2 (p^0)^2 \int_{\frac{1}{\sqrt{1-z}}}^1 d u{\rm Disc}_z  {\cal K}_d (u,z)\Pi^{(GR)}(u)^2,
\ee 
where for simplicity we have set $p^\mu=(p^0,\vec{0})$. Moreover, the discontinuity of the kernel is simpler than the kernel itself and vanishes for $u<\frac{1}{\sqrt{1-z}}$. See Appendix \ref{app:DiscKernel} for more details on the discontinuity of the kernel. Taking the discontinuity of the EEC therefore effectively removes ambiguities from small impact parameters. Given $\Pi^{(GR)}(u)$ in \eqref{eq:PHCT} and \eqref{eq:finalformulaDisc}, we need to perform the integral over the impact parameter. Performing this integral we find the leading quantum-gravity correction to the EEC in theories dual to pure Einstein gravity in $AdS_{d+1}$: 
\be\label{eq:finalGR}
&\langle {\cal E}(n_1){\cal E}(n_2) \rangle =-\gamma_d{\log c_T\over c_T}\Big(\frac{p^0}{\Omega_{d-1}}\Big)^2\Big[{\cal P}_d(z)\log z+{\cal R}_d(z)\Big],\cr
&{\cal P}_d(z)=\,{}_3F_2\Big(\frac{1-d}{2},d-1,d-1 ;  d-\frac{1}{2},1 ;z\Big),\cr
&\gamma_d =\frac{2^{3 d-3} d(d-2)\Gamma \left(\frac{d-1}{2}\right)^2 \Gamma \left(\frac{d+3}{2}\right) \Gamma \left(\frac{3 d}{2}-1\right)}{\pi ^{\frac{3}{2}} \Gamma
   \left(\frac{d}{2}\right)^2 \Gamma (2 d-2)},
\ee 
where ${\cal R}_d(z)$ denotes analytic terms in $z$ and will be computed below. It is easy to verify that this reproduces the explicit results in $d=3,4$ obtained above. In particular, for odd $d$, ${\cal P}_d(z)$ truncates and the non-analytic terms are given by $\log z$ times a polynomial of degree $(d-1)/2$:
\be
\begin{aligned}
&\langle {\cal E}(n_1){\cal E}(n_2)\rangle_{AdS_4} = -\frac{480}{\pi^2}{\log c_T \over c_T}\left(\frac{p^0}{2\pi}\right)^2(1-\frac{8z}{5})\log z+\ldots\\
&\langle {\cal E}(n_1){\cal E}(n_2)\rangle_{AdS_6} = -\frac{21120}{\pi^2}{\log c_T\over c_T}\left(\frac{p^0}{2\pi^2}\right)^2(1-\frac{64 z}{9}+\frac{800 z^2}{99})\log z+\ldots.
\end{aligned}\ee 

More generally, from \eqref{eq:finalGR}, we see that the discontinuity of the EEC grows at large $|z|$ as $|z|^{\frac{D-2}{2}}$ where $D$ is the number of large bulk dimensions. Here we suggestively write it in terms of the bulk dimension $D$ instead since we will see in Section \ref{Sec:N4SYM} that this is consistent with $D=10$ in the case of $AdS_5\times S^5$.

Having obtained the general $d$-result \eqref{eq:finalGR} for the non-analytic term, we can reconstruct the full result by a dispersive integral with subtractions; the result for ${\cal R}(z)$ in \eqref{eq:finalGR} is  
\be
\label{eq:GRres}
&{\cal R}_d(z) = c_d^{(0)} + c_d^{(1)} z\\
&- \Big. \sum_{k=0}^\infty
z^k \frac{\Gamma \left(d-\frac{1}{2}\right)  \left(-H_{-\frac{d}{2}+k-\frac{1}{2}}-2
   H_{d+k-2}+H_{d+k-\frac{3}{2}}+2 H_k\right) \Gamma
   \left(-\frac{d}{2}+k+\frac{1}{2}\right) \Gamma (d+k-1)^2}{\Gamma
   \left(\frac{1}{2}-\frac{d}{2}\right) \Gamma (d-1)^2 \Gamma (k+1)^2 \Gamma
   \left(d+k-\frac{1}{2}\right)}\nonumber
\ee
where the analytic part refers to the terms that are polynomial at small $z$ and $H_k$ are the harmonic numbers. The coefficients $c_d^{(0)}$ and $c_d^{(1)}$ are determined by imposing the momentum conservation Ward identities \eqref{eq:WI1} and \eqref{eq:WI2}. The analytic term in \eqref{eq:GRres} further has simple poles when $d$ is odd, these should be removed and the finite piece will be a polynomial of degree $(d-1)/2$. 

For $d>4$ there is a polynomial ambiguity in the result above related to the small impact parameter physics coming from the fact that
\be 
{\cal K}_d(u,z)(\Pi^{(GR)}(u))^2\sim u^{d-2} ( p_2(z) + u^2 p_3(z)+...)\frac{1}{u^{2(d-3)}}\sim \frac{( p_2(z) + u^2 p_3(z)+...)}{u^{d-4}},
\ee 
for small $u$ where $p_k(z)$ stands for a degree $k$ polynomial. In the computation above we regulated this divergence by analytically continuing in $d$. In the case of pure Einstein gravity, the ambiguity is a polynomial of degree $\lfloor \frac{d-1}{2}\rfloor$ for $d\geq 5$. More generally, for a $D$-dimensional bulk we will have $u^{d-2} {( p_2(z) + u^2 p_3(z)+...) \over u^{2(D-4)}}$ in the formula above. As we discuss in Section \ref{sec:corrections}, the divergence at small impact parameters can be regulated by the stringy effects, which leads to further stringy enhanced terms polynomial in $z$.

\section{$AdS_5 \times S^5$ supergravity}\label{Sec:N4SYM}
In the section above, we analyzed a universal non-analytic contribution to the energy-energy correlation in a theory with the Einstein gravity dual in $AdS_{d+1}$. In this section, we consider instead the canonical example of ${\cal N}=4$ SYM, where the dual geometry is $AdS_{5} \times S^5$, and the phase shift interpolates between the universal form used in the previous section at large impact parameters, and the $10d$ flat space one at small impact parameters. It is instructive to repeat the analysis of the previous section in this case. 

We proceed in the following steps: a) we use SUSY Ward identities to relate the energy-energy correlator to the scalar four-point function as in \cite{Belitsky:2013xxa}, see also \cite{Belitsky:2013bja,Belitsky:2013ofa,Henn:2019gkr}; b) we write down the OPE in the $\textbf{source} \times \textbf{detector}$ channel; c) we use the results for the diagonalization of the dilatation matrix of the double trace operators obtained in \cite{Aprile:2017bgs,Alday:2017xua} to effectively `eikonalize' them. The result of these steps is an OPE, which we can use to compute the universal ${\log c_T \over c_T}$ contribution to the energy-energy correlator. Moreover, we will see explicitly how the representation of the EEC in terms of the bulk phase shift arises from this point of view. 

Let us recall that in the case of ${\cal N}=4$ SYM and with our conventions reviewed in Appendix \ref{App:Conventions}
\be
\label{eq:centralchargeN4}
c_T = 40 (N^2 - 1),
\ee
where $N$ is the rank of the gauge group which we take to be $SU(N)$. It will be convenient for us, following \cite{Aprile:2017bgs}, to introduce instead
\be
a = {1 \over N^2 - 1} ,
\ee
and switch to $c_T$ at the very end.

We consider the four-point function of $20'$ half-BPS operators and we choose the $R$-symmetry polarizations as described in \cite{Belitsky:2013xxa}. We choose for the source operators $y_{14}^2=1$ so that
\be
\la O^\dagger(x_1) O(x_4) \ra = {1 \over x_{14}^4} \ . 
\ee
For detector operators, we choose $y_{23}^2=0$. We also choose $y_{12}^2 = y_{34}^2 = y_{13}^2 = y_{24}^2 = \left( {1 \over 8 (2 \pi)^4 a}\right)^{1/4}$ which simplifies the relationship to the energy-energy correlator.

With this choice, we have the following nonperturbative relationship due to superconformal symmetry
\be
\label{eq:SWI}
\la {\cal E}(n_2) {\cal E}(n_3) \ra = {4 p^4 \over (n_2 \cdot n_3)^2} \la \tilde O(n_2) \tilde O(n_3) \ra , 
\ee
which is strictly speaking valid only for $n_2^\mu \neq n_3^\mu$, see also \cite{Kologlu:2019mfz} for the complete relationship. At strong coupling this subtlety is not relevant because the energy-energy correlator does not contain any contact terms at $n_2^\mu = n_3^\mu$.

\subsection{The $\text{source} \times \text{detector}$ OPE}

We are interested in writing the OPE expansion of this correlator in the $\textbf{source} \times \textbf{detector}$ channel. Focusing on the relevant contribution of long (non-protected) superconformal multiplets, we get
\be
\label{eq:OPEsourcedet}
\la O^\dagger(x_1) \tilde O(x_2) \tilde O(x_3) O(x_4) \ra ={1 \over 8 (2 \pi)^4 a} {1 \over (x_{12}^2 x_{34}^2)^2} \sum_{t, \ell} \lambda^2_{t, \ell} u^2 v^2 F_{t, \ell}(z, \bar z) \ ,
\ee
where as usual $u = z \bar z$ and $v = (1-z)(1-\bar z)$. To get the formula above, we noticed that the projections described above correspond to $y=\bar y=1$ for the R-symmetry cross-ratios in \cite{Aprile:2017bgs}. Here $t ={\Delta - \ell \over 2}$ is half the twist of exchanged superconformal primaries. The superconformal blocks $F_{t, \ell}(z, \bar z)$ that appear in the formula above are simply related to the usual conformal blocks
\be
u^2 v^2 F_{t, \ell}(z, \bar z) = {v^2 \over u^2} G_{\Delta+4, \ell}(z, \bar z) \ . 
\ee

In \cite{Aprile:2017bgs} the matrix of double trace operators in the OPE expansion above was diagonalized using one-loop supergravity. We will use it in the computation of the energy-energy correlator. Hence, using the diagonalization result of \cite{Aprile:2017bgs}, we can rewrite the formula above as follows
\be
\la O^\dagger(x_1) \tilde O(x_2) \tilde O(x_3) O(x_4) \ra ={1 \over 8 (2 \pi)^4 a} {1 \over (x_{14}^2 x_{23}^2)^2} \sum_{t=2}^\infty \sum_{\ell = 0}^\infty \sum_{i=1}^{t-1} \lambda^2_{t, \ell , i} {v^4 \over u^4} G_{\Delta_{t, \ell, i}+4, \ell}(z, \bar z) \ ,
\ee
where $\Delta_{t, \ell, i} = 2 t + \ell + a \gamma_{t, \ell, i}$ and $i$ correspond to a sum over families of operators that are degenerate to leading order in $a$.

The contribution of interest comes from high energy scattering in the bulk closely related to the Regge limit. To approach the Regge limit we need to continue $ z \to e^{- 2 \pi i} z$ around the origin and then take $z, \bar z \to 1$ with $(1-z)/(1-\bar z)$ kept fixed. Upon doing this continuation, the s-channel conformal blocks acquire a simple but important phase
\be
&\la O^\dagger(x_1) \tilde O(x_2) \tilde O(x_3) O(x_4) \ra_{z \to e^{- 2 \pi i} z} \cr
&= {1 \over 8 (2 \pi)^4 a} {1 \over (x_{14}^2 x_{23}^2)^2} \sum_{t=2}^\infty \sum_{\ell = 0}^\infty \sum_{i=1}^{t-1} \lambda^2_{t, \ell , i} {v^4 \over u^4} e^{- i \pi a \gamma_{t, \ell, i}} G_{\Delta_{t, \ell, i}+4, \ell}(z, \bar z) \ .
\ee
This phase is directly related to the phase shift in the previous section. 

As discussed earlier, the combination relevant for the computation of the energy-energy correlation is $2 {\rm Re}\Big(e^{- i \pi a \gamma_{t, \ell, i}}  - 1 \Big)$. To proceed, it is convenient to first take the detector limit $x_{23}^2 \to -2 r_2 r_3 (n_2 \cdot n_3)$, see \cite{Belitsky:2013xxa} for details. In this way we get
\be
&\la O^\dagger(x_1) \tilde O(u_2, n_2) \tilde O(u_3, n_3) O(x_4) \ra_{\text{Regge}} \cr
&={1 \over 16 (2 \pi)^4 a} {1 \over x_{14}^4} {1 \over (n_2 \cdot n_3)^2} \sum_{t=2}^\infty \sum_{\ell = 0}^\infty \sum_{i=1}^{t-1} \lambda^2_{t, \ell , i} {v^4 \over u^4} {\rm Re}\Big(e^{- i \pi a \gamma_{t, \ell, i}}  - 1 \Big) G_{\Delta_{t, \ell, i}+4, \ell}(z, \bar z),
\ee
and the cross-ratios become
\be
u &= {((-x_1 \cdot n_2)-u_2)(u_3 - (-x_4 \cdot n_3)) \over ((-x_1 \cdot n_3)-u_3)(u_2 - (-x_4 \cdot n_2)) }, ~~~v = {x_{14}^2 (n_2 \cdot n_3) \over 2 ((-x_1 \cdot n_3)-u_3)(u_2 - (-x_4 \cdot n_2)) } . 
\ee
The Regge limit then corresponds to $u_2, u_3 \to \infty$ with $u_2 u_3 <0$. The same limit with $u_2 u_3 > 0$ corresponds to the Euclidean OPE limit and is the source of $-1$ in the formula above.

To leading order in the Regge limit we have
\be
u &= 1 - {(-x_{14} \cdot n_2) \over u_2} +  {(-x_{14} \cdot n_3) \over u_3} + \dots, ~~~ v ={x_{14}^2 (n_2 \cdot n_3) \over (-2 u_2 u_3) }+ \dots . 
\ee
Consider for example $u_3<0$ and $u_2>0$. We then introduce $u_2 = {1 \over \alpha_2}$ and $u_3 = - {1 \over \alpha_3}$. In this way we get to the light-ray transform $\int_0^\infty {d \alpha_2 d \alpha_3 \over \alpha_2^2 \alpha_3^2}$, where
\be
u &= 1 -(-x_{14} \cdot [\alpha_2 n_2 + \alpha_3 n_3]) , ~~~v ={x_{14}^2 (\alpha_2 n_2 \cdot \alpha_3 n_3) \over 2} \ . 
\ee
Our next computation step is to derive the `eikonal' representation for the conformal blocks. The Regge limit now is $\alpha_i \to 0$ with $\alpha_2/\alpha_3$ held fixed. The integration measure becomes
\be
\label{eq:measureN4}
\int_0^\infty {d \alpha_2 d \alpha_3 \over \alpha_2^2 \alpha_3^2} {1 \over x_{14}^4} {1 \over (n_2 \cdot n_3)^2} {v^4 \over u^4} ={x_{14}^4 (n_2 \cdot n_3)^2 \over 16} \int_0^\infty d \alpha_2 d \alpha_3 \ \alpha_2^2 \alpha_3^2 . 
\ee

The key aspect of the computation of interest is noticing that the quantum numbers that dominate as $\alpha_i \to 0$ are the following $i, t, \ell \sim {1 \over \sqrt{\alpha}}$. Let us introduce the following variables
\be
h+ \bar h &= \Delta, ~~~h - \bar h = \ell \ . 
\ee
We then have the following representation for the blocks, see e.g. \cite{Kulaxizi:2017ixa},
\be
\lim_{z,\bar z \to 1}\lambda^2_{GFF}(h,\bar h) G_{h, \bar h}(z, \bar z) &=(2 \pi^3)^2 \int_{M^+} {dp \over (2 \pi)^4} {d \bar p \over (2 \pi)^4} e^{i p \cdot x- i \bar p \cdot \bar x} \delta({p^2 \bar p^2 \over 16} - h^2 \bar h^2) \delta({p \cdot \bar p \over 2} + h^2 + \bar h^2) \nn \\ 
&\times 4 h \bar h (h^2 - \bar h^2),
\ee
where effectively we kept $(1-z)h^2$ and $(1-z)\bar h^2$ fixed.

In the formula above, we chose the three-point function in the generalized free field theory of non-identical operators of dimension $2$ (which means that the sum goes effectively over both odd and even spins).

Because in our case \eqref{eq:OPEsourcedet} the sum goes over even spins only we need to consider the ratio
\be
\lambda(h,\bar h, i) = {1 \over 2}{\lambda^2_{t, \ell , i} \over \lambda_{GFF}(t+\ell+2, t+2)} ={(16)^2 \over 6 \pi} {(h+i)^{3/2}(\bar h - i)^{3/2} i^{3/2}(h-\bar h+i)^{3/2} (h - \bar h + 2 i) \over h^4 \bar h^4} ,
\ee
where $\lambda_{GFF}(t, \ell)$ is the square of the three-point function in the generalized free field theory of nonidentical operators \cite{Bros:2011vh,Fitzpatrick:2011dm}. The shift $t+2$ in the denominator is due to the abovementioned shift $\Delta+4$ in the conformal block expansion.

In the language of the bulk phase shift analysis in \cite{Kulaxizi:2017ixa}, the formulas above arise if we identify
\be
x^\mu &= x_{14}^\mu , ~~~ \bar x^\mu = {1 \over 2} \left( \alpha_2 n_2^\mu + \alpha_3 n_3^\mu \right) \ .
\ee
The utility of the eikonal representation is that the dependence on $x$ and $\bar x$ is completely factorized. 

As an immediate result, we can trivially perform the Fourier transform with respect to $x_{14}^\mu$. Similarly, the integral over $\alpha_i$ trivializes and we get
\be
&\la \tilde O(n_2) \tilde O(n_3) \ra = -{1 \over 8 \pi a} (n_2 \cdot n_3)^2 \int_0^\infty d h \int_0^h d \bar h \int_0^{\bar h} d i \lambda(h,\bar h, i) \\
&\int_{M^+} {d \bar p \over (2 \pi)^4} { {\rm Re}\Big(e^{- i \pi a \gamma_{t, \ell, i}}  - 1 \Big)  \over (\bar p \cdot n_2)^3 (\bar p \cdot n_3)^3 } 4 h \bar h (h^2 - \bar h^2) \partial_p^4 \Big[ \delta({p^2 \bar p^2 \over 16} - h^2 \bar h^2) \delta({p \cdot \bar p \over 2} + h^2 + \bar h^2) \Big] , \nonumber
\ee
where $\partial_p^4$ comes from $x_{14}^4$ in the measure \eqref{eq:measureN4} that we conveniently re-expressed via the derivatives over $p^\mu$ and integrated by parts.

Next by noticing that $\delta$-functions only depend on $p^2 \bar p^2$ and $p \cdot \bar p$, we can exchange $\partial_p^4 = {\bar p^4 \over p^4} \partial_{\bar p}^4$ and again integrate by parts to get
\be
&\la \tilde O(n_2) \tilde O(n_3) \ra =- {1 \over 8 \pi a} {(n_2 \cdot n_3)^2 \over p^4} \int_0^\infty d h \int_0^h d \bar h \int_0^{\bar h} d i \lambda(h,\bar h, i) \\
&\int_{M^+} {d \bar p \over (2 \pi)^d} \partial_{\bar p}^4 \Big( { \bar p^4  \over (\bar p \cdot n_2)^3 (\bar p \cdot n_3)^3 } \Big) {\rm Re}\Big(e^{- i \pi a \gamma_{t, \ell, i}}  - 1 \Big)  4 h \bar h (h^2 - \bar h^2) \delta({p^2 \bar p^2 \over 16} - h^2 \bar h^2) \delta({p \cdot \bar p \over 2} + h^2 + \bar h^2) . \nonumber
\ee
We can take the derivative, and we find that
\be
\partial_{\bar p}^4\Big( { \bar p^4  \over (\bar p \cdot n_2)^3 (\bar p \cdot n_3)^3 } \Big) = {576 \over (\bar p \cdot n_2)^3 (\bar p \cdot n_3)^3} \left({2 \over 3} - 4 \chi + 4 \chi^2 \right),
\ee
where $\chi={\bar p^2 (n_2 \cdot n_3) \over 2 (\bar p \cdot n_2) (\bar p \cdot n_3)}$. 

We next consider the following change of variables
\be
h &= {1 \over 2} \sqrt{S} e^{L/2}, ~~~
\bar h = {1 \over 2} \sqrt{S} e^{-L/2}, ~~~
i = {1 \over 2} \sqrt{S} e^{-L/2} \alpha \  ,
\ee
so that $S=\sqrt{p^2 \bar p^2}$, $\cosh L = -{p \cdot \bar p \over \sqrt{p^2 \bar p^2}}$, and we then get for the anomalous dimension
\be
\gamma(S,L,\alpha) = -  {e^{3 L} S \over (e^{L}-1+2 \alpha)^{6}}. \ 
\ee

After performing straightforward but slightly tedious algebra and using the identity \eqref{eq:SWI}, we finally get the following result
\be
\la {\cal E}(n_2) {\cal E}(n_3) \ra =- {36 c_T \over 5 \pi} {p^4 \over (p \cdot n_2)^3 (p \cdot n_3)^3} \int_0^1 d u {\cal K}_4(u,z) {\rm Re}\langle e^{i \delta} - 1 \rangle \ , 
\ee
where we have defined
\be
\langle e^{i \delta} \rangle &= {64 \over 3 \pi} \int_0^1 d \alpha {e^{-5L} \over \sinh L} \left(e^{2 L}-1\right) \left((1-\alpha ) \alpha 
   \left(\alpha +e^L-1\right) \left(\alpha +e^L\right)\right)^{3/2}
   \left(2 \alpha +e^L-1\right) \nn \\
   &\times \int_{S_0}^\infty {d S \over S^3} e^{i \delta(S,L)}|_{\tanh L = u} \ .
\ee
The phase shift in the formula above is taken to be
\be
\delta(S,L,\alpha) = - {40 \pi \over c_T} \gamma(S,L,\alpha) , 
\ee
where we used \eqref{eq:centralchargeN4}. Let us notice the presence of the extra $\alpha$ integral in the formula above compared to a more naive representation \eqref{eq:psaver} used in the previous sections. Its origin is in the fact that we have five extra dimensions due to the fact that geometry is $AdS_5 \times S^5$. This leads to the fact that we have to consider the eigenstates of the eikonal operator which are different from a simple graviton state. The sum over these eigenstates becomes the $\alpha$ integral in the formula above.

We can now perform the integral over energies to get 
\be
\int_{S_0}^\infty {d S \over S^3}  {\rm Re}(e^{i \delta(S,L)}-1) = - {1 \over 2} {\log c_T \over c_T^2} (40 \pi)^2 \Big[{e^{3 L} \over (e^{L}-1+2 \alpha)^{6}} \Big]^2 + ... \ . 
\ee
Notice that for $\alpha=0$ the phase shift behaves as $1/L^6$ at small impact parameters. This is behavior is expected because in this limit we reproduce ten-dimensional flat space geometry. 

Next, we compute the $\alpha$ integral to get
\be\label{eq:PHN4}
{\rm Re}\langle e^{i \delta} - 1 \rangle = - {1 \over 2} {\log c_T \over c_T^2} (40 \pi)^2 
{(1-u^2)^3 \over 64 u^7} \ . 
\ee
We finally perform the $u$-integral to get 
\be
\la {\cal E}(n_2) {\cal E}(n_3) \ra =-{24 \over \pi^2} {\log c_T \log z \over c_T} {p^4 \over (p \cdot n_2)^3 (p \cdot n_3)^3} (1-36z+216 z^2-400 z^3 + 225 z^4) + \dots \ . 
\ee
When performing the $u$-integral we encounter divergence at small $u$ which as we discuss in the next section is regularized by the stringy effects and leads to stringy enhanced polynomial contributions to the expression above. 

As expected to leading order at small $z$ our ${\cal N}=4$ SYM result agrees with the computation in the previous section because it comes from the universal graviton exchange, however, the subleading corrections probe the geometry of the $S^5$. One diagnostic of large extra dimensions in the bulk in terms of CFT data was studied in \cite{Alday:2019qrf} and we here see that the enhanced contribution to the EEC provides another sharp sign of internal manifolds in the bulk dual.  

\subsection{The light-ray OPE}
In the previous section, we used the OPE in the $\textbf{source} \times \textbf{detector}$ channel to obtain the EEC in ${\cal N}=4$ SYM at strong coupling. In this section, we study the same EEC instead using the OPE in the $\textbf{detector} \times \textbf{detector}$ channel following \cite{Kologlu:2019mfz} and utilizing the known expression for the OPE coefficients analytically continued in spin $C^+(\Delta,J)$ obtained \cite{Alday:2017vkk}. The key point is that the ${\cal E} (n_1)\times {\cal E} (n_2)$ OPE is well defined when the Regge intercept $J_0<3$ which holds non-perturbatively in $c_T^{-1}$ but is violated perturbatively at one-loop. This is equivalent to the divergence at large energies that one gets from expanding the eikonal result $\sim e^{i SG_N\Pi(L)}$. From the point of view of ${\cal E} (n_1)\times {\cal E} (n_2)$ OPE this is reflected by a divergence of the one-loop OPE data at $J=3$.\footnote{In ${\cal N}=4$ SYM we use supersymmetry as before to relate the EEC to SSC of the scalar $20'$ operators which effectively shifts $J=3$ to $J=-1$.} The aim of this section is to explore this divergence and by regulating it reproduce the result obtained above.

The EEC can be written as \cite{Kologlu:2019mfz}
\be\label{eq:OPEeec}
&\la {\cal E} (n_2) {\cal E}(n_3) \ra=\frac{(-p^2)^4}{(- p\cdot n_2)^3(-p\cdot  n_3)^3} \int_{2-i\infty}^{2+i\infty}\frac{d\Delta}{2\pi i }C^+(\Delta,-1)\frac{\pi^2\Gamma(\Delta-2)}{2 \Gamma(\frac{\Delta-1}{2})^3\Gamma(\frac{3-\Delta}{2})}f_\Delta(z) .
\ee 
Here $C^+(\Delta,-1)$ are spin $J=-1$ conformal partial waves analytically continued from even spins (that is the $+$ signature), and the celestial blocks are given by \eqref{eq:celblock}. We close the contour to the right picking up the poles of $C^+(\Delta,-1)$ which are given by 
\be
C^+(\Delta,J) \sim -\frac{a_i}{\Delta-\Delta_i}
\ee 
where $a_i$ and $\Delta_i$ can be expanded in $1/c_T$\footnote{Here we use convention where $a_{i}^{(0)}\sim c_T$ which is natural when relating the scalar correlator to the stress tensor correlator.}
\be 
a_i &= a_i^{(0)}(1+c_T^{-1}a_i^{(1)}+c_T^{-2}a_i^{(2)}+\ldots),\\
\Delta_i &= \Delta_i^{(0)}+c_T^{-1}\gamma_i^{(1)}+c_T^{-2}\gamma_i^{(2)}+\ldots.
\ee 
Here $\Delta_i^{(0)}=4+2i+J$ correspond to the double-trace operators  ${\cal O}_p\square^n\partial_{\mu_1}\ldots\partial_{\mu_J}{\cal O}_p$ where again mixing will be important. 

Let us focus on the part that is non-analytic in $z$. At strong coupling, it first arises from the term $\langle a^{(0)} (\gamma^{(1)})^2\rangle$ which contributes $\partial_\Delta f_{\Delta}(z) |_{\Delta = 4+2i-1} \sim \log z$ to the energy-energy correlator, where the averaging goes over the degenerate operators. This quantity has been computed in supergravity in \cite{Alday:2017vkk} with the following result close to the relevant spin $J=-1$
\be
\label{eq:polethreep}
\langle a^{(0)} (\gamma^{(1)})^2\rangle_{n, J}  \sim {1 \over c_T} {(n+1)_4^3 \over (J+1)} + ... ,
\ee
where $(a)_n$ is the Pochammer symbol. Compared to the tree-level supergravity computation, the new feature of the result above is a pole at $J=-1$, which effectively makes the formula \eqref{eq:OPEeec} inapplicable. The origin of this pole is the same enhanced perturbative Regge behavior that we discussed before. This pole is, therefore, an artifact of the large $c_T$ perturbative expansion.

Compared to the phase shift analysis, we do not have a first principle way to see how this pole gets regularized at finite $c_T$. We proceed by noting the following simple fact. Let us assume that the pole is regularized as ${1 \over J+1} \to c_0 \log c_T$ at finite $c_T$. We then get the following prediction
\be
{\cal F}(z) &= {c_0 \log c_T \log z \over 8 \pi^2 c_T} \left({-1 \over 6} \right)\sum_{n=0}^\infty {(-1)^n (n+1)_4^3 \Gamma(n+3)^2 \over \Gamma(5+2n)} f_{7+2n}(z) \cr
&=- 768 c_0 {\log c_T \log z \over  (4 \pi)^2  c_T} \left( 1-36 z +216 z^2 -400 z^3 + 225 z^4 \right) \ ,
\ee
where recall that $\la {\cal E} (n_2) {\cal E}(n_3) \ra=\frac{(-p^2)^4}{(- p\cdot n_2)^3(-p\cdot  n_3)^3} {\cal F}(z)$ and $(a)_m$ is the standard Pochhammer symbol. Our result agrees with the analysis in the dual channel. By matching the overall coefficient we can fix $c_0 = 1/2$.

The complete structure of the OPE data in the $1/c_T$ perturbation theory involves many more finite spin singularities beyond the simple pole (and its higher order in $1/c_T$ analogs) discussed in this section, see e.g. \cite{Aprile:2017bgs,Alday:2017vkk,Bissi:2020wtv,Drummond:2022dxw} for explicit results. It would be very interesting to understand in detail how all these singularities get resolved at finite coupling. Among other possible effects, it would require understanding the unitarization of the zero impact parameter scattering.

\subsection{The $\omega$-deformation}
In this section we review yet another way to regulate the divergence at large energies, or equivalently large detector times, in the $c_T^{-1}$ expansion of the EEC, again considering the specific example of ${\cal N}=4$ SYM. Instead of considering energy correlators which measure the total energy flux deposited on the celestial sphere in \cite{Korchemsky:2021okt} a generalized energy detector was introduced
\be 
{\cal E}(\hat \omega,n) ={1 \over 4}\int_{-\infty}^{\infty} d u e^{-i\hat{\omega}u} \lim_{r \to \infty} r^{d-2} T_{\mu \nu}(u,r \vec n) \bar n^\mu \bar n^{\nu}
\ee 
See also \cite{Belin:2020lsr,Besken:2020snx} where similar operators are studied. The generalized energy detectors have an effective time resolution $\tau\sim\frac{1}{\hat{\omega}}$ and when $\hat{\omega}\to 0$ the measurement over all times as in the original energy detectors are recovered. 

In \cite{Korchemsky:2021okt} the one-loop generalized energy-energy correlation in ${\cal N}=4$ SYM at strong coupling was calculated at one-loop\footnote{We took the formula (7.6) in \cite{Korchemsky:2021okt} and noticed that $c_T^{there} = {c_T \over 160}$.}
\be 
\label{eq:n4omreg}
\la {\cal E}(\omega_1,n_1){\cal E}(0,n_2)\ra = \frac{25}{c_T} {(p^0)^2 \over \pi^2}\Big[\frac{1-6z+6 z^2}{6\omega_1^2}+\frac{3}{2}\frac{2-15z+24z^2-10z^3}{\omega_1}\cr
+\frac{12}{25}(1-36z+216z^2-400z^3+225z^4)\log|\omega_1|\log(z^2|\omega_1|)\Big],
\ee 
where the dimensionless frequency $\omega_1=\frac{2\hat{\omega}_1(p\cdot n_1)}{p^2}$ was introduced. This diverges when taking the energy detector limit $\omega\to 0$. Cutting of the detector time at times $\tau\sim c_T\sim\frac{1}{\omega}$ one obtains in particular the non-analytic in $z$ term 
\be 
\la {\cal E}(\omega_1,n_1){\cal E}(0,n_2)\ra= -{24 (p^0)^2\over \pi^2} {\log c_T \log z \over c_T} (1-36z+216z^2-400z^3+225z^4),
\ee 
in agreement with the previous results. Other divergences are related to the small impact parameter physics and we expect these to get resolved by the stringy effects as we discuss in the next section.

\section{Bulk phase shift and the light-ray OPE}\label{Sec:LightRayOPE}

In the context of the energy-energy correlator, its small angle expansion and the discontinuity \eqref{eq:DiscDef} are conveniently captured by the light-ray OPE \cite{Kologlu:2019mfz}, which we can schematically write as
\be
\label{eq:LROPE}
{\cal F}(z)  = \sum_i  \lambda_i f_{\Delta_i(3)}(z),
\ee
where the sum goes over signature-plus operators of spin $J=3$ (for spinning targets, higher spin operators contribute as well \cite{Chang:2020qpj}), $\lambda_i$ are related to the corresponding three-point functions, and we recall that the celestial block $f_\Delta(z)$ was given in \eqref{eq:celblock}. Notice that the sum in the formula above goes both over the single- and double-trace operators of spin three. It is interesting to ask how this connects to the regime of the EEC studied in this paper.

\subsection{Planar correlator}
Let us first consider the case of the planar correlator, the context in which the formula \eqref{eq:LROPE} has been applied and analyzed in detail. Notice that our analysis based on the phase shift and approximations we made when deriving the formula \eqref{eq:eecRep1} are centered around the high-energy enhanced contributions. In the planar theory, the contribution of the high-energy/Regge limit is not enhanced compared to the rest of the integral over the retarded time. Therefore we do not expect our phase shift formulas to capture the full result. It is still interesting to proceed and see what we get.

In the planar theory, the phase shift formula for the energy-energy correlator takes the following form
\be
\label{eq:StringyEEC}
{\cal F}(z)&= 8\rho_T c_T\Gamma(d+1)^2  \int_0^1 d u {\cal K}_d (u,z) \int_{S_0}^\infty {d S \over S^3}  {\rm Im} \delta(S,L(u) )+\ldots,
\ee
where the ellipses refer to the contribution not captured by the phase shift formula.

Next, we consider the contribution of a single-trace Regge pole to the phase shift $\delta(S,L)$. To this extent, we introduce the single Regge pole model (see \cite{Costa:2012cb,Costa:2017twz}):
\be 
\label{eq:phaseshiftRegge}
\delta_{\text{stringy}}(S,L) &= {1 \over {\cal N}}\frac{i\pi^{d/2}}{4}\int_{-\infty}^\infty d\nu \frac{e^{-i\pi j(\nu)/2}}{\sin\pi\frac{j(\nu)}{2}}r(\Delta(j(\nu)),j(\nu))j'(\nu)S^{j(\nu)-1}{\cal H}_{i\nu+\frac{d}{2}-1}(L), \nn \\
r(\Delta,J)&= \lambda_{TT{\cal O}_{\Delta,J}}\lambda_{{\cal O}_s{\cal O}_s{\cal O}_{\Delta,J}}K_{\Delta,J},
\ee
where, as throughout this paper, we only consider the simplest diagonal in the polarization phase shift. Notice that at finite $\lambda$ other tensor structures are expected to appear, see e.g.\ \cite{Costa:2017twz}, and it would be interesting to analyze the phase shift including them. This however goes beyond the scope of the discussion in the present section, which we expect to be qualitatively the same in the presence of these extra tensor structures. For strongly coupled theories the expected leading correction to the Regge intercept takes the form
\be\label{eq:ReggeTraj}
j(\nu) \simeq 2-2\frac{\nu^2+(\frac{d}{2})^2}{\Delta_{\text{gap}}^2} \ ,
\ee 
In this formula $\Delta_{\text{gap}}$ encodes the mass of the lightest higher-spin operator so that $j(i \Delta_{\text{gap}}) \simeq 4$. In \eqref{eq:phaseshiftRegge} $\lambda$'s are the OPE coefficients and $K_{\Delta,J}$ is a ratio of gamma functions and can be found e.g.\ in Eq.\ (41) in \cite{Costa:2012cb}. Moreover we have used the symmetry $\nu\to-\nu$ to write it in terms of the propagator ${\cal H}_{i\nu+\frac{d}{2}-1}(L)$.

Performing the integral over $S$ in \eqref{eq:StringyEEC} produces a factor $-{S_0^{j(\nu)-3} \over j(\nu)-3}$. Let us now close the $\nu$-contour. We are interested in the contributions from the pole $j(\nu)=3$. Computing the residue, we see that the dependence on $S_0$ drops out and we get
\be
\int_{- \infty}^\infty d \nu \left( -{S_0^{j(\nu)-3} \over j(\nu)-3} \right) j'(\nu) = 2 \pi i \ \Big|_{j(\nu_*)=3} \ . 
\ee
In other words, the spin-$3$ operator on the single-trace Regge trajectory contributes. Extra poles coming from $\frac{e^{-i\pi j(\nu)/2}}{\sin\pi\frac{j(\nu)}{2}}$ do not contribute to the imaginary part of the phase shift and therefore we can neglect them. 

Let us now discuss the $u$ integral. On the pole above we are left with an integral of the following form
\be
\label{eq:Reggeblock}
R_\Delta(z) \equiv \int_0^1 du {\cal K}_d (u,z) {\cal H}_{\Delta - 1}(L(u)) .
\ee
This integral is computed in \eqref{eq:resHInteg} and the result is that up to analytic in $z$ terms it reproduces precisely the celestial block that appeared in the light-ray OPE \cite{Kologlu:2019mfz}
\be
R_\Delta(z) \sim f_\Delta(z) + (\text{analytic in } z),
\ee
where the second term describes the contribution which is analytic around $z=0$ and can be thought of as the contribution of the double-trace operators to the light-ray OPE. Moreover, $R_\Delta(z)$ is regular around $z=1$ as opposed to $f_\Delta(z)$. Another interesting difference is the behavior when $\Delta = 2 d - 1 + 2 n$ in which case $f_{\Delta}(z)$ becomes analytic in $z$, whereas $R_\Delta(z) \sim \log z$.

On the energy-energy correlator side there is no reason to expect that the result is analytic around $z=1$. The opposite is true, and there is a finite coupling formula for the expected behavior of the energy-energy correlator close to $z=1$ in the planar ${\cal N}=4$ SYM \cite{Korchemsky:2019nzm}. In the $\textbf{detector} \times \textbf{source}$ OPE channel these arise from twist-two operators as opposed to the double twist operators studied in the present paper. In this regime, the correlator is controlled by the light-like Wilson loop and the detectors are localized close to the light cone of the source far from the endpoints for which we expect the phase shift formula to work.

To summarize, we get that the Regge pole expansion of the phase shift produces the following contribution to the energy-energy correlator
\be
\label{eq:LROPEregge}
{\cal F}(z)  = \sum_{i, \text{single-trace}}  \tilde \lambda_i R_{\Delta_i(3)}(z) + ... \ .
\ee
Comparing this formula to the exact result \eqref{eq:LROPE}, we see that the phase shift formula correctly captures terms which are non-analytic around $z=0$, but misses some analytic terms. In particular, it does not capture the $z=1$ non-analyticity as discussed above.  

\subsection{Non-planar correlator}
Let us consider next non-planar correlators for which we expect the high-energy enhanced terms to be correctly captured by our phase shift formula. A good toy model for the unitarization discussed in the present paper is the following integral
\be
I(J) =\int_1^\infty {dS \over S^J} \sin^2 {S \Pi \over c_T}.  
\ee
This integral is analytic for ${\rm Re}\, J > 1$. We can rewrite it as follows
\be
\label{eq:mellintransform}
I(J) =-{1\over 2}\sin {\pi J \over 2} \Gamma(1-J) \left({2 \Pi \over c_T}\right)^{J-1} -\int_0^1 {dS \over S^J} \sin^2 {S \Pi \over c_T} .
\ee
The second term generates the familiar $1/c_T^n$ perturbative expansion with poles at $J=3,5,7,\ldots$. The first term is a `background' integral, which cancels these poles. 

It is also instructive to write the inverse of the integral above, namely 
\be
\theta(S>1) \sin^2 {S \Pi \over c_T} = \int_{1- i \infty}^{1+ i \infty} {d J_L \over 2 \pi i} S^{J_L - 1} I(J_L) ,
\ee
where the contour traverses the singularity at $J=1$ from the right.
For $S<1$ we can close the contour to the right and we get zero. For $S>1$, we close the contour to the left in the second term of \eqref{eq:mellintransform} and to the right in the first term, reproducing the LHS.

We can now model ${\rm Re}(e^{i \delta}-1)$ by the formula above. Doing the integral over $S$ we get the following representation for the energy-energy correlator
\be
\label{eq:nonplanarEEC}
{\cal F}(z)&= 8\rho_T c_T\Gamma(d+1)^2  \int_0^1 d u {\cal K}_d (u,z) \nn \\
&\int_{1- i \infty}^{1+ i \infty} {d J_L \over 2 \pi i} {1\over 2}\sin {\pi J_L \over 2} \Gamma(1-J_L) \left({2 \Pi(L(u)) \over c_T}\right)^{J_L-1} {1 \over J_L-3}  .
\ee
To perform the $u$ integral we can use the following formula
\be
( \Pi(L) )^{J_L - 1} = \sum_{n=0}^\infty c_n(J_L) {\cal H}_{[(J_L-1)(d-1)+2n+1]-1}(L) ,
\ee
where the $c_n$'s can be iteratively found by expanding both sides at large $L$.

We can now perform the $u$ integral using \eqref{eq:Reggeblock} to finally get
\be
\label{eq:nonplanarEECb}
{\cal F}(z)= 4\rho_T c_T\Gamma(d+1)^2 \sum_{n=0}^\infty &\int_{1- i \infty}^{1+ i \infty} {d J_L \over 2 \pi i} \sin {\pi J_L \over 2} \Gamma(1-J_L) \left({2 \over c_T}\right)^{J_L-1} \nn \\
&\times {c_n(J_L) R_{[(J_L-1)(d-1)+2n+1]-1}(z) \over J_L-3}  .
\ee
As expected we get a finite result at finite $c_T$. If we now however try to perform the large $c_T$ expansion and close the contour to the right, we see that we get a double pole at $J_L=3$ whose residue contains the term
\be
\label{eq:nonplanarEECc}
{\cal F}(z)= 8\rho_T \Gamma(d+1)^2 {\log c_T \over c_T}\sum_{n=0}^\infty c_n(3) R_{[2(d-1)+2n+1]-1}(z)  ,
\ee
which is the focus of the present paper. 

Let us consider \eqref{eq:nonplanarEECc} explicitly in $d=4$. Decomposing the square of the phase shift in terms of ${\cal H}_{(7+2n)-1}(L)$ we find
\be\label{eq:d4DecompText}
(\Pi_{d=4}^{(GR)}(L))^2 = 3200\pi^2\sum_{n=0}^\infty {\cal H}_{(7+2n)-1}(L).
\ee
Combining the decomposition \eqref{eq:d4DecompText} together with \eqref{eq:nonplanarEECc} we reproduce the expected result. While analytic terms in $z$ get contribution from all $n$ in \eqref{eq:d4DecompText}, the non-analytic terms $z^k\log z$ are only sensitive to terms in \eqref{eq:d4DecompText} with $n\leq k$.  

In a related context, the representation of the correlator in the Regge limit as an integral over complex spins is very natural when discussing the unitarization of gravitational scattering. It was discussed in the context of the conformal partial wave expansion in \cite{Caron-Huot:2020nem}. 

\section{Corrections}\label{sec:corrections}
In this section we discuss various corrections to the results obtained in the previous sections. Let us recapitulate its basic logic. We consider the energy-energy correlator at strong coupling. 
It is convenient to write it using the $\textbf{source} \times \textbf{detector}$ OPE channel. The leading non-planar correction, as captured by the one-loop computation in gravity, is encoded in the contribution of the double-trace operators to the OPE. They, however, produce a divergent answer at order $1/c_T$. This divergence is an artifact of the perturbative expansion and can be regularized by `unitarizing' (or eikonalizing \cite{Fitzpatrick:2015qma}) the anomalous dimensions of the double trace operators, which effectively changes $\gamma_{\text{tree}}^2 \to \sin^2 {\pi \gamma_{\text{tree}} \over 2}$. This produces a finite result computed in the paper (up to additional subtleties related to scattering at small impact parameters).

In this section, we discuss various corrections to this simple universal picture. From the point of view of the OPE, it is related to the presence of infinitely many other operators in the OPE, both single- and multi-traces. Our goal here is to provide some rationale and physical intuition for why we believe that the computation done in the previous sections does capture the leading quantum-gravity correction to the energy-energy correlator.

Let us start by recalling the general OPE formula
\be\label{eq:OPERep2}
{\cal F}(z) &= {\rm Re} \sum_{\Delta, J} {|\lambda_{\Delta,J}|^2 \over |\lambda_{\Delta,J}^{GFF}|^2} \left(1 - e^{- i \pi \gamma_{\Delta,J}} \right) \left( |\lambda_{\Delta,J}^{GFF}|^2 {\cal P}_{\Delta,J}(z) \right).
\ee
This paper concerns with the high-energy or large $\Delta$ part of the sum --- the only part that could produce $\log c_T$ enhancement in the energy-energy correlator. The reason is that it is that part that corresponds to high-energy scattering in the bulk and it is that part that produces the contribution which gets enhanced. The behavior of the CFT three-point functions at large $\Delta$ is universal \emph{on average} in any CFT \cite{Pappadopulo:2012jk,Qiao:2017xif,Mukhametzhanov:2018zja}, as it is responsible to correctly reproducing the unit operator in the dual channel. In other words, we have $\overline{{|\lambda_{\Delta,J}|^2 \over |\lambda_{\Delta,J}^{GFF}|^2}}=1$, where the averaging can be made very precise. The behavior of the phase, on the other hand, is non-universal and depends on the nature of the Regge limit of the four-point function, see e.g. \cite{Caron-Huot:2020ouj} and the discussion of opaque versus transparent theories there. The phase shift description should capture the large $(\Delta,J)$ behavior of the sum on average: $\left( |\lambda_{\Delta,J}^{GFF}|^2 {\cal P}_{\Delta,J}(z) \right)$ becomes the kinematical kernel against which the phase shift is integrated; $\overline{{|\lambda_{\Delta,J}|^2 \over |\lambda_{\Delta,J}^{GFF}|^2}}$ is related to mixing between the particle dual to the source and other degrees of freedom; $e^{- i \pi \gamma_{\Delta,J}}$ becomes the phase shift itself. We then use the phase shift description to characterize the large $\Delta$ behavior of the OPE sum.

Our basic conclusion is that extra contributions produce subleading terms compared to the ${\log c_T \over c_T}$ term computed in the previous sections, up to polynomial in $z$ ambiguities due to zero impact parameter scattering. 
Recall the relationship between $S$ and $u=\tanh L$ and the quantum numbers of operators that appear in the $\textbf{source} \times \textbf{detector}$ OPE channel
\be\label{eq:SLrel}
S &=\Delta^2 - J^2 \ , ~~~\cosh L = {\Delta^2+J^2 \over \Delta^2 - J^2} \ .
\ee
In the following sections, we discuss the contribution from various regions of $(\Delta,J)$, or using \eqref{eq:SLrel}, $(S,L)$.

\begin{figure}
    \centering
    \includegraphics[width=\textwidth]{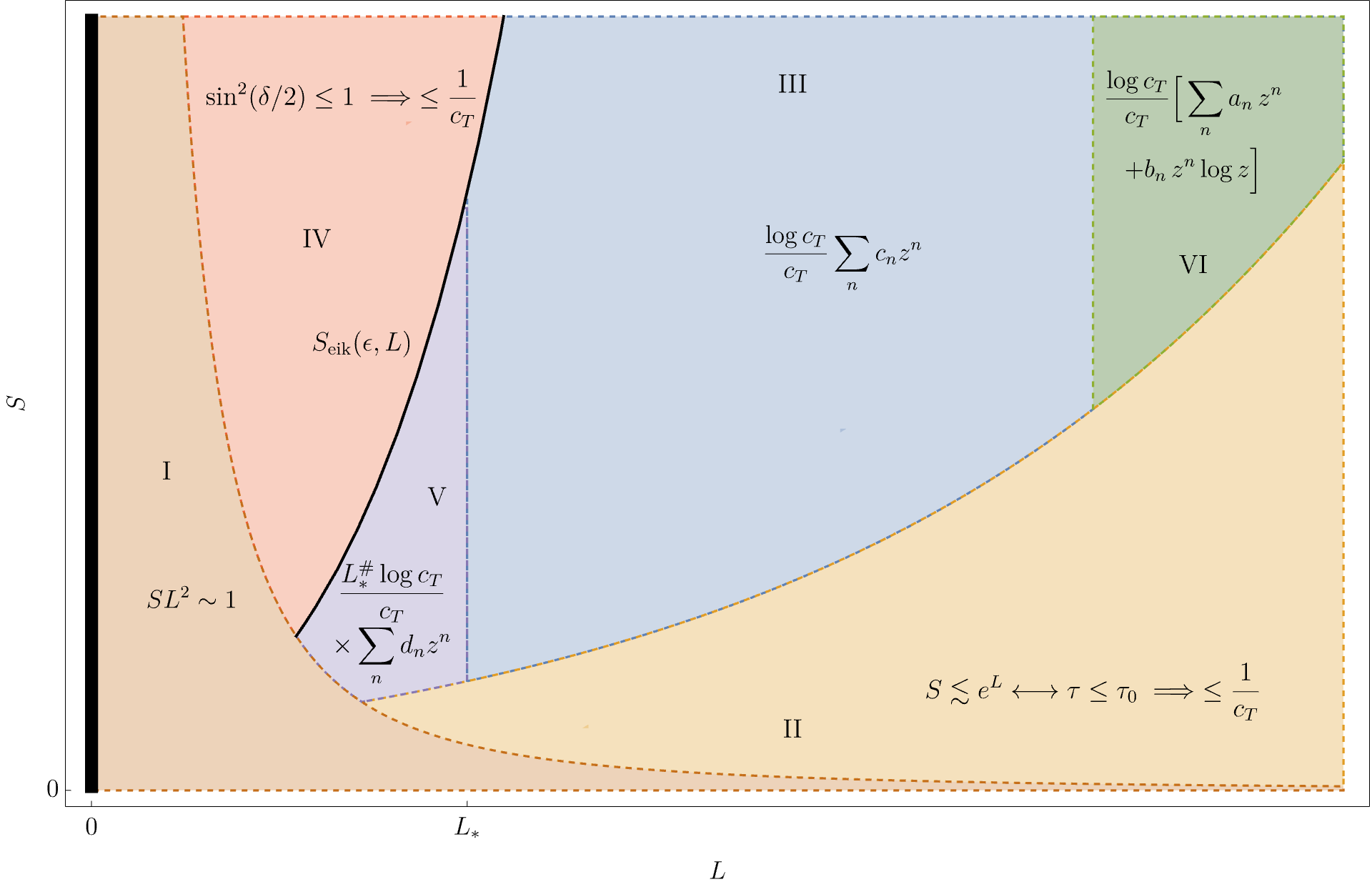}
    \caption{Schematic view over the $(S,L)$ plane and the various contributions when $L_*\ll 1$, where $L^2_*=\frac{\log c_T}{\Delta_{\text{gap}^2}}$.  Region I (brown) contains fixed spin operators which are UV-sensitive and lead to polynomial ambiguities, see Sec.\ \ref{sec:fixspin}. Region II (yellow) contains fixed twist operators and does not lead to enhanced contributions, see Sec.\ \ref{sec:fixtwist}. Region III (blue) denotes Planckian energies and it is the region that contributes to the enhanced corrections mostly discussed throughout this paper, see Sec.\ \ref{sec:Planckian}. Note that the enhancement comes from integrating $S$ all the way to $S_{\text{eik}}(\epsilon,L)\sim c_T$, if we were to introduce an upper-cutoff $S_1\sim O(1)$ in this region this would not contribute to the enhanced term. We hope this does not cause confusion. The large impact parameter regime of this region is further depicted by region VI (green) which leads to the non-analytic terms in $z$. The Trans-Planckian region IV (red) $S\geq S_{\rm eik}(\epsilon,L)$ (thick black line), we bound the contribution using unitarity, and no enhanced terms are produced, see Sec.\ \ref{sec:TransPlanck}. Region V (purple) denotes the regime of further stringy enhancement at small impact parameters, see Sec.\ \ref{Sec:Stringy}. Finally, the black solid line refers to zero-impact parameter physics.}
    \label{fig:regions}
\end{figure}

\subsection{Scattering at small impact parameters: fixed spin operators}\label{sec:fixspin}

From the formulas above, we immediately see that the contribution of fixed $J_0$ and arbitrary $\Delta$ operators to the OPE corresponds to the zero impact parameter scattering in the limit $S \to \infty$. A slightly more refined statement using the formulas above is that $S L^2 \simeq 4 J_0^2$ as we take $S \gg 1$. The phase shift discussion in terms of the impact parameters only makes sense when the Compton wavelength is much smaller than the impact parameter $\lambda_C \sim {1 \over \sqrt{S}} \ll L$ which translates to $J_0 \gg 1$.

The contribution of finite $J_0$ operators to the energy-energy correlator is given by a degree $J_0$ polynomial in $z$. Without having an explicit UV completion of a given gravitational theory, we do not have any control over such contributions. Therefore, the $S L^2 \lesssim 1$ small impact parameter region of the $(S,L)$ plane is UV sensitive, leading to polynomial in $z$ ambiguity of our results, which we cannot control. At leading order in $1/c_T$ this ambiguity is related to the one-loop renormalization of the four-point correlator ambiguity which is fixed in the complete theory. We mark this region by I (brown) in \Fig{fig:regions}.

In holographic theories, such terms arise from polynomial contributions to the Mellin amplitude underlying the four-point function. They have been explicitly computed using localization in both string/M-theory UV completions of gravity, see e.g.  \cite{Binder:2019mpb,Chester:2019pvm,Chester:2020dja}.

At high enough energies we expect black hole formation to control the small impact parameter scattering \cite{Eardley:2002re,Giddings:2004xy}. We can model this process by a simple black disc model such that ${\rm Re} (e^{i \delta(S,L)}-1) = - 1$ for $L<L_{Sch}(S) \sim \log {\sqrt{S} \over c_T}$, where $L_{Sch}(S)$ is the Schwarzschild radius in AdS, see e.g.\ \cite{Cornalba:2010vk,Brower:2006ea}. Plugging this into \eqref{eq:eecRep1}, we get the following formula for the contribution of black hole production at order ${1 \over c_T^3}$ 
\be 
\label{eq:eecRep1BH}
{\cal F}_{\text{BH}}(z) = 
8\rho_T c_T\Gamma(d+1)^2\int_0^1 d u {\cal K}_d (u,z) \int_{c_T^2 e^{2 (d-1) L}}^\infty {dS \over S^3} (-1) \sim {z^{d-1} \log z \over c_T^3} ,
\ee
where we only kept the leading non-analyticity at small $z$. It would be interesting to understand if this contribution can be unambiguously isolated from other effects. 

\subsection{Scattering at low energies: fixed twist operators}\label{sec:fixtwist}

Next, we can consider scattering of fixed twist and possibly large spin operators. In terms of $(S,L)$ we have
\be
\tau \equiv \Delta - J = \sqrt{S} e^{-L/2} . 
\ee
The contribution of such operators can be computed by simply expanding the OPE data in ${1 \over c_T}$ and in agreement with the usual perturbation theory of local operators, we will get only integer powers of ${1 \over c_T^n}$. 
To complete the argument we need to make sure that summing over spins these perturbatively expanded results does not produce a divergence. To leading order in $1/c_T$ relevant for this paper, the operators contributing are the double-twist operators for which one can use light-cone bootstrap techniques \cite{Fitzpatrick:2012yx,Komargodski:2012ek} to analyze the large spin tails explicitly. In this case, the sum over spins corresponds to scattering at very large impact parameters, and while it can produce enhanced terms in the energy-energy correlator at small $z$ after integration against our kernel, its $c_T$ dependence is not affected. We expect that the same statement applies to the multi-twist operators at higher orders in $1/c_T^{n>1}$, and it would be interesting to check it explicitly using the higher-point version of the light-cone bootstrap \cite{Antunes:2021kmm}, see also \cite{Fardelli:2024heb,Kravchuk2024}. Therefore, we conclude that operators with a bounded twist cannot produce an enhanced $\log c_T/c_T$ contribution and instead produce a subleading contribution, which is $O(1/c_T)$. We mark this region by II (yellow) in \Fig{fig:regions}.

\subsection{Non-phase shift contributions: outlier Regge trajectories}\label{sec:NotPh}

In the picture above we assumed that the large $(\Delta,J)$ limit of the OPE sum that controls the correlator is smooth and exists, an assumption that we certainly cannot rigorously justify. If we imagine an isolated family of operators, which is not part of this smooth limit our formulas will certainly miss it. We can call such contributions \emph{non-phase shift contributions}. To be concrete, we can imagine the contribution of the twist-two operators to the sum above which are responsible for non-analytic behavior of the energy-energy correlator around $z=1$.\footnote{In \cite{Korchemsky:2019nzm,Chen:2023wah}, it has been shown that the discontinuity of energy-energy correlator in the $z \to 1$ limit corresponds to the double lightcone limit of the underlying four-point function of local operators, which is controlled by the large spin tail of the twist-2 operators in the OPE.
} 
As we discussed in the previous section, we do not expect it to be captured by the approximation used to derive the phase shift formula. We however do not expect such contributions to be enhanced, and instead follow the usual $1/c_T$ perturbative pattern. For that reason we expect them to be sub-leading compared to the enhanced contribution analyzed in the present paper.

\subsection{Planckian energies}\label{sec:Planckian}

We would like next to understand the contribution from Planckian energies $S \leq S_{\text{eik}}(\epsilon, L)$, where we define $S_{\text{eik}}(\epsilon, L)$ through the following relation
\be
\delta_{\text{tree}}( S_{\text{eik}},L) = \epsilon ,
\ee
where $\epsilon$ is a fixed number which we can also take to be small;  $\delta_{\text{tree}}(S,L)$ is the tree-level phase shift $\sim {1 \over c_T}$. We mark this region by III (blue) in \Fig{fig:regions}.

Here comes our main assumption: in the regime where $S L^2 \gg 1$ and energies such that $\delta_{\text{tree}}(S,L) \leq \epsilon \ll 1$ the full phase shift can be bounded as
\be
\label{eq:basicassumption}
| \delta(S,L) | \leq c_0 \delta_{\text{tree}}(S,L) ,
\ee
where $c_0$ is an $O(1)$ numerical coefficient that does not depend on $S$ and $L$. Physically, we think of the condition \eqref{eq:basicassumption} as the statement that as long as the tree-level gravitational interaction is very weak $\delta_{\text{tree}}(S,L) \leq \epsilon \ll 1$ quantum-gravity corrections stay small as well. 

Let us first discuss this statement in $D$-dimensional asymptotically flat spacetime.\footnote{It was used in \cite{Haring:2022cyf} to derive an analogue of the Froissart-Martin bound on the growth of the amplitude with energy in gravitational theories.} In this case we can rewrite the tree-level phase shift as follows $\delta_{\text{tree}}(S,L) \sim (\sqrt{S}L) ({L_{Sch}(S) \over L})^{D-3}$, where we used that in flat space the Schwarzschild radius $L_{Sch}(S)^{D-3} \sim G_N \sqrt{S}$. The theory is expected to become strongly coupled for the collisions with $L \sim L_{Sch}(S)$ when a black hole is expected to form \cite{Eardley:2002re,Giddings:2004xy}. Given that $S L^2 \gg 1$ and that we keep $\delta_{\text{tree}}(S,L) \leq \epsilon \ll 1$, we see that the scattering takes place at $L \gg L_{Sch}(S)$, therefore we expect it to be weakly coupled and well-controlled by the tree-level result.

Consider next the situation in $AdS_{d+1}$. The condition for the black hole formation above could have been obtained by imposing that the transferred momentum as measured by $\partial_L \delta_{\text{tree}}(S,L)$ becomes $\sim \sqrt{S}$. Applying the same condition for $L \gg 1$ we get that in AdS we expect that the black hole formation corresponds to $e^{(d-1)L_{Sch}(S)} \sim G_N \sqrt{S}$. Numerical analysis of the formation of a trapped surface in the shock wave collisions in AdS was performed in \cite{Lin:2009pn,Duenas-Vidal:2010xhp}, see also \cite{Gubser:2009sx} for analytic results, and it supports this conclusion, see formula (3.19) in \cite{Duenas-Vidal:2010xhp}. We can therefore rewrite the tree-level phase shift as $\delta_{\text{tree}}(S,L) \sim (\sqrt{S}L) e^{(d-1)(L_{Sch}(S) - L)}$ and thus our regime of interest is again $L \gg L_{Sch}(S)$.

Using the inequality \eqref{eq:basicassumption}, we can estimate
\be
\label{eq:estintermen}
&\Big| \int_{S_0}^{S_{\text{eik}}(\epsilon, L)} {d S \over S^3} {\rm Re} ( 1 + i \delta - {\delta^2 \over 2} - e^{i \delta}) \Big| \leq \sum_{k=3}^\infty {c_0^k \over k!} \int_{S_0}^{S_{\text{eik}}(\epsilon, L)} {d S \over S^3} \delta_{\text{tree}}^k (S,L) \nn \\
&\leq g(c_0,\epsilon) {\Pi_{d-1}^2(L) \over  c_T^2} ,
\ee
where $g(c_0,\epsilon)$ can be easily computed explicitly but does not play any role, and we also used the Einstein gravity result for $\delta_{\text{tree}}(S,L)$. As a result, the higher-order terms produce non-enhanced $1/c_T$ contributions to the EEC.

We are thus left with the following terms
\be
\int_{S_0}^{S_{\text{eik}}(\epsilon, L)} {d S \over S^3} \left( {\rm Im} \delta + {1 \over 2} \left( {\rm Re} \delta^2 - {\rm Im} \delta^2   \right) \right) . 
\ee

To discuss the enhanced terms in the integral above we can consider the perturbative expansion of the phase shift
\be
{\rm Re} \delta(S,L) = \delta_{\text{tree}}(S,L) + \sum_{i=2}^\infty \sum_{k = 0}^\infty \left({S \over c_T} \right)^i (\log S)^k c_{i,k}(L) ,
\ee
where we only kept the corrections that could produce the enhanced terms, namely $\sim c_{i,k}(L) {\log^k c_T \over c_T^2}$. Notice, however, that the terms with $k>0$ necessarily violate our basic assumption \eqref{eq:basicassumption}. 

\begin{figure}
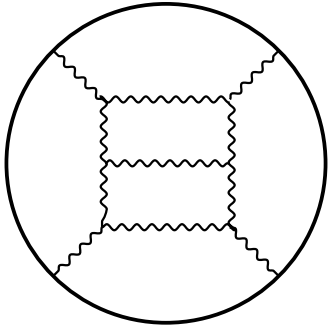

    \centering
    \HDiagram
    \caption{The H-diagram in AdS. The simplest AdS diagram that corresponds to the three-graviton state in the $s$-channel cut of the four-point graviton correlator.}
    \label{fig:hDiagram}
\end{figure}

Similarly, we get for the imaginary part that the only term that is consistent with \eqref{eq:basicassumption} that can produce the enhanced contribution is 
\be
{\rm Im} \delta(S,L) = \left({S \over c_T} \right)^2 \tilde c(L) .
\ee
From the bulk diagrammatics, we only expect the imaginary part of the phase shift to start at two loops in the bulk or, equivalently, starting from ${1 \over c_T^3}$, see \Fig{fig:hDiagram} for an example of the H-diagram that produces the leading gravitational waves production effect. Therefore $\tilde c(L)=0$. We do not expect the stringy corrections to change this basic picture, see also the discussion below.

Let us comment also on the expected form of the higher-order corrections that do appear in the expansion of the phase shift. The simplest one is related to the emission of gravity waves, see \Fig{fig:hDiagram}. This effect has been studied in flat space and we expect that the $G_N$ and $S$ dependence in AdS are the same and are given by ${\rm Im} \delta(S,L) \sim c_T^{-3} S^2 \log S$, see also formula $(7.5)$ in \cite{Amati:1987uf}. A similar diagram has been discussed in \cite{Meltzer:2019nbs}.
This contribution to the phase shift would produce terms $\sim {\log^2 c_T\over c_T^2}$ in the expansion of the energy-energy correlator, which is subleading compared to the ones studied in the present paper. It would be very interesting to understand them in greater detail.

We therefore conclude that \emph{only} $\delta_{\text{tree}}^2(S,L)$ used in the bulk of the paper produces the enhanced ${\log c_T \over c_T}$ contribution at intermediate energies.

A careful reader might have noticed the following interesting subtlety of the argument above which relied on setting $\delta_{\text{tree}}(S,L)=\delta_{\text{GR}}(S,L)$. If we plug the estimate \eqref{eq:estintermen} into the expression for the energy-energy correlator we will get a divergence coming from small impact parameters for $d>4$. In the present paper, we address this issue by considering $\delta_{\text{tree}}(S,L)=\delta_{\text{stringy}}(S,L)$, and leave the exploration of more general scenarios to the future. As we will see explicitly below, the stringy phase shift stays finite at small impact parameters and therefore this problem does not arise.

\subsection{Trans-Planckian energies}\label{sec:TransPlanck}

Finally, we need to address the case of high-energy scattering at trans-Planckian energies $S > S_{\text{eik}}(\epsilon, L)$. We mark this region by IV (red) in \Fig{fig:regions}. To estimate this contribution we will use AdS scattering unitarity \cite{Kulaxizi:2017ixa,Costa:2017twz}, which states that ${\rm Im} \delta(S,L) \geq 0$, and therefore  $| 1 - e^{i \delta} | \leq 2$.

We then get
\be\label{eq:TransplanckianBound}
\Big| \int_{S_{\text{eik}}(\epsilon, L)}^\infty {d S \over S^3} {\rm Re} ( 1 - e^{i \delta}) \Big| \leq 2 \int_{S_{\text{eik}}(\epsilon, L)}^\infty {d S \over S^3} = {\Pi_{d-1}^2(L) \over \epsilon^2 c_T^2},
\ee
where we used $\delta_{\text{tree}}(S,L)=\delta_{\text{GR}}(S,L)$. When combined with the pre-factor it produces a ${1 \over c_T}$ contribution to the energy-energy correlator. Notice that in performing the exercise above we assumed that there is no mixing between the graviton and other possible eigenstates of the eikonal operator. Including mixing, however, does not affect the estimate above. 

Another subtlety that was already discussed in the previous section is that the integral over impact parameters of the estimate above diverge in $d>4$. As before, in this paper we assume that the small impact parameter physics in this case can be regularized using the stringy effects, or by taking $\delta_{\text{tree}}(S,L)=\delta_{\text{stringy}}(S,L)$, which we describe next.

\subsection{Stringy corrections}\label{Sec:Stringy}

In this section, we consider the corrections due to stringy effects. It also helps to understand the role of strong coupling in our discussion. 

At large impact parameters the phase shift \eqref{eq:phaseshiftRegge} matches with the tree-level gravity phase shift used in the previous sections
\be
\label{eq:largeimpactSTR}
\delta(S,L) = \delta^{(GR)}(S,L) + ... , ~~~~ L \gg 1 . 
\ee
When $L \sim  {\log S \over \Delta_{\text{gap}}^2}$, the integral develops a saddle point at $\nu_* = - i {L \over 4 {\log S \over \Delta_{\text{gap}}^2}}$. The saddle point is a good approximation when ${\log S \over \Delta_{\text{gap}}^2} \gg 1$. To reproduce the behavior \eqref{eq:largeimpactSTR} at large $L$ we deform the $\nu$ contour into the lower half-plane and pick the stress tensor pole $j(\nu)=2$. This is always the correct asymptotic at large $L$ (independently of the validity of the saddle point approximation and the value of ${\log S \over \Delta_{\text{gap}}^2}$).

The stringy phase shift, however, has a very different small impact parameter asymptotic. To see this, we can expand
\be
\label{eq:smallexp}
{\cal H}_{i\nu+\frac{d}{2}-1}(L(u)) &= \Big(\frac{2^{2-d} \pi ^{1-\frac{d}{2}} u^{3-d} \Gamma (d-3)}{\Gamma \left(\frac{d}{2}-1\right)}+\ldots\Big)\cr &+\frac{2^{1-d} \pi ^{\frac{1-d}{2}} \Gamma \left(\frac{3}{2}-\frac{d}{2}\right) \Gamma \left(\frac{d}{2}+i \nu -1\right)}{\Gamma \left(-\frac{d}{2}+i \nu
   +2\right)}+ O(u^2), 
\ee
where the first line contains all the singular terms and they are even in $\nu\to-\nu$. These integrate to zero because the rest of the integrand is odd. The regularity of the phase shift \eqref{eq:phaseshiftRegge} can equivalently be seen by writing it in terms of $\Omega_{i\nu}(L)=\frac{i\nu}{2\pi}({\cal H}_{i\nu+\frac{d}{2}-1}(L)-{\cal H}_{-i\nu+\frac{d}{2}-1}(L))$ which is regular as $L\to 0$:
\be\label{eq:OmegasmallL}
\Omega_{i\nu}(L)=&\frac{\Gamma\Big(\frac{d-2}{2}+i\nu\Big)\Gamma\Big(\frac{d-2}{2}-i\nu\Big)\nu}{2^{d-1} \pi ^{\frac{d+1}{2}} \Gamma \left(\frac{d-1}{2}\right)}\sinh(\pi\nu)\Big[1-L^2\frac{\nu^2+\frac{(d-2)^2}{4}}{2(d-1)}+\ldots\Big],
\ee 
at small $L$. Below we further need the large $\nu$ limit of \eqref{eq:OmegasmallL} which is given by
\be\label{eq:largenuLim}
\Gamma\Big(\frac{d-2}{2}+i\nu\Big)\Gamma\Big(\frac{d-2}{2}-i\nu\Big)\nu\sinh(\pi\nu)\sim \nu^{d-2}.
\ee 
Inserting the small $L$-behaviour \eqref{eq:OmegasmallL} in \eqref{eq:phaseshiftRegge}, together with \eqref{eq:largenuLim}, we get 
\be\label{eq:smallLexpPS}
\delta(S,L(u))&\sim r(d,2)S^{1-\frac{d^2}{2\Delta^2_{\text{gap}}}}\int_{0}^{\infty} d\nu e^{-\frac{\log S}{\Delta_{\text{gap}}^2}\nu^2}\nu^{d-4}(1+\# \nu^2u^2+\ldots)\cr 
&\sim r(d,2)S^{1-2{({d \over 2})^2 \over \Delta_{\text{gap}}^2}} \left( {\Delta_{\text{gap}}^2 \over \log S} \right)^{{d-3 \over 2}} \left(1+ \#' {\Delta_{\text{gap}}^2\over \log S}u^2 + ... \right)
\ee 
where we used $\Delta_{\text{gap}}^2\gg \nu^2\gg 1$ and $r(\Delta(j(\nu)),j(\nu))\approx r(d,2)$. The latter condition was discussed in \cite{Costa:2017twz,Meltzer:2017rtf} and follows from assuming that $r(\nu)$ has a well-defined limit as $\Delta_{\text{gap}}^2\to\infty$ keeping $\nu^2/\Delta_{\text{gap}}^2$ fixed such that 
\be 
r(\nu) = r(0)+\sum_{n=1}^{\infty} \frac{f_n(\nu^2)}{\Delta_{\text{gap}}^{2n}},
\ee
where $r(\nu)\equiv r(\Delta(j(\nu)),j(\nu))$ and $f_n(\nu^2)$ is a polynomial with a maximum degree of $2n$ and $f_n(0)=0$. To leading order in $\Delta_{\text{gap}}^{-2}$ this is determined by the stress tensor $r(0)\approx r(d,2)$. 

We now want to compute the leading enhanced correction to the energy-energy correlator using the phase shift \eqref{eq:phaseshiftRegge}.
The basic effect is that after doing the integral over $S$, the phase shift is evaluated at energies $S \sim c_T$. By comparing the small $L$ expansion \eqref{eq:smallLexpPS} to the large $L$ expansion \eqref{eq:largeimpactSTR}, we find that the interpolation between two different regimes takes place at a scale $L_*$ defined as follows
\be
L_*^2 \equiv {\log c_T \over \Delta_{\text{gap}}^2} . 
\ee
When $L_* \ll 1$, stringy corrections become important only for $u \lesssim u_* \ll 1$. On the other hand, if $L_* \gg 1$, stringy corrections dominate for $u\lesssim 1-e^{-16L_*^2}$ in $d=4$. The transition between the stringy phase shift and the pure gravity phase shift as a function of $L_*$ can be seen explicitly in $d=4$ by evaluating the integral over $\nu$; this is done in Appendix \ref{App:StringyPhaseShift}.

We consider now $L_*\ll 1$. In this case we see that the naive IR divergence is substituted by the following integral when we include the string corrections
\be
&{\log c_T \over c_T} L_*^{2(3-d)}\int_0^{L_*} d u {\cal K}_d (u,z) \left( c_0 + c_1 {1 \over L_*} u^2 + ... \right) \ \nn \\
&=\left(L_*^{5-d} {\rm Poly_2}(z) + L_*^{7-d}{\rm Poly_3}(z) + ... \right),
\ee
where $c_i \sim O(1)$ are coefficients and ${\rm Poly_n}(z)$ are polynomials in $z$ of degree $n$. 

If we approach the stringy regime from the large $L$ direction we get the same estimate for the contribution to the energy-energy correlator
\be
\label{eq:polyenhanced}
\int_{L_*}^1 d u {\cal K}_d (u,z) \langle \Pi^2 (L(u)) \rangle = \left(L_*^{5-d} {\rm Poly_2}(z) + L_*^{7-d}{\rm Poly_3}(z) + ... \right) ,
\ee
where here in $d=5$ we get $\log L_*$.

Applying this prescription to the ${\cal N}=4$ SYM computation, using \eqref{eq:PHN4} for $\langle \Pi^2 (L(u)) \rangle$, we get the following enhanced terms
\be\label{eq:enhN4}
&{\log c_T \over c_T} \Big( L_*^{-4} (1-6z + 6 z^2) + L_*^{-2} {\rm Poly}_3(z)+{\rm Poly}_4(z)  \Big. \nn \\
&\Big. + \log L_* (1-36 z +216 z^2 -400 z^3 + 225 z^4) \Big)   \ , ~~~ L_* \ll 1 .
\ee
This is precisely the structure that we observed in the $\omega$-regularization of the energy-energy correlator \eqref{eq:n4omreg}. It would be interesting to compute these enhanced polynomial corrections exactly. Comparing the enhancement \eqref{eq:enhN4} to that in \eqref{eq:polyenhanced} the presence of the large $S^5$ is reflected in the leading power of $L_*$ which coincides with $d=9$ in \eqref{eq:polyenhanced}.

The conclusion is that the small impact parameter divergence of the phase shift in $d>4$ observed in the previous sections is regulated by stringy corrections which leads to enhanced polynomial contributions in the regime $L_* \ll 1$. 

Finally, let us comment on the expected contribution from the zero impact parameter physics. The complete form of the one-loop bulk correlator, including polynomial terms in the Mellin amplitude, in ${\cal N}=4$ SYM was computed using localization in \cite{Chester:2019pvm,Chester:2020dja}. It includes a constant term ${\sqrt{\lambda} \over c_T}$ which assuming stringy unitarization leads to an additional term ${\lambda \over c_T}(1-6 z+6 z^2)$ in the energy-energy correlator. In the language of the formulas above it corresponds to the contribution ${\Delta_{\text{gap}}^4 \over c_T}\sim {(\log c_T)^2 \over c_T} L_*^{-4}$. We notice that at this order, the zero impact parameter contribution in string theory is more enhanced compared to the non-zero impact parameter contribution computed above.

\subsection{Stringy unitarization}

Consider next the case $L_* \gg 1$ and fixed. In this regime, the tree-level gravitational computation used in the previous sections is still valid at large impact parameters $L \gtrsim L_*^2$, which means that the non-analytic part of the answer $\sim \log z$ is not affected. However, the part which is analytic in $z$ changes. 

To see what happens more explicitly, we plug the stringy phase shift into the integral $\int {d S \over S^3} \delta^2(S,L)$ to get
\be\label{eq:stringyS}
\int_{S_0}^{S_{eik}(\epsilon,L)} {d S \over S^3} S^{j(\nu) + j(\nu')-2} = {\left(S_{eik}(\epsilon,L) \right)^{j(\nu) + j(\nu')-4} - S_0^{j(\nu) + j(\nu')-4} \over j(\nu) + j(\nu')-4} ,
\ee
where recall that $S_{eik}(\epsilon,L) \sim c_T$.

One can easily check that the computations of the previous sections arise from considering \eqref{eq:stringyS} and evaluating the $\nu$ and $\nu'$ integral by taking the residue at the graviton $j(\nu)=j(\nu')=2$ pole. The factor ${1 \over j(\nu) + j(\nu')-4}$ which is singular when $j(\nu) = j(\nu') = 2$ is responsible for an extra factor of $\log c_T$.

To understand the analytic in $z$ part for $L_* \gg 1$, we can simply take $L_* \to \infty$. In this case, the $\nu$ and $\nu'$ integrals in the first term are effectively localized close to the origin. As a result, we get an extra suppression factor $S_{\text{eik}}^{j(0)}\sim e^{- {1 \over 2}d^2 L_*^2} \to 0$ for the first term in \eqref{eq:stringyS}. The main contribution therefore comes from the term $- S_0^{j(\nu) + j(\nu')-4}$ which is again dominated by the graviton pole and produces  ${\Delta_{gap}^2 \over c_T}$ contributions to the energy-energy correlator. Effectively the enhancement by the $\Delta_{gap}^2$ factor happens because in this case the unitarization happens at the stringy energy scale $\log S_0\sim \Delta_{\text{gap}}^2$.\footnote{This is the energy at fixed $L$ at which the saddle point $\nu_*$ approach the graviton pole.}

\section{Discussion}\label{sec:discussion}

Unitarization of high-energy gravitational scattering requires all-loop re-summation. In flat space, unitarity of the four-point amplitude implies that at finite $G_N$, all partial waves must satisfy $|S_J| \leq 1$: the scattering probability cannot be larger than one. In a unitary CFT, a somewhat analogous statement is that the nonperturbative Regge intercept of the four-point correlator is $J_{\text{np}}\leq 1$ \cite{Caron-Huot:2017vep}. This statement contrasts with perturbative gravity computations in AdS, which exhibit a Regge intercept that grows with the number of gravitational loops: $J_{\text{tree}}=2$, $J_{\text{1-loop}}=3,\ldots$\ .\footnote{Here we only talk about the part which is non-local in the impact parameter space. Contact terms exhibit an even larger Regge intercept.} The same phenomenon also leads to a rich structure of poles at integer $J>1$ in the conformal partial waves and the OPE data, see e.g.\ \cite{Aprile:2017bgs,Alday:2017vkk,Drummond:2022dxw,Bissi:2020wtv}. Understanding in detail how these singularities disappear at finite $G_N$ and the nonperturbative Regge bound is satisfied is an interesting open problem. 

In this paper, we explored imprints of unitarization of high-energy gravitational scattering in the energy-energy correlator at strong coupling. For the leading quantum-gravity correction, this in particular amounts to understanding the finite-$c_T$ resolution of the universal ${1 \over c_T} {\# \over J-3}$ pole in the one-loop gravity OPE data of double trace operators \cite{Aprile:2017bgs,Alday:2017vkk}. We restated this problem in the language of high-energy fixed impact parameter scattering in AdS \cite{Amati:1987uf,Cornalba:2006xk,Cornalba:2006xm,Cornalba:2007zb}. We then argued that the eikonalization of the tree-level gravitational scattering removes the pole and leads to an enhanced contribution to the energy-energy correlator $\sim {\log c_T \over c_T}$. We also showed that for $D$-dimensional bulk theories with $D>5$, there is an additional enhancement coming from small impact parameter scattering which we modeled using stringy effects. We computed the ${\log c_T \over c_T}$ correction to the energy-energy correlator explicitly in various examples. 

The result for the ${\log c_T \over c_T}$ correction to the energy-energy correlator encodes the bulk geometry via the dependence of the phase shift $\delta(S,L)$ on the impact parameter $L$, see \eqref{eq:phaseshift1loop} and  \eqref{eq:finalformula}. The leading small angle contribution $\log c_T \log z/c_T$ is universal \eqref{eq:universalAsy}, whereas higher-order terms encode the full bulk geometry, including the internal dimensions. We have explicitly computed the $\log c_T/c_T$ correction to the energy-energy correlator in momentum eigenstates of pure Einstein gravity in $AdS_{d+1}$, see \eqref{eq:GRres}, and ${\cal N}=4$ SYM dual to $AdS_5\times S^5$ \eqref{eq:N4res}. In the latter case, the graviton mixes with KK modes at high energies, and we used the results of \cite{Aprile:2017bgs} to identify the eigenstates of the eikonal operator. More generally, we have observed that \emph{non-local} scattering in AdS (scattering at large impact parameter) maps to \emph{non-analyticity} in the angle variable $z$ on the energy correlation side.

An important part of our argument is the analysis of various corrections in Section \ref{sec:corrections} based on the ${\bf source} \times {\bf detector}$ OPE channel. While eikonalization (or exponentiation of anomalous dimensions) is completely automatic in this language, we used gravity-based arguments to estimate possible corrections to our result. In particular, we assumed that when the tree-level interaction is weak and the impact parameter is much larger than the Schwarzschild radius, loop corrections to the tree-level phase shift stay small, see \eqref{eq:basicassumption} for the precise statement. It would be, of course, much better to have a rigorous CFT proof of this fact.  Our analysis is blind to UV-sensitive zero impact parameter terms in the bulk scattering amplitude. These cannot be computed on general grounds and lead to an extra polynomial contribution to our results.\footnote{For example, in ${\cal N}=4$ SYM it leads to an extra correction $\sim {\lambda \over c_T}(1-6z+6z^2)$ to the energy-energy correlator.} While most of the paper focused on computations in pure gravity, in Section \ref{Sec:Stringy} we also discussed the effect of stringy corrections on our discussion. We modeled the phase shift using the full leading Regge trajectory while not changing the simple tensor structure expected in pure general relativity. We have found that the presence of stringy corrections does not modify our discussion when $\Delta_{\text{gap}}^2/\log c_T \gg 1$, where $\Delta_{\text{gap}} \sim m_{\text{HS}} R_{AdS}$ is related to the mass of the lightest spin-four stringy excitation. In addition to that, when the number of bulk dimensions $D$ is larger than $5$, we get enhanced by $\Delta_{\text{gap}}^2/\log c_T$ polynomial corrections in $z$ coming from small impact parameter scattering in the bulk. It would further be interesting to incorporate the additional tensor structures.

In this paper, we focused on `light' states, namely we assumed that their scaling dimension does not scale with $c_T$. It would be interesting to understand energy-energy correlators in `heavy' states, e.g., dual to black holes, for which $\Delta \sim c_T$. In this case, we can use the Eigenstate Thermalization Hypothesis (ETH) \cite{Srednicki:1999bhx,DAlessio:2015qtq,Lashkari:2016vgj,Delacretaz:2020nit} in the ${\bf source} \times {\bf detector}$ channel. The diagonal term is then responsible for the homogeneous distribution of energy, whereas the connected thermal two-point function related to the off-diagonal terms produces corrections on top of it. At high energies of the probe particle, the gravitational phase shift, in this case, encodes the propagation of a particle on top of a classical black hole background \cite{Kulaxizi:2018dxo,Karlsson:2019qfi,Karlsson:2019txu,Parnachev:2020zbr}. It would be interesting to explore this connection in more detail and compute the leading correction explicitly, see \cite{Chicherin:2023gxt,Firat:2023lbp} for a related discussion.

We have also focused on the energy-energy correlators but similar effects should be possible to observe in other event shapes. In general we expect the detectors of spins $J_1$ and $J_2$ to probe the $\int^\infty {d S \over S^{J_1 + J_2 -1}}$ moment of the phase shift $e^{i \delta}-1$. In this way different event shapes measure different high-energy moments of the gravitational scattering in AdS. Similarly, the connection between high-energy gravitational scattering and energy correlators should generalize beyond two detectors. In this case, we expect that the multi-point Regge theory recently developed in the CFT context in \cite{Costa:2023wfz} could be useful. The three- and higher-point energy correlators have been recently explored in~\cite{Chen:2019bpb,Chen:2020vvp,Chang:2022ryc,Chen:2022jhb,Yan:2022cye,Chen:2023zlx,Yang:2022tgm,Yang:2024gcn,Chicherin:2024ifn}. It would be also interesting to generalize our discussion to the case of holographic gapped theories \cite{Polchinski:2000uf,Polchinski:2001tt}.

Finally, it would be very interesting to explore the connection between energy correlators and high-energy AdS scattering using non-perturbative bootstrap methods \cite{Poland:2018epd}. For example, if we schematically write
\be
\la {\cal E}(n_1) {\cal E}(n_2)  \ra = \left(\frac{p^0}{\Omega_{d-1}} \right)^2 \Big[ 1 + \text{enhanced}(z) + ... \Big],
\ee
one can wonder if the average null energy condition $\la {\cal E}(n_1) {\cal E}(n_2)  \ra  \geq 0$ imposes interesting bounds on the enhanced term and the underlying theory.\footnote{See \cite{Zhiboedov:2013opa,Cordova:2017zej,Meltzer:2017rtf,Belin:2019mnx} for related work.} Indeed, since the enhanced correction to $1$ satisfies the momentum conservation Ward identities (which state that it integrates to zero with a pair of non-negative kernels, see \eqref{eq:WardIntro}), it has to become negative somewhere, and therefore by making the enhancement big enough we risk violating the ANEC condition. For example, in the context of the present paper, this argument suggests the existence of a bound ${\Delta_{\text{gap}}^{\#} \over c_T} \lesssim 1$.\footnote{See also a discussion in \cite{Caron-Huot:2017vep}.} To the best of our knowledge the condition $\la {\cal E}(n_1) {\cal E}(n_2)  \ra  \geq 0$ has not yet been explored in the bootstrap studies of the CFT correlators of stress tensors, see e.g. \cite{Beem:2013qxa,Beem:2016wfs,Dymarsky:2017yzx}. We leave it here to the reader's imagination and future work.

\appendix

\section*{Acknowledgments}
We thank Alex Belin, James Drummond, Gregory Korchemsky, Petr Kravchuk, Vasco Goncalves, Andrei Parnachev, Hynek Paul, Joao Penedones, Emery Sokatchev, Gabriele Veneziano and Zahra Zahraee for helpful discussions. This project has received funding from the European Research Council (ERC) under the European Union’s Horizon 2020 research and innovation program (grant agreement number 949077). 

\section{Conventions}\label{App:Conventions}
The vacuum stress-tensor two-point function in position space is given by 
\be 
\langle T_{\mu\nu}(x)T_{\rho\sigma}(0)\rangle_{\text vac} &=\frac{c_T}{\Omega_d^2}\frac{1}{ x^{2d}}\left(\frac{1}{2!}(I_{\mu\rho}(x)I_{\nu\sigma}(x)+I_{\mu\sigma}(x)I_{\nu\rho}(x))-\frac{1}{d}\eta_{\mu\nu}\eta_{\rho\sigma}\right)\\ 
I_{\mu\nu}(x)&=\eta_{\mu\nu}-2\frac{x_\mu x_\nu}{x^2},
\ee 
where $\Omega_d=2\pi^{\frac{d}{2}}/\Gamma(\frac{d}{2})$ is the volume of a $(d-1)$-dimensional unit sphere. Performing the Fourier transform of the expression above, we get for the Wightman function
\be
\langle T_{\mu\nu}(q)T_{\rho\sigma}(-q)\rangle = \rho_Tc_T\theta(q^0)\theta(-q^2)(-q^2)^{\frac{d}{2}}\Pi_{\mu\nu,\rho\sigma}(q),
\ee 
where $\Pi_{\mu\nu,\rho\sigma}(q)$ is given in terms of $\pi_{\mu\nu}(q)=\eta_{\mu\nu}-\frac{q_\mu q_\nu}{q^2}$ as follows
\be
    \Pi_{\mu\nu,\rho\sigma}(q) = \frac{1}{2}(\pi_{\mu\rho}\pi_{\nu\sigma}+\pi_{\mu\sigma}\pi_{\nu\rho})-\frac{1}{d-1}\pi_{\mu\nu}\pi_{\rho\sigma}
\ee
and 
\be 
\rho_T = \frac{\pi^{\frac{d}{2}+1}}{2^{d-1}(d+1)\Gamma(d-1)\Gamma(\frac{d+2}{2})\Omega_d^2}.
\ee 

Defining $\chi=\frac{q^2 n_1\cdot n_2}{2 (q\cdot n_1)(q\cdot n_2)}$ we have $n_1^\mu\pi_{\mu\nu} n^\nu_2=\frac{n_1\cdot n_2}{2\chi}(2\chi-1)$ and $n^\mu_i\pi_{\mu\nu}n^\nu_i=-\frac{(q\cdot n_i)^2}{q^2}$ and therefore
\be\begin{aligned}
\langle T(q;n_1)T(-q;n_2)\rangle=&\rho_Tc_T\theta(q^0)\theta(-q^2)(q\cdot n_1)^2 (q\cdot n_2)^2 (-q^2)^{\frac{d-4}{2}}\\
&\times\Big[4\chi^2-4\chi+\frac{d-2}{d-1}\Big].
\end{aligned}\ee 

\section{Poincaré shifts and correlation functions}
\label{app:fold}

Here we review some standard material regarding CFT correlation functions in the Lorentzian signature. Lorentzian correlation functions in conformal field theory are naturally defined on the Lorentzian cylinder $\mathbb{R} \times S^{d-1}$. To get to it, one can start from the usual Euclidean correlators on $\mathbb{R}^d$ and map them to the Euclidean cylinder $x^\mu = e^{\tau} \mathbf{e}^\mu$, with $\mathbf{e} \cdot \mathbf{e}=1$, such that
\be
d s^2 = \delta_{\mu \nu}dx^\mu d x^\nu = e^{2 \tau} ( d \tau^2 + d\mathbf{e}^2) = e^{2 \tau} d s_{cyl}^2 . 
\ee
Under this transformation, we have
\be
\langle {\cal O}_1(x_1) ... {\cal O}_n(x_n) \rangle_{\mathbb{R}^d} = \prod_{i=1}^n e^{-\Delta_i \tau} \langle {\cal O}_1(\tau_1, \mathbf{e}_1) ... {\cal O}_n(\tau_n, \mathbf{e}_n) \rangle_{\mathbb{R} \times S^{d-1}} .
\ee
We can then perform the Wick rotation $\tau \to - i t$ and arrive at the Lorentzian cylinder. 

The Lorentzian cylinder can be thought of as a universal cover of the null cone in $\mathbb{R}^{2,d}$ modded by rescaling $\mathbb{R}_+$
\be
X^2 = -(X^{-1})^2 - (X^0)^2 + \sum_{i=1}^d (X^i)^2 = 0 \ . 
\ee
Indeed, introducing the coordinates
\be
X^{-1} = \cos t, ~~~ X^0 = \sin t, ~~~ X^i = \mathbf{e}^i , ~~~ i = 1, .. , d ,
\ee
we see that the equation above is satisfied. 

With each point on the Lorentzian cylinder we can associate a Poincaré patch. To achieve it, we split $\mathbb{R}^{2,d} = \mathbb{R}^{1,1} \times \mathbb{R}^{1,d-1}$, where $\mathbb{R}^{1,1}$ is spanned by a pair of orthonormal null vectors $K^A$ and $\bar K^A$ that satisfy $K^2=
\bar K^2 =0$ and $-2 K \cdot \bar K = 1$. We then write
\be
X^A = \bar K^A + K^A y^2 + y^A ,
\ee
where $y^A$ are the usual coordinates in Minkowski space $\mathbb{R}^{1,d-1}$. Indeed, with this choice $X^2 = 0$.
The above choice corresponds to the gauge condition $-2 X \cdot K = 1$ for rescaling of $X^A$.

For example, consider the gauge condition $X^{-1}+X^d = 1$ for which the Minkowski space coordinates are
\be
y^\mu = {X^\mu \over X^{-1}+X^d }, ~~~ \mu = 0, ..., d-1.
\ee
In terms of the global coordinates this becomes
\be
y^0 &= {\sin t \over \cos t + \mathbf{e}^d} , ~~~ y^i = {\mathbf{e}^i \over \cos t + \mathbf{e}^d} \ . 
\ee

We can then write the following relationship between the correlation functions in the Minkowski space Poincaré patch and the Lorentzian cylinder
\be
\langle {\cal O}_1(y_1) ... {\cal O}_n(y_n) \rangle_{\mathbb{R}^{1,d-1}} = \prod_{i=1}^n (\cos t_i + \mathbf{e}_i^d)^{\Delta_i} \langle {\cal O}_1(t_1, \mathbf{e}_1) ... {\cal O}_n(t_n, \mathbf{e}_n) \rangle_{\mathbb{R} \times S^{d-1}} ,
\ee
such that, for example, the two-point function of scalar primary operators takes the form $\langle {\cal O}(t_1, \mathbf{e}_1) {\cal O}(t_2, \mathbf{e}_2) \rangle = 2^{-\Delta} \left(\cos(t_1-t_2 - i \epsilon ) - \mathbf{e}_1 \cdot \mathbf{e}_2 \right)^{-\Delta}$. 

We next introduce the operation ${\cal T}$ of translating a point one Poincaré patch forward, such that ${\cal T}(t,\mathbf{e}){\cal T}^{-1} = (t+\pi, -\mathbf{e})$. In the embedding space coordinates, it is simply $X^A \to - X^A$. Applying this to the two-point function, we immediately get
\be
e^{i \pi \Delta}\langle {\cal O}(t_1+\pi, -\mathbf{e}_1) {\cal O}(t_2, \mathbf{e}_2) \rangle=e^{-i \pi \Delta} \langle {\cal O}(t_1, \mathbf{e}_1) {\cal O}(t_2+\pi, -\mathbf{e}_2) \rangle=\langle {\cal O}(t_1, \mathbf{e}_1) {\cal O}(t_2, \mathbf{e}_2) \rangle .
\ee
In other words, translating a point one Poincaré patch forward produces a trivial overall phase. 

Consider next four points on the Lorentzian cylinder, out of which we can build a cross-ratio ${d_{i j}^2 d_{kl}^2 
\over d_{i k}^2 d_{jl}^2}$. If we now translate a given point one Poincaré patch forward, for example, $i$, it sends $d_{ij}^2 \to - d_{ij}^2$, $d_{ik}^2 \to - d_{ik}^2$ leaving the cross-ratio invariant. This, however, does not yet mean that the monodromy of this transformation is trivial. It could be that the cross-ratio has been continued around the origin in a nontrivial way. To avoid this let us consider an operator that acts on the in- or out-vacuum inside the Wightman function. In this case, all the distances have the same $i 
\epsilon$ prescription, e.g. $t_i - i \epsilon$ if we have $\langle \Omega| {\cal O}(t_i,\mathbf{e}_i) ...$. Therefore, the phase acquired by the numerator and the denominator of the cross-ratio precisely cancel so that the monodromy of the cross-ratio is trivial. 

We, therefore, conclude that Poincaré translation applied to operators acting on the vacuum acts trivially on cross-ratios. To see its overall effect, we need only to check its effect on the pre-factor and possibly tensor structures. We already have seen above that it produces a nontrivial phase when acting on the two-point function. 

Let us next consider the four-point function of symmetric tensor operators relevant for this paper. We want to act with the Poincaré translation on a spin $J$ operator. In the embedding space, it acts on the polarization tensor as $Z^A \to - Z^A$. The general form of the correlator was considered in \cite{Costa:2011mg,Costa:2011dw} and it involves the basic tensor structures $H_{ij}$ and $V_{i,jk}$. These structures are invariant under $(X_i,Z_i) \to -(X_i, Z_i)$. Taking into account the pre-factor, we therefore conclude that parity-even tensor conformal structures transform as
\be
\label{eq:Ttransform}
\langle \Omega | {\cal O}_J(-X,-Z) ... &= e^{- i \pi (\Delta+J)} \langle \Omega | {\cal O}_J(X,Z) ... , \nn \\
... {\cal O}_J(-X,-Z) | \Omega \rangle &=e^{i \pi (\Delta+J)} {\cal O}_J(X,Z) | \Omega \rangle \ . 
\ee

Including the parity odd structures containing the $\epsilon$-tensor slightly changes this conclusion. Consider, for example, the following parity-odd tensor structure in $d=3$ $\epsilon(Z_1, Z_2, X_1, X_2, X_3)$.\footnote{For a detailed discussion of the embedding formalism in this context, see \cite{Costa:2011mg}.} This structure is odd under the ${\cal T}$-transformation $Z_i \to - X_i$, $X_i \to - X_i$. More generally, it is clear that in an odd number of dimensions, parity-odd conformal structures that involve the $\epsilon$-tensor acquire an additional minus sign under this transformation. In an even number of dimensions \eqref{eq:Ttransform} holds.

\section{Details on the kernel ${\cal K}_{d}(u,z)$\label{App:Kernel}}

In this Appendix, we provide details and useful expressions for the kernel ${\cal K}_{d}(u,z)$. The definition of the kernel $\mathcal{K}_d(u,z)$ is
\be
\mathcal{K}_d(u,z) \equiv \int_{M^+} {d^dq \over (2\pi)^d} {G_0(q;n_1,n_2)\over (q\cdot n_1)^{d+1}(q\cdot n_2)^{d+1}} \delta(1-\sqrt{-q^2}\sqrt{-p^2}) \delta\left(u - \sqrt{1-{p^2 q^2 \over (p\cdot q)^2}}\right)\,, \label{eq:kernel_def}
\ee
where $G_0(q;n_1,n_2)$ is the vacuum stress-tensor two-point function
\be\label{eq:vacuumapp}
		G_0(q;n_1,n_2)	&=(q \cdot n_1)^2 (q \cdot n_2)^2 (-q^2)^{{d-4 \over 2}} \Big( 4\chi^2-4\chi+\frac{d-2}{d-1} \Big) .
\ee
To eliminate the delta functions in \eqref{eq:kernel_def}, we use the Baikov representation~\cite{Baikov:1996iu} to rewrite the integral measure as
\be
\int_{M^+} d^dq  ={\Omega_{d-3} \over 2 } \int_{\cal R} d(q\cdot n_1) d(q\cdot n_2) d(q\cdot p) d(q^2) {[-G(q,n_1,n_2,p)]^{d-5 \over 4} \over [-G(n_1,n_2,p)]^{d-4 \over 2}}
\ee
where $G(k_1,\cdots, k_n)$ is the Gram determinant $\det(k_i\cdot k_j)$ and the integral region $\mathcal{R}$ is defined by the algebraic constraint $G(q,n_1,n_2,p)<0$.
As a consequence, the kernel $\mathcal{K}_d(u,z)$ can be decomposed into
\be
\mathcal{K}_d(u,z)={1 \over (- p^2)^{{d-2 \over 2}}}{\Omega_{d-3} \over (2\pi)^d}&{u (1-u^2)^{-3/2} \over [-G(n_1,n_2,p)]^{d-4 \over 2}}\Big[{(n_1\cdot n_2)^2 \over (-p^2)^2} I^{d-5\over 2}_{d+1,d+1}\nn \\
&+2{(n_1\cdot n_2) \over -p^2} I^{d-5\over 2}_{d,d}+ {d-2 \over d-1} I^{d-5\over 2}_{d-1,d-1}\Big],
\ee
by introducing the integral family
\be
I^\gamma_{\alpha_1, \alpha_2} = \int_{\cal R} d(q\cdot n_1) d(q\cdot n_2) {[-G(q,n_1,n_2,p)]^\gamma \over (-q\cdot n_1)^{\alpha_1} (-q\cdot n_2)^{\alpha_2}}.
\ee

The Gram determinants are
\begin{eqnarray}
G(n_1,n_2, p) &=& -4z(1-z) {(p \cdot n_1)^2 (p \cdot n_2)^2 \over -p^2}\,,\\
G(q,n_1,n_2,p) &=& {(p \cdot n_1)^2 (p \cdot n_2)^2 \over p^4} \cr &&\Big(y_1^2+y_2^2-2(1-2z) y_1 y_2 + {4 z \over \sqrt{1-u^2}}(y_1+y_2)+4z(1 + {u^2 z \over 1-u^2}) \Big) \,,\nn
\end{eqnarray}
where $y_i={p^2 \over (p \cdot n_i)} q\cdot n_i$. For convenience, we can choose a special configuration for scalar products
\be
p^2 = p\cdot n_1 = p\cdot n_2=-1\,,\qquad n_1\cdot n_2 = -(1-\cos \theta)=-2 z\,.
\ee
The full dependence is then easy to restore using the dimensionality and the transformation properties of the object. 

We map the integration region to a unit disk using the change of variables:
\be
y_1 = \sqrt{u^2 \over 1-u^2} x_1 - {1 \over \sqrt{1-u^2}}\,,\quad
y_2 = \sqrt{u^2 \over 1-u^2} (\cos \theta x_1 + \sin \theta x_2) - {1 \over \sqrt{1-u^2}}\,,
\ee
followed by the polar coordinate parametrization $x_1+i x_2 = r e^{i\phi}$. The integral $I^{\gamma}_{\alpha_1,\alpha_2}$ then becomes
\be
I^\gamma_{\alpha_1, \alpha_2} =& [4z(1-z)]^{\gamma+1/2} \left({u^2 \over 1-u^2}\right)^{\gamma+1-{\alpha_1+\alpha_2 \over 2}}\nn\\
&\qquad \times \int_0^1 r dr \int_0^{2\pi}d\phi {(1-r^2)^\gamma \over ({1\over u}-r\cos\phi)^{\alpha_1} ({1\over u}- r\cos(\phi-\theta))^{\alpha_2}}\,,
\ee
which belongs to the same category as the Mandelstam kernel in the massive $2\to 2$ scattering process \cite{Correia:2020xtr}
\be\label{eq:MasterIntegral}
\tilde{I}_\gamma(\eta_1,\eta_2) = \int_0^1 r dr \int_0^{2\pi}d\phi {(1-r^2)^\gamma \over (\eta_1-r\cos\phi) (\eta_2- r\cos(\phi-\theta))}.
\ee
For $\alpha_i$ taking positive integer values, $\tilde{I}_\gamma$ can serve as a generating function through taking derivatives with respect to $\eta_1$ and $\eta_2$:
\be
I^\gamma_{\alpha_1, \alpha_2} = [4z(1-z)]^{\gamma+1/2} \left({u^2 \over 1-u^2}\right)^{\gamma+1-{\alpha_1+\alpha_2 \over 2}} {(-1)^{\alpha_1+\alpha_2} \over (\alpha_1-1)!(\alpha_2-1)!} \tilde{I}_\gamma^{(\alpha_1-1, \alpha_2-1)}(u^{-1}, u^{-1})\,.
\ee

We can now restore the dependence on $p^2$ and $(p \cdot n_i)$, which produce the expected factor ${(-p^2)^d \over (-p \cdot n_1)^{d-1} (-p \cdot n_2)^{d-1}}$. In this way, we get the following expression for the kernel
\be
\label{eq:kerneld}
\mathcal{K}_d(u,z) &={\Omega_{d-3} \over (2\pi)^d} u (1-u^2)^{-3/2}  \nonumber \\ 
&\times\Big[ 4 z^2  \tilde{I}^{d-5\over 2}_{d+1,d+1}(u,z) - 4 z \tilde{I}^{d-5\over 2}_{d,d}(u,z) + {d-2 \over d-1} \tilde{I}^{d-5\over 2}_{d-1,d-1}(u,z) \Big] ,
\ee
where we defined the integral $\tilde{I}^\gamma_{\alpha_1, \alpha_2}$ as follows
\be
\tilde{I}^\gamma_{\alpha_1, \alpha_2}(u,z) &= \left({u^2 \over 1-u^2}\right)^{\gamma+1-{\alpha_1+\alpha_2 \over 2}}\int_0^1 r dr \int_0^{2\pi}d\phi {(1-r^2)^\gamma \over ({1\over u}-r\cos\phi)^{\alpha_1} ({1\over u}- r\cos(\phi-\theta))^{\alpha_2}}\,,\nonumber\\
&= \left({u^2 \over 1-u^2}\right)^{\gamma+1-{\alpha_1+\alpha_2 \over 2}} {(-1)^{\alpha_1+\alpha_2} \over (\alpha_1-1)!(\alpha_2-1)!} \tilde{I}_\gamma^{(\alpha_1-1, \alpha_2-1)}(u^{-1}, u^{-1})\,.
\ee

The differential representation is useful to calculate the full kernel $\mathcal{K}_d(u,z)$ at small integer-valued $d$, while the integral representation is easier to study in generic $d$. One example is the small $u$ expansion, where the expansion and integration commute.
Therefore, at small $u$ the kernel can be expanded as follows
\be\label{eq:small_u}
&\mathcal{K}_d(u,z) =\frac{2 \pi  u^{d-2} \left(d (1-2 z)^2-4 z^2+4 z-2\right)}{(d-3) (d-1)} \nn \\
&-\frac{\pi  u^d \left(d^2 (4 z-3) (1-2 z)^2+d \left(32 z^3-36
   z^2+6\right)+8 z \left(2 z^2-4 z+1\right)\right)}{(d-3) (d-1)} + O(u^{d+2}) \ . 
\ee

\subsection{Kernel in four dimensions}

Here we provide some explicit detail on the kernel in $d=4$. The basic ingredient that we need to compute in $d=4$ is~\footnote{For the generic case, $\tilde{I}_{d-5\over 2}$ can be expressed in terms of AppellF1 hypergeometric functions.}
\be
\tilde I_{-1/2}(\eta_1, \eta_2, z) = \frac{\pi  \log \left(-\frac{i \eta _1 \eta _2+\sqrt{-\left(\eta _1-\eta _2\right){}^2-4 z^2+\left(4-4 \eta _1 \eta _2\right) z}+2 i
   z-i}{-i \eta _1 \eta _2+\sqrt{-\left(\eta _1-\eta _2\right){}^2-4 z^2+\left(4-4 \eta _1 \eta _2\right) z}-2 i
   z+i}\right)}{\sqrt{\left(\eta _1-\eta _2\right){}^2+4 z^2+4 \left(\eta _1 \eta _2-1\right) z}} . 
\ee
After this the kernel can be obtained using \eqref{eq:kerneld}. 

In this way we get the 4d expression for the kernel
\be
\mathcal{K}_4(u,z) &={(-p^2)^4 \over (-p \cdot n_1)^{3} (-p \cdot n_2)^{3}}{1 \over 8 \pi^4} \tilde K(u,z) \ , 
\ee
where
\be
\tilde K(u,z) = \frac{4}{3} \pi  u^2 \left(6 z^2-6 z+1\right)-\frac{8}{3} \pi  u^4 \left(50 z^3-78 z^2+33 z-3\right) +O(u^6)\ . 
\ee
This is consistent with the small $u$ expansion \eqref{eq:small_u} from the integral representation.

\subsection{Ward identities}
\label{app:WardId}
Based on the definition \eqref{eq:kernel_def}, we can show the kernel $\mathcal{K}_d(u,z)$ satisfies the Ward identities
\be
0&=\int_0^1 dz\, (z(1-z))^{d-4 \over 2} \mathcal{K}_d(u,z)\,, \label{eq:kernel_ward_1}\\
0&=\int_0^1 dz\, z(z(1-z))^{d-4 \over 2} \mathcal{K}_d(u,z)\,. \label{eq:kernel_ward_2}
\ee
These Ward identities are satisfied by perturbative corrections to the EEC as we explain below, which is the consequence of energy-momentum conservation. 

Let us first explain the reason for the EEC case. To measure the total energy (as the eigenvalue of the Hamiltonian $\hat{H}$) at future null infinity $\mathcal{I}_+$, we need to integrate the energy detector operator $\mathcal{E}(\vec{n})$ over the whole $S^{d-2}$ sphere
\be
\hat{H} = \int_{S^{d-2}} d^{d-2}\vec{n} \mathcal{E}(\vec{n})\,.
\ee
As a result, the EEC satisfies the following Ward identity
\be\label{eq:eec_ward_id1}
{1\over (p^0)^2}\int_{S^{d-2}}d^{d-2}\vec{n}_1 d^{d-2} \vec{n}_2 \langle \mathcal{E}(\vec{n}_1) \mathcal{E}(\vec{n}_2) \rangle_p = 1\,,
\ee
where $p^\mu$ denotes the momentum of the scalar source $\mathcal{O}$. For convenience, we choose the center of mass frame $p^\mu = (Q,\vec{0})$ so that $\langle \mathcal{E}(\vec{n}_1) \mathcal{E}(\vec{n}_2) \rangle_p=Q^2 f(-{n_1\cdot n_2 \over 2})$. The Ward identity \eqref{eq:eec_ward_id1} reduces to a constraint on $f(z)$
\be\label{eq:eec_ward_id1_f}
1=\int_0^1 dz \int_{S^{d-2}}d^{d-2}\vec{n}_1 d^{d-2} \vec{n}_2 \delta(z+\frac{n_1\cdot n_2}{2})f(z) =2^{d-3}\Omega_{d-1}\Omega_{d-2} \int_0^1 dz\, (z(1-z))^{d-4 \over 2} f(z)\,.
\ee
This result holds non-perturbatively and is independent of any coupling constant. Therefore, if we consider a perturbative expansion $f(z)=\sum_{n=0}^{\infty} g^n f^{(n)}(z)$, all higher order corrections must obey the identity
\be
\label{eq:WI1}
\int_0^1 dz\, (z(1-z))^{d-4 \over 2} f^{(n\geq 1)}(z) =0\,.
\ee
To derive the second Ward identity recall that the spatial momentum generator takes the following form
\be
\hat P^i = \int_{S^{d-2}} d^{d-2}\vec{n} n^i \mathcal{E}(\vec{n}) \ .
\ee 
Let us consider next the source with zero momentum $\vec p=0$. We can then write 
\be\label{eq:eec_ward_id1_p}
{1\over (p^0)^2}\int_{S^{d-2}}d^{d-2}\vec{n}_1 d^{d-2} \vec{n}_2 (\vec{n}_1 \cdot \vec{n}_2) \langle \mathcal{E}(\vec{n}_1) \mathcal{E}(\vec{n}_2) \rangle_{p^0, \vec p=0} = 0\, .
\ee
Repeating the same manipulations as above and remembering that in the center-of-mass frame $\vec{n}_1 \cdot \vec{n}_2 = 1 - 2 z$ we get
\be
\int_0^1 dz\, (z(1-z))^{d-4 \over 2} (1-2z)f(z) = 0 .
\ee
When applied to the perturbative expansion together with \eqref{eq:WI1} we get
\be
\label{eq:WI2}
\int_0^1 dz\, (z(1-z))^{d-4 \over 2} z f^{(n\geq 1)}(z) =0\,.
\ee

Now, we turn back to the discussion of $\mathcal{K}_d(u,z)$ and mathematically derive the corresponding Ward identities \eqref{eq:kernel_ward_1} and \eqref{eq:kernel_ward_2}. Instead of directly integrating over variable $z$, it is more convenient to consider the integration of $\vec{n}_1$ and $\vec{n}_2$ on the celestial sphere $S^{d-2}$. Again, we emphasize that we have followed the previous convention to choose $p^\mu =(Q, \vec{0})$. From the definition \eqref{eq:kernel_def}, we notice that all the $\vec{n}_1$ and $\vec{n}_2$ dependence is contained in $((q\cdot n_1)(q\cdot n_2))^{-(d-1)}(4\chi^2 - 4\chi + {d-2 \over d-1})$. Our aim is to prove the following identities
\be
0&=\int_0^1 d\chi\, \left(4\chi^2 - 4\chi + {d-2 \over d-1}\right)
\int_{S^{d-2}}d^{d-2}\vec{n}_1 d^{d-2} \vec{n}_2 {\delta(\chi - {(n_1\cdot n_2) q^2 \over 2(q\cdot n_1)(q\cdot n_2)}) \over ((q\cdot n_1)(q\cdot n_2))^{d-1}} \,,\\
0&=\int_0^1 d\chi\, \left(4\chi^2 - 4\chi + {d-2 \over d-1}\right)
\int_{S^{d-2}}d^{d-2}\vec{n}_1 d^{d-2} \vec{n}_2 
{\delta(\chi - {(n_1\cdot n_2) q^2 \over 2(q\cdot n_1)(q\cdot n_2)}) \over ((q\cdot n_1)(q\cdot n_2))^{d-1}} (-n_1\cdot n_2)\,.
\ee
If $q^\mu$ has vanishing spatial components, these will reduce to the previous EEC discussion in the center-of-mass frame because polynomial $4\chi^2 - 4\chi + {d-2 \over d-1}$ satisfies the Ward identities. However, $q$ is an integration variable in $\mathcal{K}_d(u,z)$ and hence these identities are not evident to be true. We can verify these identities by directly performing the spherical integration and we find the following results
\be
&\int_{S^{d-2}}d\vec{n}_1 d \vec{n}_2 {\delta(\chi - {(n_1\cdot n_2) q^2 \over 2(q\cdot n_1)(q\cdot n_2)}) \over ((q\cdot n_1)(q\cdot n_2))^{d-1}}
={2^{d-3}\Omega_{d-1}\Omega_{d-2} \over (-q^2)^d}\left((q^0)^2 + \frac{1-2\chi}{d-1}|\vec{q}|^2\right)(\chi(1-\chi))^{d-4 \over 2}\,, \label{eq:w1_original}\\
&\int_{S^{d-2}}d\vec{n}_1 d \vec{n}_2 
{\delta(\chi - {(n_1\cdot n_2) q^2 \over 2(q\cdot n_1)(q\cdot n_2)}) \over ((q\cdot n_1)(q\cdot n_2))^{d-1}} (-n_1\cdot n_2)
=2^{d-2}\Omega_{d-1}\Omega_{d-2} {\chi (\chi (1-\chi))^{d-4 \over 2} \over (-q^2)^{d-1}}\,.\label{eq:w2_original}
\ee
They are linear combination of $(\chi (1-\chi))^{d-4 \over 2}$ and $\chi (\chi (1-\chi))^{d-4 \over 2}$ and therefore annihilate the $\chi$ integral.

We can also obtain \eqref{eq:w1_original} and \eqref{eq:w2_original} by promoting these integrals in a Lorentzian invariant way
\be
&W_1(\chi, {(p\cdot q)^2 \over p^2 q^2}) = \frac{(-q^2)^{d-1}}{-p^2}\int D^{d-2}n_1 D^{d-2}n_2 {(p\cdot n_1)(p\cdot n_2) \over ((q\cdot n_1)(q\cdot n_2))^{d-1}} \delta(\chi - {(n_1\cdot n_2) q^2 \over 2(q\cdot n_1)(q\cdot n_2)})\,,\\
&W_2(\chi) = (-q^2)^{d-1}\int D^{d-2}n_1 D^{d-2}n_2 {-n_1\cdot n_2 \over ((q\cdot n_1)(q\cdot n_2))^{d-1}} \delta(\chi - {(n_1\cdot n_2) q^2 \over 2(q\cdot n_1)(q\cdot n_2)})\,,
\ee
where the measure $D^{d-2}n_i \equiv {d^d n_i \theta(n_i^0) \delta(n_i^2) \over 2\mathrm{vol}\, \mathbb{R}_+}$ (see e.g.\ Eq.\ (2.27) in \cite{Kravchuk:2018htv} for a detailed discussion, but with a different normalization). For $W_2(\chi)$, we can choose the center of mass frame $\vec{q}=0$ so that it is similar to \eqref{eq:eec_ward_id1_f}
\be
W_2(\chi) = 2\chi \int_{S^{d-2}} d^{d-2}\vec{n}_1 d^{d-2}\vec{n}_2 \delta(\chi - \frac{1-\vec{n}_1\cdot \vec{n}_2}{2})=2^{d-2}\Omega_{d-1}\Omega_{d-2} \chi (\chi (1-\chi))^{d-4 \over 2}\,.
\ee

$W_1$ is more complicated due to the presence of an extra variable $y={(p\cdot q)^2 \over p^2 q^2}$. By acting with the differential operator $(p^\mu \partial_p^\nu - p^\nu \partial_p^\mu)^2$ on $W_1$, we can construct a differential equation
\be
(2(1-y)y \partial_y^2 + (1-d y)\partial_y + d)W_1(\chi,y) =  W_2(y)\,.
\ee
A general solution is 
\be
W_1(\chi, y) = \frac{W_2(\chi)}{d} + c_1(\chi) {y d-1 \over 2d - 1} + c_2(\chi) (y-1)^{3-d \over 2}\sum_{k=0}^\infty \#_k (y-1)^k\,.
\ee
The simplest boundary condition is that $W_1(\chi,y)$ should be regular at $y=1$ (or equivalently, $q=p$)
\be
W_1(\chi,1)= \int_{S^{d-2}} d^{d-2}\vec{n}_1 d^{d-2}\vec{n}_2 \delta(\chi - \frac{1-\vec{n}_1\cdot \vec{n}_2}{2})=2^{d-3}\Omega_{d-1}\Omega_{d-2} (\chi (1-\chi))^{d-4 \over 2} \,,
\ee
which constrains the coefficient functions
\be
c_2(\chi)=0\,,\qquad c_1(\chi) = {(2d-1)(d-2\chi) \over 2d(d-1)\chi} W_2(\chi)\,.
\ee
The final result for $W_1(\chi, y)$ is
\be
W_1(\chi,y) = 2^{d-3}\Omega_{d-1}\Omega_{d-2} (\chi (1-\chi))^{d-4 \over 2} 
{y d -1 -2(y-1)\chi \over d-1}\,,
\ee
which is the same as \eqref{eq:w1_original}.

\subsection{Asymptotic expansion}\label{app:AsymptoticExpansion}

The large impact parameter physics in AdS high-energy scattering is imprinted in the non-analytic terms at small $z$ in the EEC. The kernel $\mathcal{K}_d(u,z)$ plays an important role of converting the $u\to 1$ limit (or large impact parameter limit $L\to \infty$) to non-analyticity at small $z$. In terms of a region expansion, the non-analytic terms are generated from the integration region $1-u \sim z$ in \eqref{eq:finalformula}. Therefore, in this subsection, we consider the leading asymptotic expansion of $\mathcal{K}_d(u,z)$ in the limit $z,1-u\to 0$ while keeping the ratio $\xi = {1-u\over z}$ fixed. This result can be used to determine the leading coefficient of the non-analytic part in \eqref{eq:finalGR}. Though using the discontinuity of the kernel is a more convenient method to extract the subleading coefficients, considering asymptotic expansion is intuitive and provides an independent check for the overall normalization.

Let us consider the integral $u(1-u^2)^{-3/2}\tilde{I}^{d-5 \over 2}_{\alpha,\alpha}$ in \eqref{eq:kerneld}. The dominant contribution in the expansion comes from the region $1-r^2\sim z$ and $\phi\sim \sqrt{z}$. To perform the asymptotic expansion, we change variables from $(r,\phi)$ to 
\be
\rho = {1-r^2\over z}\,,\quad \Phi = \phi/\sqrt{z}\,,
\ee
such that the leading term of the integral simplifies
\be
& u(1-u^2)^{-3/2}\tilde{I}^{d-5 \over 2}_{\alpha,\alpha} \nonumber\\
=& z^{-1-\alpha}\left[
2^{3\alpha-d/2-1}\xi^{\alpha-d/2}\int_0^\infty d\rho \int_{-\infty}^\infty d\Phi {\rho^{d-5\over 2}\over (\Phi^2+\rho+2\xi)^{\alpha} ((\Phi-2)^2+\rho+2\xi)^{\alpha}}
+\mathcal{O}(z)\right]\,.
\ee
Using the Feynman parametrization 
\be
&(\Phi^2+\rho+2\xi)^{-\alpha} ((\Phi-2)^2+\rho+2\xi)^{-\alpha}\nonumber\\
=&{\Gamma(2\alpha)\over \Gamma(\alpha)^2}\int_0^1 dx {(x(1-x))^{\alpha-1}\over ((\Phi-2(1-x))^2+\rho+2\xi + 4x(1-x))^{2\alpha} }\,,
\ee
the integration becomes straightforward and we obtain
\be
& u(1-u^2)^{-3/2}\tilde{I}^{d-5 \over 2}_{\alpha,\alpha} \nonumber\\
=&{1\over z^{\alpha+1}}\left[
{2^{\alpha-2}\sqrt{\pi}\Gamma({d-3\over 2})\Gamma(2\alpha+1-{d\over 2})
\over
\xi^{\alpha+1} \Gamma(2\alpha)
} {}_2F_1(\alpha,2\alpha+1-{d\over 2},\alpha+\frac{1}{2};-{1\over 2\xi})
+\mathcal{O}(z)\right]\,.
\ee
Therefore, the asymptotic expansion of the $d$-dimensional kernel in this limit is 
\be
& {\mathcal{K}_d(u=1-z\xi,z) \over {(-p^2)^d \over (-p \cdot n_1)^{d-1} (-p \cdot n_2)^{d-1}} } = {\Omega_{d-3}\over (2\pi)^d}{2^d\sqrt{\pi}\Gamma({d-3\over 2}) \over z^d \xi^d } \left[
\frac{(d-2) \Gamma \left(\frac{3 d}{2}-1\right) \, _2F_1\left(d-1,\frac{3 d}{2}-1;d-\frac{1}{2};-\frac{1}{2 \xi }\right)}{4 \Gamma (2 d-1)}    \right. \nonumber \\
& \left. +\frac{2 \Gamma \left(\frac{3 d}{2}+3\right) \, _2F_1\left(d+1,\frac{3 d}{2}+3;d+\frac{3}{2};-\frac{1}{2 \xi  }\right)}{\xi ^2 \Gamma (2 d+2)} -\frac{\Gamma \left(\frac{3 d}{2}+1\right) \, _2F_1\left(d,\frac{3 d}{2}+1;d+\frac{1}{2};-\frac{1}{2 \xi }\right)}{\xi  \Gamma (2 d)}   \right] \label{eq:asym_Kd}
\ee

\subsection{Discontinuity of the kernel}\label{app:DiscKernel}

The integral over $\phi$ in the master integral $\tilde{I}_\gamma(\eta_1,\eta_2)$ in \eqref{eq:MasterIntegral} can be done explicitly and it is given by
\be 
\tilde{I}_\gamma(\eta_1,\eta_2) = 2\pi\int_0^1rdr\frac{(1-r^2)^{\frac{d-5}{2}}\Big[\frac{\eta_1}{\sqrt{\eta_1^2-r^2}}+\frac{\eta_2}{\sqrt{\eta_2^2-r^2}}\Big]}{\eta_1\eta_2+\sqrt{(\eta_1^2-r^2)(\eta_2-r^2)}-(1-2z)r^2}.
\ee 
and changing variables to $\eta=r^{-2}(\eta_1\eta_2+\sqrt{(\eta_1^2-r^2)(\eta_2^2-r^2)})$ one obtains the representation \cite{Correia:2020xtr}
\be\label{eq:tildeIResc}
\tilde{I}_\gamma(\eta_1,\eta_2)&=2\pi\int_{\eta_+}^\infty \frac{d\eta}{\eta-(1-2z)}\frac{(\eta^2-1)^{\frac{4-d}{2}}}{(\eta-\eta_+)^{\frac{5-d}{2}}(\eta-\eta_-)^{\frac{5-d}{2}}},\cr
\eta_\pm &= \eta_1\eta_2\pm\sqrt{\eta_1^2-1}\sqrt{\eta_2^2-1}.
\ee
We now rescale further $\eta\to \eta_+ s$ and inserting \eqref{eq:tildeIResc} into the representation of the kernel \eqref{eq:kerneld} we obtain
\be 
 \mathcal{K}_d(u,z)=\int_1^\infty ds\,{\cal \hat{K}}_d(u,z,s),
\ee 
with a computable function ${\cal \hat{K}}_d(u,z,s)$ given the equations above. 

The singularity is then obtained by taking the residue at $s=\frac{u^2 (1-2 z)}{2-u^2}$ and we can write it as follows
\be 
{\rm Disc}_z \mathcal{K}_d(u,z) =- \pi \theta(u-\frac{1}{\sqrt{1-z}}){\text{ Res}}_{s=\frac{u^2 (1-2 z)}{2-u^2}}{\cal \hat{K}}_d(u,z,s) .
\ee 
We did not include the contribution of higher order poles because they produce terms localized at $u={1 \over \sqrt{1-z}}$. Their effect can be correctly included by applying dimensional regularization to $\int_{\frac{1}{\sqrt{1-z}}}^1 du {\rm Disc}_z \mathcal{K}_d(u,z)$ which is formally divergent at $u=\frac{1}{\sqrt{1-z}}$. Equivalently, the discontinuity of the kernel is given by $\oint_{\frac{1}{\sqrt{1-z}}}^1 du$, where the integral is taken along the `keyhole' contour, see e.g. \cite{Correia:2020xtr}, which makes it manifestly regular.

\section{Relating the Regge expansion and the light-ray OPE}\label{app:ReggeLightRayOPE}

In this Appendix, we give a derivation of the structure
\be\label{eq:R_def}
R_\Delta(z)\equiv \int_0^1du \mathcal{K}_d(u,z) \mathcal{H}_{\Delta-1}(L(u))\sim f_\Delta(z)+(\text{analytic in $z$})\,,
\ee
where $f_\Delta(z)$ is the celestial block for the EEC \cite{Kologlu:2019mfz} and it is an eigenfunction of the celestial quadratic Casimir $C_2=-\frac{1}{2} M_{\mu\nu} M^{\mu\nu}$, where $M_{\mu\nu}$ are Lorentz generators.

Let us briefly review the concept of celestial blocks. In conformal theories, based on dimensionality and homogeneity, the EEC has the following functional form
\be
\langle \mathcal{E}(n_1) \mathcal{E}(n_2) \rangle = {(-p^2)^{d} \over (-p\cdot n_1)^{d-1} (-p\cdot n_2)^{d-1}} f(z)\,.
\ee
Now let us define the differential operator
\be
\mathcal{L}_x^{\mu\nu}=x^\mu \partial_x^\nu - x^\nu \partial_x^\mu\,,
\ee
and apply the celestial Casimir differential operator
\be
\mathcal{C}_2=-\frac{1}{2}(\mathcal{L}_{n_1}^{\mu\nu}+\mathcal{L}_{n_2}^{\mu\nu})^2\,,
\ee
to the energy correlator $\langle \mathcal{E}(n_1) \mathcal{E}(n_2) \rangle$. The result is
\be
\mathcal{C}_2 \langle \mathcal{E}(n_1) \mathcal{E}(n_2) \rangle
= &{(-p^2)^{d} \over (-p\cdot n_1)^{d-1} (-p\cdot n_2)^{d-1}}\big[
-4 (z-1) z^2 f''(z) \nn \\ 
& + 2 z (d (3-4 z)+2 z) f'(z)-2 (d-1) (2 (d-1) z-d) f(z)\big]
\ee
A light-ray operator $\mathbb{O}_{\Delta,J}(n)$, corresponding to a local (fictitious) operator with dimension $\Delta$ and spin $J$, is an eigenfunction to the celestial Casimir operator
\be
[\mathcal{C}_2, \mathbb{O}_{\Delta,J}(n)] = (\Delta-1)(\Delta-d+1) \mathbb{O}_{\Delta,J}(n)\,.
\ee
Therefore, the block structure associated with $\mathbb{O}_{\Delta,J}(n)$ in the light-ray OPE of $\mathcal{E}(n_1)\mathcal{E}(n_2)$ should satisfy the celestial Casimir equation
\be
-4 (z-1) z^2 f_\Delta''(z) + 2 z (d (3-4 z)+2 z) f_\Delta'(z)-2 (d-1) (2 (d-1) z-d) f_\Delta(z) \nn\\
=(\Delta-1)(\Delta-d+1)f_\Delta(z)\,,
\ee
whose solution is
\be
f_{\Delta}(z) = z^{{\Delta-(2d-1) \over 2}} \ _2 F_1({\Delta -1 \over 2}, {\Delta -1 \over 2}, \Delta - {d \over 2}+1, z) \ .
\ee

To show \eqref{eq:R_def}, we follow the same trick and apply the celestial Casimir differential operator $\mathcal{C}_2$ to $R_\Delta(z)$
\be
R_\Delta(z)=\int_{M^+} {d^d q\over (2\pi)^d} {G_0(q;n_1,n_2) \over (q\cdot n_1)^{d+1} (q\cdot n_2)^{d+1}} \delta(1-\sqrt{-q^2}\sqrt{-p^2})\mathcal{H}_{\Delta-1}\left(\mathrm{arccosh}({-p\cdot q \over \sqrt{p^2 q^2}})\right)\,.
\ee
Notice that all the $n_1,n_2$ dependence is contained in the Lorentz-invariant integrand ${G_0(q;n_1,n_2) \over (q\cdot n_1)^{d+1} (q\cdot n_2)^{d+1}}$. So the action of $\mathcal{C}_2$ is equivalent to
\be
\mathcal{C}_2 \left({G_0(q;n_1,n_2) \over (q\cdot n_1)^{d+1} (q\cdot n_2)^{d+1}}\right) = -\frac{1}{2} (\mathcal{L}_q^{\mu\nu})^2\left({G_0(q;n_1,n_2) \over (q\cdot n_1)^{d+1} (q\cdot n_2)^{d+1}}\right)\,.
\ee
Given that the integration region is also Lorentz-invariant, we can use integration by parts and move the action of $-\frac{1}{2} (\mathcal{L}_q^{\mu\nu})^2$ onto the function $\mathcal{H}_{\Delta-1}$:
\be
\mathcal{C}_2 R_\Delta(z) 
= -\frac{1}{2}\int_{M^+} {d^d q\over (2\pi)^d} {G_0(q;n_1,n_2) \over (q\cdot n_1)^{d+1} (q\cdot n_2)^{d+1}} \delta(1-\sqrt{p^2 q^2})  (\mathcal{L}_q^{\mu\nu})^2\mathcal{H}_{\Delta-1}\left(\mathrm{arccosh}({-p\cdot q \over \sqrt{p^2 q^2}})\right)\,.
\ee
On the other hand, the propagator $\mathcal{H}_{\Delta-1}$ satisfies
\be
\left(-\frac{1}{2}(\mathcal{L}_q^{\mu\nu})^2 -(\Delta-1)(\Delta-d+1)\right)
\mathcal{H}_{\Delta-1} = -\delta(p-q)\,,
\ee
which leads to the differential equation
\be
[\mathcal{C}_2 - (\Delta-1)(\Delta-d+1)]R_\Delta(z) = - {1\over (2\pi)^d} {G_0(p;n_1,n_2) \over (p\cdot n_1)^{d+1} (p\cdot n_2)^{d+1}}\,.
\ee
Again, we set $p^2=p\cdot n_i=-1$ for  convenience and then the differential equation becomes
\be\label{eq:diffeq_for_R}
\mathcal{D}_2 R_\Delta(z)= -{1\over (2\pi)^d} \left(4z^2 - 4z +{d-2 \over d-1}\right)\,,
\ee
where $\mathcal{D}_2$ is given by
\be 
{\cal D}_2=&4(1-z) z^2\partial^2_z-2z (2 (2 d-1) z-3 d)\partial_z\cr
&+ \left((-d-\Delta +1) (-2 d+\Delta +1)-4 (d-1)^2 z\right)\,.
\ee 

Neglecting the inhomogeneous term on the r.h.s,  \eqref{eq:diffeq_for_R} coincides with the celestial Casimir equation. Therefore the solution for $R_\Delta(z)$ is
\be\label{eq:resHInteg}
R_\Delta(z) =& \frac{\left((d-\Delta )^2-1\right) \Gamma \left(\frac{\Delta -1}{2}\right) \Gamma \left(\frac{\Delta +3}{2}\right)
   \Gamma \left(d-\frac{\Delta }{2}-\frac{1}{2}\right) \Gamma \left(\frac{1}{2} (d+\Delta -1)\right)}{2^{d+2} \pi^d \Gamma (d+1)^2 \Gamma
   \left(-\frac{d}{2}+\Delta +1\right)} f_{\Delta}(z) \cr
   &\quad +\frac{(d-2) (2 \pi )^{-d}}{(d-1) (-2 d+\Delta +1) (d+\Delta -1)}g_{\Delta}(z) \ .
\ee
Here the coefficient of the celestial block $f_\Delta(z)$ cannot be determined by \eqref{eq:diffeq_for_R}. However, this term fully captures non-analytic terms as $z \to 0$, so the coefficient can be calculated from the asymptotic expansion using \eqref{eq:asym_Kd}. The function $g_\Delta(z)$ is analytic at $z=0$ and the explicit result is 
\be\label{eq:defg}
g_{\Delta}(z) &= 1-z \frac{4 (d-1) \left(d^2+d \Delta -\Delta ^2-1\right)}{(d-2) (2 d-\Delta +1) (d+\Delta +1)} \\
&+ z^2 \frac{4 (d-1) \left(\Delta ^2-1\right)\left( (d-\Delta)^2-1\right) \, _3F_2\left(1,d+1,d+1;d-\frac{\Delta
   }{2}+\frac{5}{2},\frac{d}{2}+\frac{\Delta }{2}+\frac{5}{2};z\right)}{(d-2) (2 d-\Delta +1) (2 d-\Delta +3) (d+\Delta +1) (d+\Delta
   +3)} \nonumber \ . 
\ee

Note that the coefficient of $f_{\Delta}(z)$ in \eqref{eq:resHInteg} has simple poles at $\Delta=2d-1+2n$ for $n=0,1,\ldots$ which cancel with the second term involving $g_{\Delta}(z)$ and ends up producing a $\log z$ term 
\be\label{eq:LogIdent}
\int_0^1& du {\cal K}_d(z) {\cal H}_{(2d-1+2n)-1}(L(u))|_{\log z} \\ 
&= \frac{(-1)^{n+1} \left((d+2 n-1)^2-1\right) \Gamma (d+n-1) \Gamma (d+n+1) \Gamma \left(\frac{3 d}{2}+n-1\right)}{2^{d+2} \pi ^{d} n! \Gamma (d+1)^2 \Gamma \left(\frac{3
   d}{2}+2 n\right)}f_{2d-1+2n,d}(z)\nonumber.
\ee 
In particular these are the relevant dimensions from decomposing $(\Pi^{(GR)}(L))^2$ in terms of blocks ${\cal H}_{(2d-1+2n)-1}(L)$. 

\section{Tensor structures of the four-point function $\la {\cal O}T_{\mu\nu}T_{\rho\sigma}{\cal O }\ra$}\label{app:TensorStructures}

In this appendix, we spell out the assumption about the possible tensor structures of the correlator $\la {\cal O}T_{\mu\nu}T_{\rho\sigma}{\cal O }\ra$ made in the paper. 

We consider the $T \times T$ OPE channel. Given that we consider as a source a scalar primary operator ${\cal O }$, only symmetric traceless operators contribute to the OPE. The relevant three-point function takes the form \cite{Costa:2011mg,Costa:2011dw}
\be 
    \langle T(P_1,Z_1)T(P_2,Z_2){\cal O}_{\Delta,J}(P_3,Z_3)\rangle = \frac{\sum_{i=1}^{10}\lambda_i Q_i}{(P_{12})^{d+2-\frac{\Delta+J}{2}}(P_{23})^{\frac{\Delta+J}{2}}(P_{13})^{\frac{\Delta+J}{2}}},
\ee 
where $P_{ij}=-2P_i\cdot P_j$ and $Q_i$ are defined in terms of the building blocks 
\be 
 H_{ij} &= -2((Z_i\cdot Z_j)(P_i\cdot P_j)-(Z_i\cdot P_j)(Z_j\cdot P_i)),\cr
 V_{i;jk} &= \frac{(Z_i\cdot P_j)(P_i\cdot P_k)-(Z_i\cdot P_k)(P_i\cdot P_j)}{P_j\cdot P_k},
\ee 
and ($V_1\equiv V_{1;23}$,$V_2\equiv V_{2;31}$,$V_3\equiv V_{3;12}$)
\be 
Q_1&=V_1^2V_2^2V_3^J\cr 
Q_2&=(H_{23}V_1^2V_2+H_{13}V_1V_2^2)V_3^{J-1}\cr 
Q_3&=H_{12}V_1V_2 V_3^J\cr
Q_4&= H_{12}(H_{13}V_2+H_{23}V_1)V_3^{J-1}\cr 
Q_5&= H_{13}H_{23}V_1V_2V_3^{J-2}\cr
Q_6&= H_{12}^2 V_3^J\cr 
Q_7&=(H_{13}^2V_2^2+H_{23}^2V_1^2)V_3^{J-2}\cr
Q_8&=H_{12}H_{13}H_{23}V_3^{J-2}\cr 
Q_9&=(H_{13}^2H_{23}V_2+H_{13}H_{23}^2V_1)V_3^{J-3}\cr 
Q_{10}&=H_{13}^2H_{23}^2V_3^{J-4}.
\ee 
Imposing the conservation condition reduces the number of structures to be at most three, see Table 1 in \cite{Costa:2011mg}.\footnote{The counting of possible structures is different in $d=3$. In this case one can use the $\epsilon$-tensor when constructing conformal three-point functions \cite{Giombi:2011rz,Costa:2011mg}.} Three structures are possible for $J \geq 4$ or for conserved spin-two current. Notice that $J$ is necessarily even in our case.

When translated to the phase shift, these structures lead to three possible polarization dependence as explained in detail in \cite{Costa:2017twz}. In this paper, we have effectively considered a simplified case when the phase shift representation of the correlator is diagonal in the polarization space. In other words, contracting the stress tensor with $\epsilon^{\mu} \epsilon^{\nu}T_{\mu \nu}(p)$ and setting $\epsilon \cdot p = 0$, we assumed that the correlator is diagonal in the polarization space $\sim \epsilon^{\mu \nu} \epsilon^*_{\mu \nu}$.

In the language of \cite{Costa:2017twz}, we set $\beta_2 = \beta_3 = 0$. This picks a unique `holographic' three-point function, or, equivalently, the one that produces a diagonal phase shift, the same as the one in semi-classical Einstein gravity. In ${\cal N}=4$ SYM, this choice is dictated by supersymmetry. When ${\cal O}_{\Delta,J} = T$, the holographic three-point function corresponds to zero value of the Hofman-Maldacena parameters $t_2 = t_4 = 0$.

\section{Stringy Phase Shift}\label{App:StringyPhaseShift}
\begin{figure}[H]
    \centering
    \includegraphics[width=\textwidth]{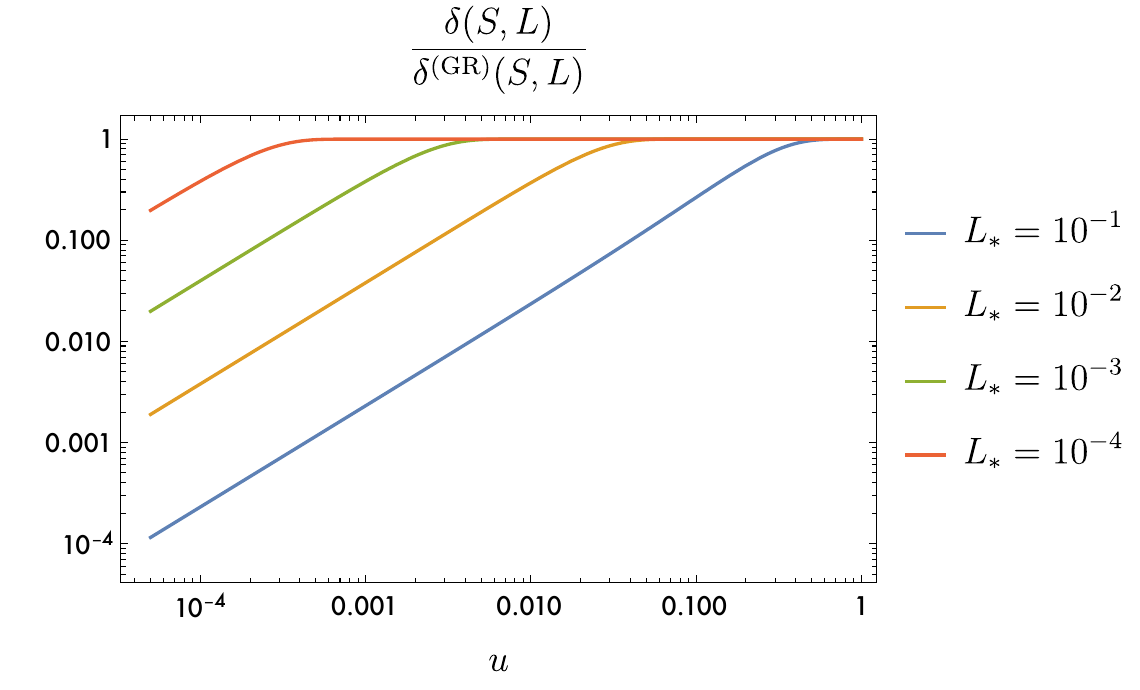}
    \caption{Here we plot the ratio $\frac{\delta(S,L(u))}{\delta^{{(\text{GR})}}(S,L)}$ from \eqref{eq:StringyPhaseExpl} for $L_*=10^{-k}$ with $k=1,2,3,4$. It is seen explicitly that the phase shift is dominated by the pure gravity result for $L\gtrsim L_*$, where we note that $u\approx L$ for small $L$.}
    \label{fig:ExplStringyPhase}
\end{figure}

Let us consider $d=4$ and compute the stringy phase shift explicitly. To leading order at high energies, we get the following integral
\be
\delta(S,L) \sim i S^{1-{8 \over \Delta_{gap}^2}} e^{-L}\int_{- \infty}^{\infty} d \nu S^{-{2 \nu^2 \over \Delta_{gap}^2}}{\nu  e^{-i \nu L}\over \nu^2 + 4} (1 + \coth L) 
\ee
Performing the integral we get
\be\label{eq:StringyPhaseExpl}
\delta(S,L) = \frac{\delta^{{(\text{GR})}}(S,L)}{2}\left(1+\text{erf}\left(\frac{L-8 L_*^2}{2 \sqrt{2} L_*}\right)-e^{4 L} \text{erfc}\left(\frac{L+8 L_*^2}{2 \sqrt{2}
  L_*}\right)\right),
\ee
where $\text{erf}(x)=\frac{2}{\sqrt{\pi}}\int_0^x e^{-t^2}dt$ and $\text{erfc}(x)=1-\text{erf}(x)$. For $L\gtrsim 8L^2_*\gg 1$ this approaches the pure gravity phase shift while for smaller impact parameters the stringy corrections are important. On the other hand, for $L_*\ll 1$ we see that for $ L \gtrsim  L_*$ the pure gravity phase shift is a good approximation but for smaller $L$, stringy corrections are important as shown in \Fig{fig:ExplStringyPhase}. 

\bibliographystyle{JHEP}
\bibliography{mybib}
  
\end{document}